\documentclass[letterpaper,usenatbib]{mn2e}
\usepackage[totalwidth=480pt, totalheight=680pt]{geometry}
\usepackage{amsmath,amssymb}
\usepackage{aas_macros}
\usepackage{graphicx}
\usepackage[compatibility=false,center]{caption} 
\usepackage{subcaption} 
\usepackage{fancyhdr}
\usepackage{url}
\usepackage{color}
\usepackage{bm}
\usepackage{placeins}

\def\beq{\begin{equation}}
\def\eeq{\end{equation}}
\def\bea{\begin{eqnarray}}
\def\eea{\end{eqnarray}}
\def\Pu{\mathbb{P}}
\def\Tu{\mathbb{T}}

\newcommand{\vecf}[1]{{\bm{#1}}}
\newcommand{\uvec}[1]{\vecf{\hat{#1}}}
\newcommand{\matf}[1]{\bm{\mathrm{#1}}}

\newcommand{\kmsMpc}{$\text{km }\text{s}^{-1}\text{Mpc}^{-1}$}

\newcommand{\dTsz}{\delta T^\text{SZ}}
\newcommand{\dTpa}{\delta T^\text{PA}}

\begin{document}
\bibliographystyle{mn2e.not_for_mnras}

\title[Optimal SZ measurement]{An Optimal and Model-Independent Measurement of the Intracluster Pressure Profile \MakeUppercase{\romannumeral1}: Methodology and First Applications}

\author[S. Aslanbeigi, G. Lavaux, A. Hajian \& N. Afshordi]{
Siavash Aslanbeigi$^{1,2}$, Guilhem Lavaux$^{1,2}$, Amir Hajian$^{3}$, Niayesh Afshordi$^{1,2}$\\
$^{1}$ Department of Physics \& Astronomy, University of Waterloo, Waterloo, ON,  N2L 3G1 Canada\\
$^{2}$ Perimeter Institute for Theoretical Physics, 31 Caroline St. N., Waterloo, ON N2L 2Y5, Canada \\
$^{3}$ Canadian Institute for Theoretical Astrophysics, University of Toronto, Toronto, ON M5S 3H8, Canada
}

\maketitle


\begin{abstract}
We present a statistically-optimal and model-independent method to extract the pressure profile of hot gas in the intracluster medium (ICM). 
Using the thermal Sunyaev-Zeldovich effect, we constrain the mean pressure profile of the ICM by appropriately considering \textit{all} primary cosmic microwave background (CMB) and instrumental noise correlations, while using the maximum resolution and sensitivity of all frequency channels. 
As a first application, we analyze 
CMB maps of WMAP $9$-year data through a study of the Meta-Catalogue of X-ray detected Clusters of galaxies (MCXC). 
We constrain the universal pressure profile out to $4R_{500}$ with $15\sigma$ confidence, though our measurements are only significant out to $R_{200}$.
Using a temperature profile constrained from X-ray observations, we measure the mean gas mass fraction out to $R_{200}$.
Within statistical and systematic uncertainties, our constraints are compatible with the cosmic baryon fraction and the expected gas fraction in halos.
While \textit{Planck} multi-frequency CMB data are expected to reduce statistical uncertainties by a factor of $\sim20$, 
we argue that systematic errors in determining mass of clusters dominate the uncertainty in  gas mass fraction measurements at the level of $\sim20$ percent.
\end{abstract}


\section{Introduction}
Clusters of galaxies have long been recognized as remarkable laboratories to test cosmological theories.
They are the largest gravitationally bound objects in the universe, 
thought to have formed from the eventual collapse of initially overdense dark matter perturbations. 
Their abundance and large scale properties are sensitive to the expansion and initial conditions  of the universe, making them excellent tools to constrain cosmological models. 
On smaller scales, the physics of clusters is dominated by complex baryonic processes such as gas cooling, star formation, and feedback from supernovae and active galactic nuclei. 
In order to use clusters as standard probes of the geometry and dynamics of the universe, it is necessary to reliably model these processes and distinguish amongst different feedback mechanisms
\citep[see][for recent reviews]{Kravtsov2012, Allen2011}. 

The main baryonic budget of clusters is a
hot plasma of ionized hydrogen and helium in the intracluster medium (ICM), making it the natural target for studying the complex astrophysical processes at play. 
This virialized plasma emits bremsstrahlung radiation in X-ray, making it possible to probe the dense regions of the ICM.
\footnote{This is because X-ray luminosity is proportional to the \textit{square} of gas density \citep[e.g.][]{weinberg2008cosmology}.}
Historically, most of our observational understanding of the ICM has come from X-ray observations, leading to 
a fairly consistent picture of the scaling and structural properties of low-redshift clusters 
\citep[see e.g.][]{xray1, vikhlinin2008chandra, arnaud2010universal, xray2, xray3, xray4}. 

%

The thermal Sunyaev-Zeldovich (tSZ) effect \citep{Sunyaev:1972eq} is another important probe of the ICM: as the Cosmic Microwave Background (CMB) photons inverse-Compton scatter off of the hot electrons in the ICM, their blackbody spectrum is distorted.
The tSZ effect has the unique property that its signal is independent of redshift, making it a powerful observational tool for detecting clusters at cosmological redshifts, and hence a promising cosmological probe of dark energy \citep[e.g.][]{birkinshaw:1999,carlstrom/holder/reese:2002}.
Within the last few years, cluster surveys exploiting
the SZ effect have
started delivering cluster samples \citep[e.g.,][]{staniszewski/etal:2009,
marriage/etal:2011, williamson/etal:2011, planck/esz:2011} as well as constraints on cosmological parameters
\citep{vanderlinde/etal:2010, sehgal/etal:2011}.
Another important feature of the tSZ effect is that it is directly proportional to the integrated pressure of free electrons along the line of sight, which makes it a 
powerful probe of the ICM in the outskirts ($r\gtrsim R_{200}$),
where X-ray emission is extremely faint.

Resolving the tSZ signal for individual clusters requires high resolution CMB measurements, which have become available only in the recent years. In fact, the only all-sky CMB survey with high enough resolution {\it and}  sensitivity to detect individual SZ clusters is \textit{Planck}. Even with \textit{Planck}'s sensitivity, it is necessary to combine the tSZ signal from many clusters to meaningfully constrain 
physical quantities of interest, such as baryonic mass fraction \citep{collaboration2012planck}.
Luckily, there is fairly strong evidence from X-ray observations and numerical simulations that clusters are self-similar to a good approximation \citep[see e.g.][]{Nagai,arnaud2010universal}. 
This fact justifies combining SZ signatures of many clusters to obtain constraints on the \textit{mean} ICM properties. There have been quite a few efforts in this direction over the past few years.
By analyzing WMAP-1 (WMAP-3 respectively) CMB data for 116 (193 respectively) X-ray detected clusters, \cite{afshordi1, afshordi2} (respectively) provided constraints on the ICM pressure profile out to $\sim R_{200}$.
Other similar works include  
WMAP-3 stacking of over 700 clusters by \cite{Atrio-Barandela}, 
WMAP-5 analysis of about 900 ROSAT NORAS/REFLEX clusters \citep{Melin},
WMAP-7 analysis of 175 \textit{Planck} ESZ clusters \citep{MaESZ},
and SZ measurements of 15 massive X-ray selected clusters obtained with the South Pole Telescope \citep{PlaggeSPT}.
Most notably, \cite{collaboration2012planck} have studied the tSZ signal of 62 low-redshift massive clusters by using CMB data from the \textit{Planck} satelite, constraining the 
mean pressure profile of the ICM out to $3R_{500}$ with unprecedented precision. 

The practice of averaging signals from many clusters goes under the title ``stacking''. 
The basic idea is the following: the main sources of uncertainty in extracting the tSZ signal are the primary CMB anisotropies and instrumental noise. 
Since these sources of noise are random in nature, they ``drop out'' if the temperature profile around many clusters is averaged over.
This procedure, in its various forms, is not statistically optimal for multiple reasons. Firstly, large-angle correlations of primary CMB fluctuations are ignored when stacking. 
Secondly, it is not clear how contributions from different clusters should be optimally weighed in the averaging process. Typically, different weighing methods are adopted to see
whether the effect on the extracted tSZ signal is significant or not \citep[e.g.][]{Atrio-Barandela}. Thirdly, when using multiple frequency channels, the final resolution of the reconstructed tSZ map is determined by the \textit{lowest} resolution of the combined frequency maps \citep[e.g.][]{collaboration2012planck}. 
Finally, the 3D pressure profiles are usually obtained a posteriori by deprojecting the tSZ signal, which may lead to noise amplification.
For these reasons, stacking procedures either result in an underestimation of error or loss of statistical information.

We believe all the aforementioned shortcomings of stacking procedures can be overcome with the methodology we have formulated in this paper, which is more in line with
optimized template fitting procedure of \cite{WMAP7}.
Our analysis includes an all-sky multi-channel fit to the mean pressure profile of the ICM which appropriately takes into account primary CMB and noise correlations on \textit{all} scales, while
using the maximum resolution and sensitivity of all channels to their full potential. 
Furthermore, following \cite{afshordi2}, our method is completely model-independent, thus eliminating any systematic uncertainty associated with theoretical modelling of the ICM.
In this paper, we apply our formalism to WMAP-9 CMB data using the Meta-Catalogue of X-ray detected Clusters of galaxies \citep{piffaretti2011mcxc}. 
In a companion paper, we will repeat our measurements using CMB data from the \textit{Planck} mission. 

Our paper is organized as follows. In Sections \ref{tSZmodel} and \ref{UPP}, we review the tSZ effect and the concept of a universal pressure profile, describing how
model-independence can be achieved. Sections \ref{Statistical Methods} and \ref{Numerical Methods} contain our main statistical and numerical methodology,
outlining in detail how the mean pressure profile of ICM can be optimally constrained. Section \ref{Data} describes the CMB data and cluster sample we use to test our methodology, and is followed by 
a discussion of the resulting pressure profiles in Section \ref{pResults}. Section \ref{fGas} presents gas mass fraction measurements of various subsamples of our cluster catalogue. We discuss
future work and how we anticipate our results to improve by using \textit{Planck} CMB data in Section \ref{discussion}, before concluding our findings in Section \ref{conclusions}.

Throughout this paper, we assume a $\Lambda$CDM cosmology with present matter density $\Omega_m=0.3$, dark energy density $\Omega_{\Lambda}=0.7$, and Hubble parameter $H_0=100~h$~\kmsMpc with $h=\frac{7}{10}h_{70}=0.7$. We also denote the normalized Hubble parameter at redshift $z$ by $E(z)\equiv\frac{H(z)}{H_0}=\sqrt{\Omega_m(1+z)^3+\Omega_{\Lambda}}$. 
\section{Extracting the Pressure Profile}

This section contains the statistical and numerical methodology we use to extract the mean ICM pressure profile from a full-sky CMB experiment. Section~\ref{tSZmodel} reviews the tSZ effect and how it is related to the electron pressure profile. In Section~\ref{UPP}, we reduce the problem of finding the exact profile of each cluster to a single, universal up to normalization, pressure profile. In Section~\ref{Statistical Methods}, we derive the maximum likelihood estimator of the profile and its covariance matrix. Finally, in Section~\ref{Numerical Methods}, we describe how the components of the estimator are in practice computed numerically.

\subsection{tSZ Effect Model}
\label{tSZmodel}
The contribution of the tSZ effect to the CMB temperature anisotropy at frequency $\nu$ and location $\uvec{n}$ on the sky is proportional to the integral of the electron pressure along the line of sight: (see $\S 2.5$ of \cite{weinberg2008cosmology} for a derivation)
\begin{align}
	\dTsz(\uvec{n}; \nu)&=\frac{\sigma_\mathrm{T} T_\mathrm{CMB}}{m_\mathrm{e} c^2}F\left(\frac{h\nu}{k_\mathrm{B} T_\mathrm{CMB}}\right)\int dl_{\uvec{n}}P_\mathrm{e}(l_{\uvec{n}}),\nonumber  \\
	F(x)&\equiv x\coth(x/2)-4,
\label{SZ_effect}
\end{align}
where $\sigma_\mathrm{T}$ is the Thomson scattering cross-section, $m_e$ is the mass of the electron, $c$ is the speed of light, $k_B$ is the Boltzmann constant, $h$ is the Planck constant, \footnote{We've used $h$ to denote Planck's constant only in Equations \eqref{SZ_effect} and \eqref{templates}. Throughout the rest of our paper, $h$ is the reduced Hubble constant.}  $T_{CMB}=2.725 \text{ K}$ is the mean CMB temperature \citep{COBE}, and $P_\mathrm{e}(l_{\uvec{n}})$ is the pressure of free electrons along the line of sight direction $\uvec{n}$.

Our task is to constrain $P_\mathrm{e}$ through the tSZ effect. We will assume that $P_\mathrm{e}$ is spherically symmetric to a good approximation and denote the pressure profile of the $a^\text{th}$ cluster by $P^{(a)}_\mathrm{e}(r)$. Furthermore, since it is not possible to constrain a continuous function without introducing model-dependence, we consider spherical bins around the centre of clusters, in each of which the pressure is assumed to be constant:
\begin{equation}
 P^{(a)}_\mathrm{e}(r)=
    \begin{cases}
      P^{(a)}_1 & \text{if } 0<r<r^{(a)}_1\\
      P^{(a)}_2 & \text{if } r^{(a)}_1<r<r^{(a)}_2\\
      \vdots\\
      P^{(a)}_{N_b} & \text{if } r^{(a)}_{N_b-1}<r<r^{(a)}_{N_b}.
    \end{cases}
\end{equation}
Here $P^{(a)}_1,\dots,P^{(a)}_{N_b}$ are all constants with units of pressure, 
$r$ is the radius away from the centre of cluster, 
and $N_b$ is the total number of bins. 
The value of the pressure in each bin may be better understood as the volume-weighed average of the pressure in that bin.
With these simplifications, Equation~\eqref{SZ_effect} may be written as
\beq
\dTsz(\uvec{n}; \nu)=\sum_{k=1}^{N_b}\sum_{a=1}^{N_c}P^{(a)}_kt^{(a)}_k(\uvec{n}; \nu),
\label{SZdisc}
\eeq
where $N_c$ is the total number of clusters and $t^{(a)}_k(\uvec{n}; \nu)$ is given by:
\begin{multline}
t^{(a)}_k(\uvec{n}; \nu)=\frac{\sigma_\mathrm{T} T_\mathrm{CMB}}{m_\mathrm{e} c^2}F\left(\frac{h\nu}{k_\mathrm{B} T_\mathrm{CMB}}\right)\times 2 \\
    \begin{cases}
      l^{(a)}_{k+1}(\uvec{n})-l^{(a)}_k(\uvec{n}) & \text{if }  w^{(a)}(\uvec{n}) \le r^{(a)}_k\\
      l^{(a)}_{k+1}(\uvec{n}) & \text{if } r^{(a)}_k\le w^{(a)}(\uvec{n}) \le r^{(a)}_{k+1}\\
      0 & \text{if } w^{(a)}(\uvec{n})\ge r^{(a)}_{k+1}.
    \end{cases}
    \label{templates}
\end{multline}
\begin{figure}
	\begin{center}
		\includegraphics[trim=20 210 20 210, clip=true, width=\hsize]{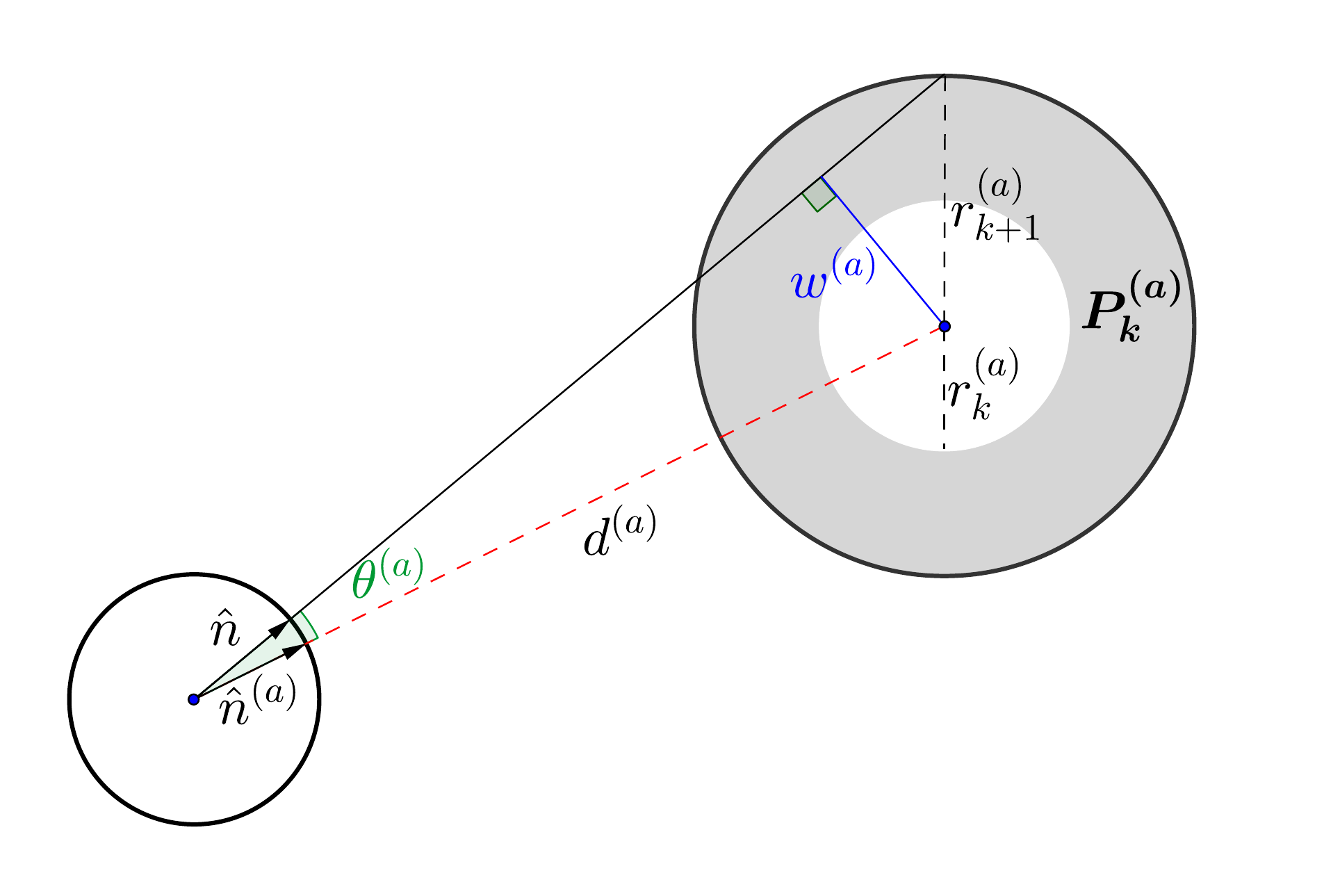}
	\end{center}
	\caption{The $k^{\text{th}}$ bin of the $a^\text{th}$ cluster. The contribution of this bin to the temperature anisotropy of the CMB at frequency $\nu$ and location $\uvec{n}$ on the sky is given by $P^{(a)}_k t^{(a)}_k(\uvec{n}; \nu)$, where $t^{(a)}_k(\uvec{n}; \nu)$ is defined in Equation~\eqref{templates}. \label{fig1}}
\end{figure}
The functions used in Equation~\eqref{templates} are defined as follows:
\begin{subequations}
  \begin{align}
	  l^{(a)}_k(\uvec{n}) &= \sqrt{[r^{(a)}_{k}]^2-[w^{(a)}(\uvec{n})]^2}, \\
	  w^{(a)}(\uvec{n}) &\equiv d^{(a)}\sin(\theta^{(a)}(\uvec{n})), \\
      \cos(\theta^{(a)}(\uvec{n}))&=\uvec{n}\cdot\uvec{n}^{(a)},
  \end{align}\label{geometry}
\end{subequations}
where $d^{(a)}$ is the angular diameter distance to the $a^\text{th}$ cluster and  $\uvec{n}^{(a)}$ is the unit vector pointing to its centre.%
\footnote{
Angular diameter distance to the $a^{\text{th}}$ cluster with redshift $z$ is given by $d^{(a)}(z)=\frac{c/H_0}{1+z}\int_{0}^{z}\frac{dy}{\sqrt{\Omega_{\Lambda}+\Omega_m(1+y)^3}}$.
}
Fig.~\ref{fig1} shows the basic geometry that underlies  Equations~(\ref{SZdisc}-\ref{geometry}).

\subsection{Universal Pressure Profile}
\label{UPP}
In principle, the analysis that will follow can be used to optimally estimate all parameters $P^{(a)}_1,\dots,P^{(a)}_{N_b}$.
Unfortunately, this is too computationally-intensive for a large sample of clusters, given the large angle correlations of primary CMB anisotropies and the current resolution of CMB experiments. However, there is fairly concrete evidence that the pressure profile of the hot gas in clusters is self-similar \citep[see e.g.][]{Nagai,arnaud2010universal}. 
This means that for a given cluster, there is a self-similarity
scale $r_c^{(a)}$ such that the pressure profile takes the form
\beq
P^{(a)}_\mathrm{e}(r)=\Pu(r/r_\mathrm{c}^{(a)})P_\mathrm{c}^{(a)},
\label{univP}
\eeq
where $P_\mathrm{c}^{(a)}$ is a constant characteristic pressure of the $a^{\text{th}}$ cluster, and $\Pu(x)$ is the so-called {\it universal pressure profile}. 
Within this context, it is natural to construct the
radial bins so that $r_k^{(a)}=n_kr_c^{(a)}$, where $\{n_k\}_{k=1-N_b}$ are positive numbers
satisfying $n_1<n_2<\cdots<n_{N_b}$. With these assumptions, discretization of the electron pressure profile amounts to $P_k^{(a)}=\Pu_kP_c^{(a)}$, and the 
tSZ contribution \eqref{SZdisc}  to the CMB anisotropy takes the form
\begin{subequations}
  \label{SZdisc2}
  \begin{align}
	\dTsz(\uvec{n}; \nu) &=\sum_{k=1}^{N_b}\Pu_kt^{(\nu)}_{k}(\uvec{n}), \\
	t^{(\nu)}_{k}(\uvec{n})&=\sum_{a=1}^{N_c}P_c^{(a)}t^{(a)}_k(\uvec{n}; \nu).
  \end{align}
\end{subequations}
If the characteristic scales  $r_c^{(a)}$ and $P_c^{(a)}$ are fixed by external observations (such as X-ray's) for each cluster, our task is simplified to finding best estimate values (and their associated uncertainties) for $N_b$ parameters:
$\Pu_1,\dots,\Pu_{N_b}$.

We use $R^{(a)}_{500}$ as the self-similarity length scale of the $a^{\text{th}}$ cluster (i.e. we set $r^{(a)}_c= R^{(a)}_{500}$).
The quantity $R^{(a)}_{\Delta}$ is defined as the radius up to which the matter density is $\Delta$ times the critical mass-density of the universe:
\begin{align}
	M^{(a)}_{\Delta}&\equiv\int_{0}^{R^{(a)}_{\Delta}}\rho^{(a)}_{m}(r)4\pi r^2dr\notag\\
	&=\Delta\times\frac{4}{3}\pi [R_{\Delta}^{(a)}]^3\times\rho_{crit}(z_a),
\label{DeltaDef}
\end{align}
where $z_a$ is the redshift of the $a^{\text{th}}$ cluster, and $\rho_{crit}(z)=\frac{3H(z)^2}{8\pi G}$. 
We consider $8$ bins with radii $r^{(a)}_{k}=0.5k\times R_{500}^{(a)}$, where $k\in\{1,2,\dots,8\}$. This is equivalent to setting $n_k=0.5k$.
We use two proposals for the characteristic cluster-dependent pressure $P_c^{(a)}$:
\begin{multline}
  P^{(a)}_c=1.65\times 10^{-3}E(z_a)^{8/3}h_{70}^2  \\ 
     \left[\frac{M^{(a)}_{500}}{3\times10^{14}h_{70}^{-1}M_\odot}\right]^{2/3+\delta}\text{keV}~\text{cm}^{-3},
     \label{Pc}
\end{multline}
with $\delta=0$ and $\delta=0.12$. The former corresponds to the mass variation expected in the standard self-similar model based purely on gravitation 
\citep[see][]{Nagai,arnaud2010universal}. The $\delta=0.12$ case is a modification to the standard self-similar model proposed by
\cite{arnaud2010universal}, which is a first approximation to quantifying how the mass scaling of $P_c^{(a)}$ changes with radius in the REXCESS sample \citep{xray1}.
Using the $\delta=0.12$ scaling makes for a meaningful comparison of our results with those of \cite{collaboration2012planck}, since this is what is used in their analysis.

\subsection{Statistical Methods}
\label{Statistical Methods}
We use the principle of maximum likelihood to find best-estimate values for the parameters $\Pu_1,\dots,\Pu_{N_b}$. \citep[For template fitting on CMB sky, see e.g.][]{Gorski1996, Jaffe2004, WMAP7}.
We assume that the only contributions to the temperature anisotropies of the CMB are the primordial anisotropies $\dTpa$, the tSZ effect $\dTsz$ (given by Equation \eqref{SZdisc2}), possible residual monopole and dipole components, and the instrumental noise. 
Furthermore, we assume that primordial anisotropies follow the statistics of an isotropic Gaussian random field, for which we know the angular power spectrum $C_\ell$. 
\footnote{
We use the CAMB code to generate the expected values of $C_\ell$'s for the WMAP concordance $\Lambda$CDM cosmology \citep{Lewis:1999bs, Howlett:2012mh}.
}

Let $L_p$ ($L_{\nu}$) denote the set of all pixels (frequency channels) we wish to use in our analysis.
In Appendix~\ref{Appendix Statistical methods} we show that the log-likelihood of the observed temperature fluctuations $\{\delta T_{i\nu}\}_{i\in L_{p}}^{\nu\in L_{\nu}}$, given the binned pressure profile $\{\Pu_k\}$, is analytic and equal to
\begin{multline}
    -\frac{1}{2}\chi^2(\{\delta T_{i\nu}\}_{i\in L_{p}}^{\nu\in L_{\nu}}|\{\Pu_k\})=\\
      -\frac{1}{2}(\vecf{\delta T}-\vecf{\dTsz})^T\matf{C}^{-1}(\vecf{\delta T}-\vecf{\dTsz}), \label{eq:log_likelihood} 
\end{multline}
where
\begin{equation}
  \matf{C} =\matf{C_S}+\matf{C_N}.
\end{equation}
Here $\matf{C_S}$ is the covariance matrix of the primary CMB fluctuations and $\matf{C_N}$ is the covariance of the instrumental noise. The former is related to the angular power spectrum $C_\ell$ through
\begin{multline}
[\matf{C_S}]_{i\nu,i'\nu'}\equiv\langle\dTpa_{i\nu}\dTpa_{i'\nu'}\rangle= \\
 \sum_{\ell=0}^{\ell_{max}}\left(\frac{2\ell+1}{4\pi}\right)C_\ell B_{\ell\nu}B_{\ell\nu'}W_\ell^2P_\ell(\uvec{n}_i\cdot\uvec{n}_{i'}),
 \label{signalCov}
\end{multline}
where $B_{\ell\nu}$ is the spherically averaged beam transfer function for the mode $\ell$ and frequency channel $\nu$, $W_\ell$ is the spherically averaged pixel transfer function, $\uvec{n}_i$ is the sky direction of the $i^{\text{th}}$ pixel, and $P_\ell$ is the $\ell$-th degree Legendre polynomial.
We use $\ell_{max}=2\times N_{side}$ throughout our analysis, where $N_{side}=512$ is set by the HEALPix \citep{HEALPIX} resolution of the WMAP sky maps.

In the case of WMAP, the instrumental noise is largely uncorrelated both between pixels and between different frequency channels. Its properties are adequately modelled by a Gaussian distribution with covariance matrix
\beq
  [\matf{C_N}]_{{i\nu,i'\nu'}}=n_{i\nu}^2\delta_{ii'}\delta_{\nu\nu'}.
  \label{C_N}
\eeq
In Equation~\eqref{eq:log_likelihood}, we have used $\dTsz_{i\nu}$ to stand for a pixelized version of Equation~\eqref{SZdisc2}:
\begin{subequations}
  \begin{align}
\dTsz_{i\nu} &=\sum_{k=1}^{N_b}\Pu_kt^{(\nu)}_{k}(i), \\
t^{(\nu)}_k(i)&=\sum_{l=0}^{l_{max}}\sum_{m=-l}^{l}(t^{(\nu)}_k)_{lm}B_{l\nu}W_lY_{lm}(\uvec{n}_i),
   \end{align}
\end{subequations}
where $(t^{(\nu)}_k)_{lm}$ are the spherical harmonic coefficients of $t^{(\nu)}_k(\uvec{n})$.
In practice, we generate the templates at HEALPix resolution 12 ($N_\text{side}=4096$), and then downgrade to resolution 9 ($N_\text{side}=512$).%
\footnote{
To be more specific, following the notation of Section \ref{UPP}, we give the value $t^{(\nu)}_k(\uvec{n}_i)$ to the $i$-th pixel of the $k$-th template in frequency channel $\nu$, where $\uvec{n}_i$ is the centre of the $i$-th pixel at a resolution 12. (At this resolution, the first radial bin of all MCXC clusters occupies at least one pixel.)
We then downgrade the templates to resolution 9 and finally convolve all templates with the instrumental beams.
}
We then convolve all templates with instrumental beams to obtain the quantities $t^{(\nu)}_{k}(i)$. 

Let us note again that $\Pu_k$ is the universal pressure of the $k^{\text{th}}$ bin and $t^{(\nu)}_k(i)$ is the coefficient that multiplies it at pixel $i$ and frequency channel
$\nu$. 
Assuming a uniform prior on $(\hat{\Pu}_1,\dots,\hat{\Pu}_{N_b})$, the posterior probability function of these variables is also a Gaussian distribution which peaks at the maximum of the log-likelihood function given in Equation~\eqref{eq:log_likelihood}, which is
\begin{subequations}
	\label{bestFit}
	\begin{align}
		\hat{\Pu}_k&=\sum_{k'=1}^{N_b}[\matf{\alpha}^{-1}]_{k,k'}\beta_{k'},\\
		\alpha_{k,k'}&=\sum_{\nu\in L_{\nu}}\sum_{i\in L_p}t^{(\nu)}_k(i)X^{(\nu)}_k(i), \\
		\beta_{k}&=\sum_{\nu\in L_{\nu}}\sum_{i\in L_p}\delta T_{i\nu}X^{(\nu)}_k(i),
	\end{align}
\end{subequations}
where we have introduced the inverse covariance weighed template
\beq
X^{(\nu)}_k(i)=\sum_{\nu'\in L_{\nu}}\sum_{i'\in L_p}[\matf{C}^{-1}]_{i\nu,i'\nu'}t^{(\nu')}_{k}(i').
\label{X}
\eeq
Finally, the covariance matrix $\matf{C_\Pu}$ of $\{\Pu_k\}$ is determined by the Hessian of the log-likelihood \eqref{eq:log_likelihood}. Its matrix elements are
\beq
[\matf{C_\Pu}]_{k,k'}=\left\langle\left(\hat{\Pu}_k-\Pu_k\right)\left(\hat{\Pu}_{k'}-\Pu_{k'}\right)\right\rangle=[\matf{\alpha}^{-1}]_{k,k'}.
\label{twopointfunc}
\eeq
This matrix does not involve further computations as it is already required to obtain the best estimates $\{\hat{\Pu}_k\}$. All the measurements on the pressure profiles and their attached error bars are obtained using only the expressions indicated in this Section. 

In Appendix \ref{monDipFit}, we have shown how any residual monopole or dipole contribution can be conveniently accounted for in this formalism.

\subsection{Numerical Methods}
\label{Numerical Methods}
It is clear from Equation~\eqref{bestFit} that all quantities of interest can be calculated once the weighed templates $\vecf{X^{(\nu)}_k}$ are known. This is impossible to achieve by direct computation, which would involve inversion of the full covariance matrix $\matf{C}$.
In this section, we will describe how we compute $\vecf{X^{(\nu)}_k}$ numerically. 

We start by establishing some notation.
The set of all masked (unmasked) pixels is denoted by 
$L_{\bar{p}}$ ($L_{p}$), so that $L=L_{\bar{p}}\cup L_{p}$ contains all pixels in the sky. 
We denote the total number of pixels (i.e. the size of $L$) by $N_T$.

In the statistical modelling described in Section~\ref{Statistical Methods}, the covariance matrices $\matf{C_S}$ and $\matf{C_N}$ are only computed on the observed pixels $L_{p}$. However, numerical manipulation of these matrices is more efficient in harmonic space, using the spherical harmonic transform, which itself requires the knowledge of all pixels. 
Therefore, it is advantageous to compute the quantities of interest 
by extending the domain of $\matf{C_S}$ and $\matf{C_N}$ to the entire sky. 
We refer the reader to Appendix \ref{masking} for details on how this can be achieved and state the final result here \citep[see also][]{WLL04,LAH13}. 

We encode information about masking of pixels into a diagonal $N_T \times N_T$ matrix $\matf{M}$, with elements $M_{ii}=0$ if $i \in L_{\bar{p}}$, and $M_{ii}=1$ if $i \in L_p$. 
Furthermore, we let $\matf{S}$ be the full pixel to pixel covariance matrix due to primary CMB fluctuations: ($i,j\in L$)
\beq
[\matf{S}]_{ij}=\sum_{\ell=0}^{\ell_{max}} C_\ell \left(\frac{2\ell+1}{4\pi}\right) P_\ell(\uvec{n}_i\cdot\uvec{n}_{j}),
\eeq
 $\matf{N}_\nu$ the pixel-to-pixel covariance matrix of the instrumental noise:
\begin{equation}
	[\matf{N}_{\nu}]_{ij}=n_{i\nu}^2\delta_{ij}, \label{N_nuij_def} 
\end{equation}
and $\matf{B}_\nu$ a model of the complete beam (pixelization and instrumental):
\begin{equation}
	\label{beamMatrix}
	[\matf{B}_{\nu}]_{ij}=A_\text{pix}\sum_{\ell=0}^{\ell_{max}}\left(\frac{2\ell+1}{4\pi}\right)B_{\ell\nu}W_\ell P_\ell(\uvec{n}_i\cdot\uvec{n}_{j}),
\end{equation}
where $A_{pix}$ is the area of one pixel (which is equal to $4\pi / N_T$ for all pixels in the HEALPix scheme). We further define the square root of $\matf{S}$:
\begin{equation}
  [\matf{S}^{1/2}]_{ij} \equiv 
\sqrt{A_{pix}} \sum_{\ell=0}^{\ell_{max}}\sqrt{C_\ell}\left(\frac{2\ell+1}{4\pi}\right) P_\ell(\uvec{n}_i\cdot\uvec{n}_{j}).
\end{equation}
Equation~\eqref{bestFit} may now be rewritten using the full covariance matrices, temperature data, and templates:
\begin{subequations}
	\label{bestFit2}
	\begin{align}
		\hat{\Pu}_k &=\sum_{k'=1}^{M}(\matf{\alpha}^{-1})_{k,k'}\beta_{k'}, \\
		\alpha_{k,k'} &=\sum_{\nu\in L_{\nu}}\sum_{i\in L}t^{(\nu)}_k(i)X^{(\nu)}_k(i), \\
		\beta_{k} &=\sum_{\nu\in L_{\nu}}\sum_{i\in L}\delta T_{i\nu}X^{(\nu)}_k(i).
	\end{align}
\end{subequations}
where 
\begin{subequations}
    \begin{align}
		X^{(\nu)}_k(i) &=\sum_{\nu'\in L_{\nu}}\sum_{i'\in L}[G_{\nu,\nu'}]_{ii'}t^{(\nu')}_{k}(i'), \\
		\matf{G}_{\nu,\nu'} &=\matf{M} \matf{N}_{\nu}^{-1}\matf{M}\delta_{\nu,\nu'}- \nonumber \\
		            &\matf{M} \matf{N}_{\nu}^{-1}\matf{M} \matf{B}_{\nu}\matf{S}^{1/2}\matf{D}^{-1}\matf{S}^{1/2}\matf{B}_{\nu'}\matf{M} \matf{N}_{\nu'}^{-1}\matf{M}, \label{Gdef}\\
		\matf{D} & =\matf{1}+\matf{S}^{1/2}\left(\sum_{\nu\in L_{\nu}}\matf{B}_{\nu}\matf{M} \matf{N}_{\nu}^{-1}\matf{M} \matf{B}_{\nu}\right)\matf{S}^{1/2}.
   \end{align}
   \label{finalEq}
\end{subequations}

The computation of $\vecf{X}^{(\nu)}_k$ is now reduced to solving for the quantities $\vecf{g}^{(\nu)}_{k}=\matf{D}^{-1}\vecf{\tilde{t}}^{(\nu)}_{k}$, where $\vecf{\tilde{t}}^{(\nu)}_{k}\equiv \matf{S}^{1/2}\matf{B}_{\nu}\matf{M}\matf{N}_{\nu}^{-1}\matf{M}\vecf{t}^{(\nu)}_{k}$. The other operations may be done trivially as all involved operators are either diagonal in pixel space or in harmonic space. Computing $\vecf{g}^{(\nu)}_{k}$ is equivalent to solving the equation $\matf{D} \vecf{g}^{(\nu)}_{k}=\vecf{\tilde{t}}^{(\nu)}_{k}$, for which a number of numerical techniques are available. We use the algorithm of the conjugate gradient method with preconditioning \citep[see e.g.][]{Shewchuk:1994:ICG:865018}, which is an iterative prescription for solving large linear systems equations of the type
\begin{equation}
	\matf{A} \vecf{x} = \vecf{y}.
\end{equation}

To speed up the convergence, we construct a preconditioner matrix $\matf{D}_0^{-1}$ (essentially an approximation of $\matf{D}^{-1}$) as follows: the block corresponding to all harmonic modes with $\ell \leq 60$ is taken to be the exact inversion of the same block in $\matf{D}$, which is computed using a Cholesky decomposition. The rest of $\matf{D}_0^{-1}$ is taken to be diagonal, the elements of which are reciprecals of the corresponding diagonal element of $\matf{D}$. This preconditioner has already been used in other works \citep[see e.g.][]{EOJW04}. We stop the conjugate gradient algorithm whenever the relative error 
\begin{equation}
	\epsilon_n = \frac{||\matf{A} \vecf{x}_n - \vecf{y}||_2}{||\vecf{y}||_2}
\end{equation}
is less than a specified threshold. In the case of this work, we take $\epsilon_n< 10^{-6}$. We have checked that changing this threshold to $10^{-5}$ does not change the results, indicating that the solution has indeed converged (see Appendix \ref{convergence} for detailed convergence tests).

Finally, we note that all $\{\vecf{g^{(\nu)}_{k}}\}$ are fully independent and thus may be computed in parallel. 
We fully employ this property. Our software, {\sc ABYSS} (the sphericAl BaYesian Statistical Sampler), runs in 25 hours and 52 minutes on seven nodes (16 cores) to solve for the 12 templates on an Intel Xeon E5620. We note that the monopole and dipole take significant more time to reach the same level of precision as the other maps.

\section{Data}
\label{Data}
\subsection{CMB Data}
We use co-added inverse-noise weighted data from nine-year maps observed by WMAP at 41 GHZ (Q-band), 62 GHZ (V-band), and 94 GHz (W-band).%
\footnote{
\url{http://lambda.gsfc.nasa.gov}
}
These maps are foreground cleaned \citep{WMAP9_MAP} and are at HEALPix
resolution $9$ ($N_{side} = 512$). The standard deviation of the pixel noise in each map is given by 
(using notation of Section \ref{Numerical Methods})
\beq
n_{i\nu}=\frac{\sigma_0^{(\nu)}}{\sqrt{N^{obs}_i}},
\eeq
where $\nu\in L_{\nu}=\{Q,V,W\}$, $i\in L$, and $\sigma_0^{(Q)}=2.188$ mK, $\sigma_0^{(V)}=3.131$ mK, $\sigma_0^{(W)}=6.544$ mK.
The number of observations $N^{obs}_i$ at pixel $i$ is included in the maps available from the LAMBDA website.
In all of our analysis, we use the `extended temperature data analysis mask' to exclude foreground-contaminated regions of the sky from the analysis.
The beam transfer function for every differencing assembly is also provided on the LAMBDA website.
For a single value of $\ell$, we average beam transfer function values for all differencing assemblies belonging to the same frequency channel. This is how we obtain the quantities $B_{\ell\nu}$ introduced in Section \ref{Statistical Methods}.

\subsection{Cluster Sample}
\label{Cluster Sample}
We use the Meta-Catalogue of X-ray detected Clusters of galaxies (MCXC) to extract the universal pressure profile of the ICM 
\citep{piffaretti2011mcxc}.
\footnote{
All information about MCXC clusters may be found here: \url{http://vizier.cfa.harvard.edu/viz-bin/VizieR?-source=J/A+A/534/A109}
}
The MCXC provides (amongst other quantities) sky coordinates, redshift, and $M_{500}$ data for all clusters.
With a few exceptions, luminosity is used as a mass proxy for all clusters \citep[see equation (2) of][]{piffaretti2011mcxc}. 

\begin{figure}
\begin{center}
	\includegraphics[width=\hsize]{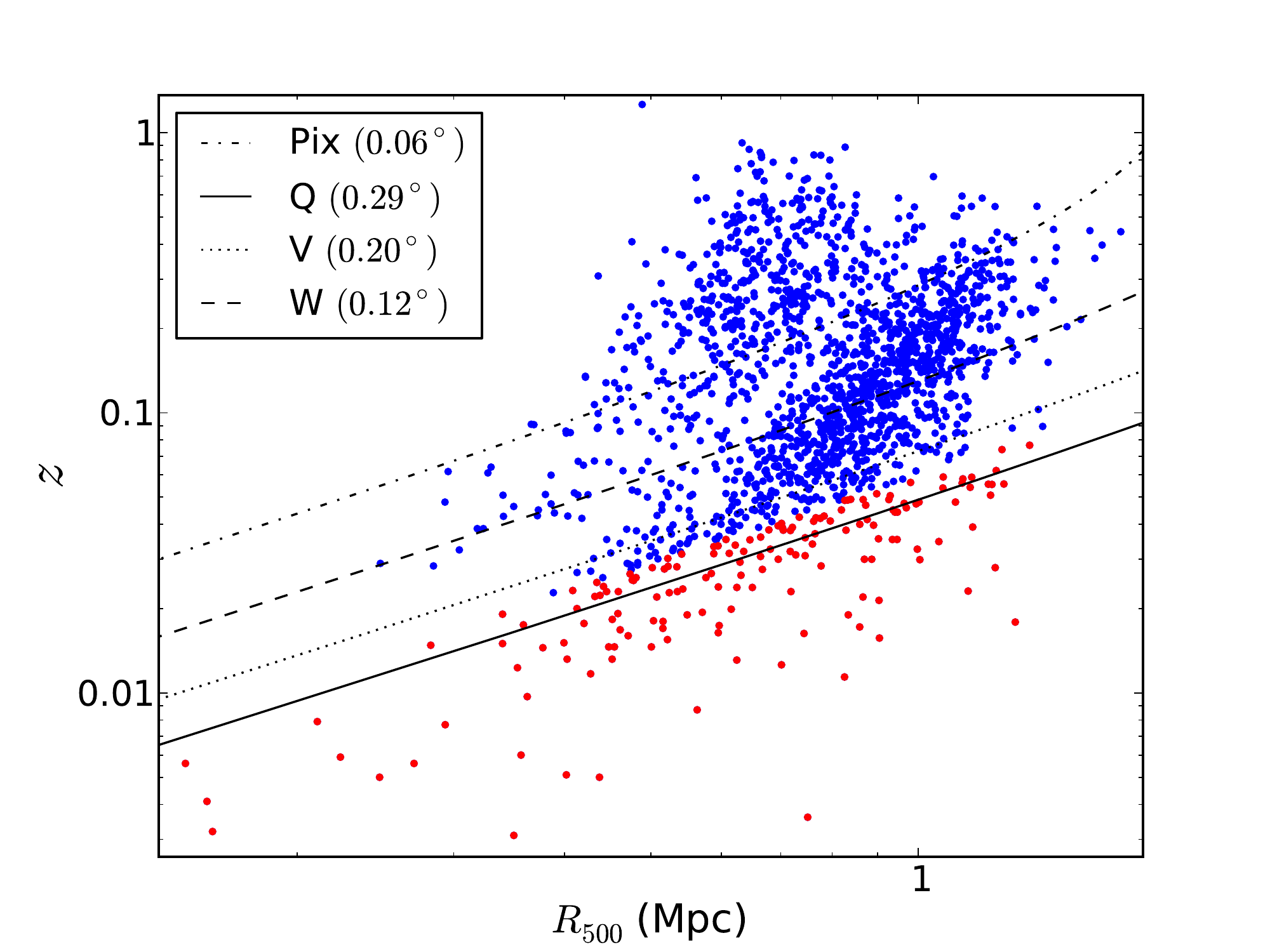}
\end{center}
\caption{
The redshift-$R_{500}$ distribution of MCXC clusters. 
The red points represent clusters whose first radial bin is resolved by the $W$ frequency channel of WMAP.
The three lines plot $d(z)\theta^{(\nu)}$ for different WMAP frequency bands, where $\theta^{(\nu)}=\sqrt{\Omega^{(\nu)}/\pi}$ is the angular radius of the disk with the same effective area as the detector beam in frequency channel $\nu$.
Here $\Omega^{(\nu)}$ is the beam solid angle of frequency channel $\nu$, which is provided on the LAMBDA website: $\Omega^{(Q)}=0.51^2$, $\Omega^{(V)}=0.35^2$, and $\Omega^{(W)}=0.22^2$ ($\text{deg}^2$). The curve labeled `Pix' is constructed similarly and reflects the resolution associated with pixelization. 
}
\label{MCXC}
\end{figure}

We perform our analysis on all 1743 MCXC clusters, as well as a subsample of 162 clusters  
whose first radial bin ($=0.5\times R_{500}$, as discussed in the next Section) is resolved by the $W$ frequency channel of WMAP. 
More specifically, we obtain this subsample by requiring $d(z)\theta^{(W)}<0.5R_{500}$, where $\theta^{(W)}=0.12^{\circ}$ is the effective angular radius of the (averaged) $W$-channel detector beam.
We will refer to clusters in this subsample as \textit{resolved MCXC clusters}. 
Fig. \ref{MCXC} shows the redshift-$R_{500}$ distribution of all MCXC clusters, differentiating between the resolved and unresolved ones.
The redshift of all (resolved) MCXC clusters
range from $0.0031-1.26$ ($0.0031-0.077$) with a median of $0.14$ ($0.028$), and their masses range from $\frac{M_{500}}{10^{14}M_{\odot}}=0.0096-22.1$ $(0.0096-7.27)$ with a median of $1.77$ ($0.86$).

Since most MCXC clusters cannot be resolved, one expects numerical uncertainties to become important. This is why we have chosen to study a subsample in which all clusters are resolved. 
However, even unresolved clusters contribute to the tSZ signal, especially in the outer bins. Therefore, the price one pays for ignoring unresolved clusters is statistical information. 
We have analyzed both samples to see how this trade-off manifests itself in practice. 

We also analyze subsamples of MCXC clusters binned according to mass. 
This allows us to study the dependence of various quantities, such as pressure and gas mass fraction, on the mass of clusters.
Table \ref{tab:massBin} shows the mass range and number of clusters in every bin. We have subdivided the resolved MCXC clusters into three mass bins, and the entire MCXC sample into four bins. These bins have been chosen so that they lead to roughly similar signal-to-noise properties, characterized by the null chi-squared of pressure measurements. We have excluded the $45$ most massive clusters because none of them are resolved, resulting in a measurement with extremely low significance and a nearly degenerate covariance matrix. 

\begin{table}
	\caption{\label{tab:massBin}  Binning MCXC clusters according to their mass.}
	\begin{subtable}{\hsize}
		\centering
		\begin{tabular}{ccc}
      			\hline
      			Bin number& $M_{500}$  range $(10^{14} M_{\odot}$) & Number of  clusters \\
     			 \hline
      			1 & 0.0096-2.41 & 1140  \\
      			2 & 2.41-4.17 & 364  \\
      			3 & 4.18-5.31 & 124 \\
			4 & 5.32-7.27 & 70 \\
      			\hline
   		 \end{tabular}
		\caption{\label{tab:massBin_all} All but the $45$ most massive MCXC clusters. }
	\end{subtable}%
	\\
	\begin{subtable}{\hsize}
		\centering
		\begin{tabular}{ccc}
      			\hline
      			Bin number& $M_{500}$  range $(10^{14} M_{\odot}$) & Number of  clusters \\
     			 \hline
      			1 & 0.0096-2.71 & 138  \\
      			2 &2.83-4.56 & 15  \\
      			3 & 5.17-7.27 & 9 \\
      			\hline
   		 \end{tabular}
		\caption{\label{tab:massBin_162} Resolved MCXC clusters.}
	\end{subtable}%
\end{table}

\section{Results}

In this section, we describe our two main results. Section~\ref{pResults} discusses WMAP constraints on the universal pressure profile $\Pu$, and Section~\ref{fGas} includes our gas mass fraction measurements. 

\subsection{WMAP Constraints on the Universal Pressure Profile of the ICM}
\label{pResults}
\begin{figure}
	\begin{subfigure}[b]{\hsize}
        		\includegraphics[width=\hsize]{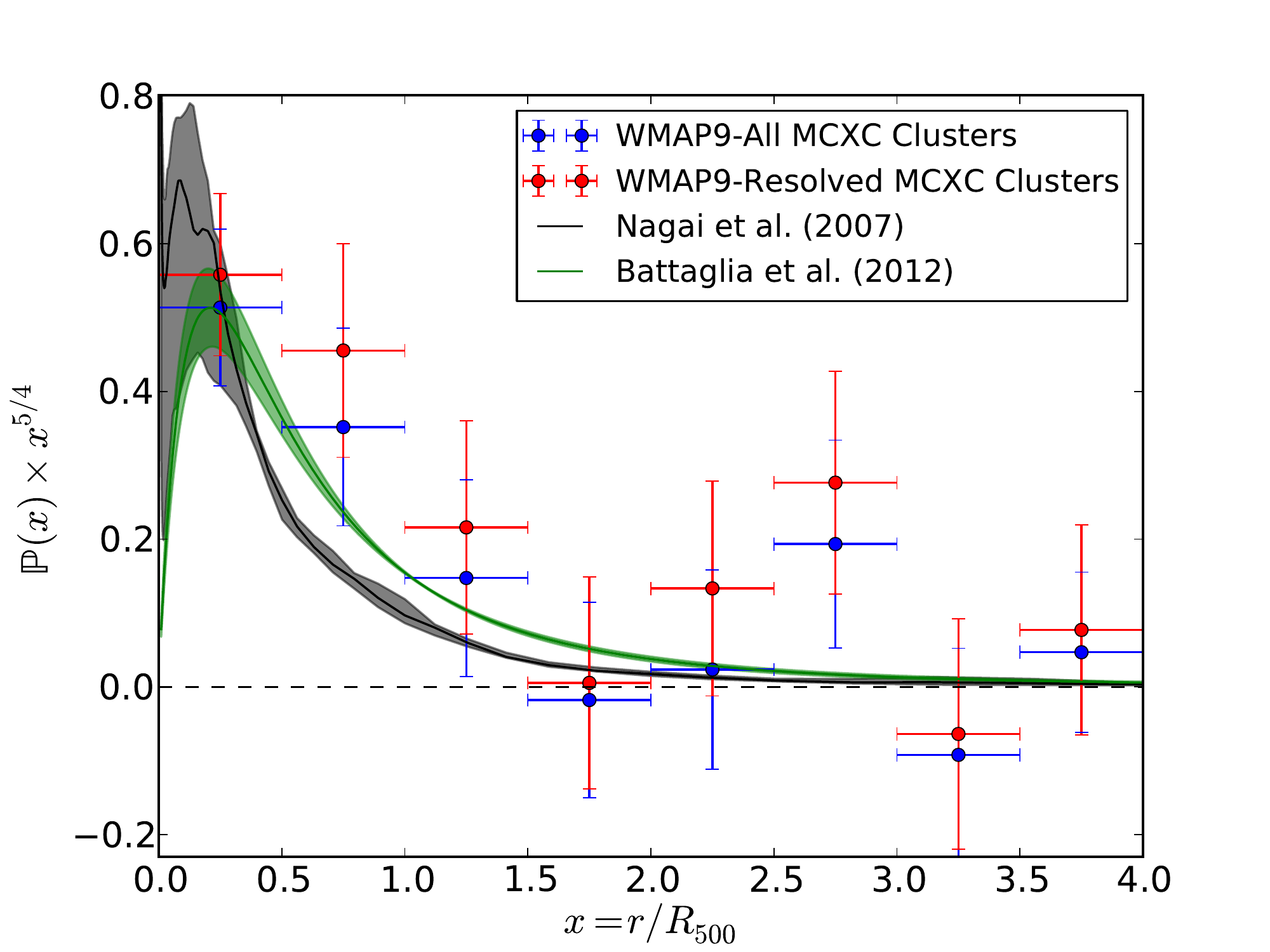}
		\caption{Standard self-similar scaling of pressure with mass ($\delta=0$ in Equation \eqref{Pc}). The shaded areas mark the dispersion about the mean profiles of simulated clusters from \cite{Nagai} (gray), and \cite{Battaglia} (green).}	
		\label{delta0}	
    	\end{subfigure} %
	\begin{subfigure}[b]{\hsize}
        		\includegraphics[width=\hsize]{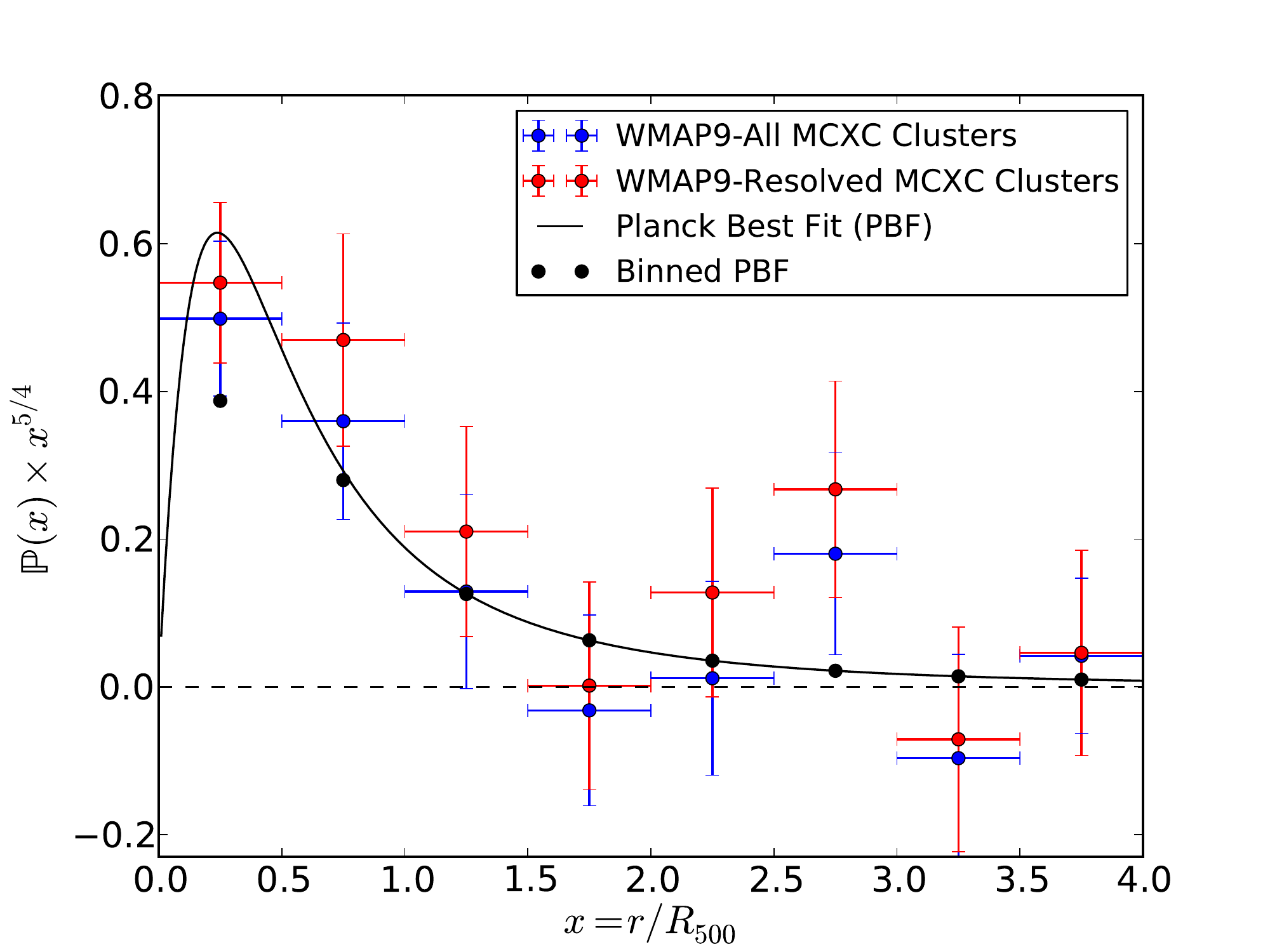}
		\caption{Modified self-similar scaling of  pressure with mass ($\delta=0.12$ in Equation \eqref{Pc}). The black curve is the best fit GNFW profile to pressure measurements of \textit{Planck}. The black points are obtained by a volume weighed average of \textit{Planck}'s best fit profile over our radial bins.
		}
		\label{delta12}	
    	\end{subfigure} %
	\caption{
	WMAP-9 constraints on the universal pressure profile $\Pu$ of the ICM. 
	The blue (red) data points are the resulting pressure profiles for all (only resolved) MCXC clusters. A cluster is considered resolved if its first radial bin subtends a solid angle larger than the effective beam area of the $W$ frequency channel (see Section \ref{Cluster Sample}). 
	Defined in Equation \eqref{Pc}, $\delta$ characterizes deviation from the standard self-similar model. $\delta=0$ corresponds to the mass variation expected in the standard 		self-similar model \citep[see][]{Nagai,arnaud2010universal}, and $\delta=0.12$ is a modification which better captures the variation of mass scaling with radius 
	in the REXCESS sample \citep{arnaud2010universal, xray1}.
	}
	\label{pressureWMAP9}
\end{figure}

Fig. \ref{pressureWMAP9} shows the result of our pressure measurements as applied to all MCXC clusters, as well as the resolved subsample defined in Section \ref{Cluster Sample}. 
As was mentioned in Section \ref{UPP}, we perform our analysis using the standard self-similar model $P_c^{(a)}\propto M_{500}^{2/3}$ ($\delta=0$ in Equation \eqref{Pc}), as well as the modified 
scaling $P_c^{(a)}\propto M_{500}^{2/3+0.12}$ ($\delta=0.12$ in Equation \eqref{Pc}). These results are shown in Fig. \ref{delta0} and Fig. \ref{delta12}, respectively.
There is essentially no signal beyond $1.5 R_{500}$. The best fit pressure values even become negative for some bins in this regime. We have decided not to impose positivity of pressure as a prior in order to keep the statistics Gaussian and not spoil the analytic results \eqref{bestFit} and \eqref{twopointfunc}. Repeating these measurements with \textit{Planck} CMB data is expected to provide a \textit{significantly} tighter constraint on the universal pressure profile (see Section \ref{discussion} below).

In Fig. \ref{delta0}, we have compared our pressure measurements with two sets of simulations \citep{Nagai, Battaglia},
which include treatment of radiative cooling, star formation and energy feedback from supernova explosions. \cite{Battaglia} also account for feedback from active galactic nuclei, 
while \cite{Nagai} consider the effect that electrons and ions are not kept in thermal equilibrium in the outskirts \citep{Rudd}.
Comparison of our measurements with the simulated profiles of \cite{Nagai} is straightforward because they compute the exact same quantity. \cite{Battaglia}, however, use $R_{200}$ as the self-similarity scale and also consider the variation of $\Pu$ with mass and redshift. In this case, we use the $c_{200}-M_{200}$ relation of \cite{bhattacharya2011dark} to compute $R_{200}$, and use the fitting formula of \cite{Battaglia} (equations (11-12) and table 1) to compute $\Pu$ for all MCXC clusters. 
In Fig. \ref{delta0} we have plotted in green the average of these profiles, as well as the standard deviation about their mean.
Where there is signal, our pressure measurements are slightly more consistent with those of \cite{Battaglia}.
Due to the large statistical uncertainties, however, we cannot meaningfully discriminate between the two. 
In an upcoming companion paper, we will return to this question in greater detail once we apply our methodology to \textit{Planck} CMB data.

\begin{table}
	\caption{\label{tab:chiSqD} Level of detection for various pressure measurements. $\delta=0$ ($0.12$) corresponds to measurements presented in Fig. \ref{delta0} (\ref{delta12}), respectively. The null chi-squared is given by $\chi_0^2=\hat{\vecf{\Pu}}^T\matf{C}_{\Pu}^{-1}\hat{\vecf{\Pu}}$, where $\hat{\vecf{\Pu}}$ are the best fit pressure measurements and $\matf{C}_{\Pu}$ is their associated covariance matrix. The level of detection is calculated for 8 degrees of freedom, i.e. number of radial bins.}	
	
	\begin{tabular}{lcc}
      		\hline
      		Measurement& $\chi^2_0$ & Detection ($\sigma$) \\
     		\hline
      		All MCXC clusters, $\delta=0$ & 259.3 & 15.1 \\
      		All MCXC clusters, $\delta=0.12$ & 262.2 & 15.2 \\
      		Resolved MCXC clusters, $\delta=0$ & 115.6 & 9.5 \\
		Resolved MCXC clusters, $\delta=0.12$ & 118.6 & 9.6 \\
      		\hline
   	 \end{tabular}	
\end{table}

In Fig. \ref{delta12}, we have compared our pressure measurements with those of \cite{collaboration2012planck}.
Given that we use different radial bins, and more importantly that \textit{Planck}'s measurements are a lot more precise, it suffices to compare our measurements with their best fit generalized Navarro-Frenk White (GNFW) profile \citep{NFW, Nagai}. We discretize this profile by a volume-weighed average over our radial bins, which makes for a more meaningful comparison with the discretized universal pressure profile we have defined, i.e. $\{\Pu_k\}$. Our measurements are in good agreement. 

\begin{table*}
	 \caption{\label{tab:normedCp} The normalized covariance matrix of the universal pressure profile for all MCXC clusters (blue or top-right), as well as only the resolved ones (red or bottom-left). The modified self-similar model is used for these measurements (i.e. $\delta=0.12$ in Equation \eqref{Pc}). To construct these matrices, let $\matf{C}_{\Pu}^{\text{all}}$ be the covariance matrix for analysis done on all MCXC clusters.  Construct a diagonal matrix $\matf{\Delta}^{\text{all}}$ such that $\matf{\Delta}_{k,k}^{\text{all}}=\sqrt{[\matf{C}_{\Pu}^{\text{all}}]_{k,k}}$, where $k$ runs over the different radial bins. We define the normalized covariance matrix via $\matf{D}^{\text{all}}=[\matf{\Delta^{\text{all}}}]^{-1}\matf{C}_{\Pu}[\matf{\Delta^{\text{all}}}]^{-1}$, which normalizes all diagonal elements of $\matf{C}_{\Pu}^{\text{all}}$ to one. By the same construction, let $\matf{D}^{\text{res}}$ be the resulting normalized covariance matrix for analysis done on resolved MCXC clusters. The blue or top-right (red or bottom-left) numbers in this table denote the off-diagonal elements of $\matf{D}^{\text{all}}$ ($\matf{D}^{\text{res}}$), respectively.}

	\begin{center}
	\begin{tabular}{c|c|c|c|c|c|c|c|c}
		\hline
		blue &  Bin 1 & Bin 2 & Bin 3 & Bin 4 & Bin 5 & Bin 6 & Bin 7 & Bin 8 \\
\hline
 Bin 1 & 1.000 &\color{blue} -0.727 &\color{blue} 0.282 &\color{blue} -0.082 &\color{blue} 0.029 &\color{blue} -0.022 &\color{blue} 0.011 &\color{blue} -0.001 \\
 Bin 2 & \color{red} -0.612 &1.000 &\color{blue} -0.671 &\color{blue} 0.214 &\color{blue} -0.051 &\color{blue} 0.005 &\color{blue} -0.003 &\color{blue} 0.006 \\
 Bin 3 & \color{red} 0.277 &\color{red} -0.497 &1.000 &\color{blue} -0.633 &\color{blue} 0.144 &\color{blue} 0.000 &\color{blue} -0.023 &\color{blue} 0.018 \\
 Bin 4 & \color{red} -0.069 &\color{red} 0.232 &\color{red} -0.445 &1.000 &\color{blue} -0.598 &\color{blue} 0.109 &\color{blue} 0.009 &\color{blue} -0.014 \\
 Bin 5 & \color{red} 0.033 &\color{red} -0.039 &\color{red} 0.147 &\color{red} -0.434 &1.000 &\color{blue} -0.594 &\color{blue} 0.117 &\color{blue} -0.004 \\
 Bin 6 & \color{red} -0.024 &\color{red} -0.007 &\color{red} -0.013 &\color{red} 0.091 &\color{red} -0.451 &1.000 &\color{blue} -0.597 &\color{blue} 0.156 \\
 Bin 7 & \color{red} -0.001 &\color{red} -0.023 &\color{red} -0.041 &\color{red} -0.015 &\color{red} 0.108 &\color{red} -0.442 &1.000 &\color{blue} -0.697 \\
 Bin 8 & \color{red} -0.011 &\color{red} -0.043 &\color{red} -0.044 &\color{red} -0.075 &\color{red} -0.053 &\color{red} 0.085 &\color{red} -0.476 &1.000 \\

		\hline
	\end{tabular}
	\end{center}
\end{table*}

Table \ref{tab:chiSqD} shows the level of confidence for our various pressure measurements. The difference between the standard and modified scalings of $P_c^{(a)}$ with mass is very small. However, the significance of detection reduces from $15.1\sigma$ to $9.5\sigma$ if we limit our sample to the resolved clusters. 
This may seem surprising because, looking at Fig. \ref{pressureWMAP9}, the uncertainties are similar in both cases and the best fit pressure values are even consistently higher in the case of resolved clusters. Note, however, that Fig. \ref{pressureWMAP9} does not compare the off-diagonal elements of the covariance matrices $-$ i.e. correlation between different bins. {\it In fact, the extra statistical information coming from unresolved clusters is encoded almost entirely in the off-diagonal correlations.}
We refer the reader to Appendix \ref{app:ExactNums}
for the full covariance matrix and numerical values of $\{\hat{\Pu}_k\}$.
To get some sense for the nature of correlations, however, we have shown a normalized version of $\matf{C}_{\Pu}$ for both samples in Table \ref{tab:normedCp}. 
Nearby bins are anti-correlated in both cases, but more so for the sample containing all clusters. The extra information contained in these anti-correlations can be quantitatively described by examining the eigenvalues and eigenvectors of the covariance matrix $\matf{C}_{\Pu}$. 
We refer the reader to Appendix \ref{EigenBusiness} for a detailed discussion of this point and state the results here.
In the case of the resolved clusters, the eigenmodes with the three largest eigenvalues are responsible for most of the contribution to $\chi_0^2$.
For the whole MCXC sample, however, all eigenmodes contribute more or less equally. Moreover, eigenvectors corresponding to larger eigenvalues carry most of their weight from the inner bins. 
Therefore, in the case of resolved clusters, mostly the inner bins are contributing to the signal, whereas for the whole MCXC sample, there is also contribution from outer bins.
This analysis reassures us that even unresolved clusters contribute to the tSZ signal in the outskirts of the ICM.

Although the unresolved clusters add to the tSZ signal in the outer bins, one expects numerical uncertainties associated with them. This is especially worrisome for those on sub-pixel scales, where certain approximations, such as a spherically averaged pixel transfer function, break down. In order to get an estimate for how large such uncertainties are, we performed our analysis on all MCXC clusters using higher resolution WMAP sky maps ($N_{side}=1024$). 
The result is shown in Appendix \ref{DiffRes}. For all radial bins, this discrepancy is at most at the 1$\sigma$ level and is random in nature.

Our pressure measurements are also affected by the uncertainty present in determining masses of clusters. 
In Appendix \ref{massChange}, we have investigated this issue by considering 62 MCXC clusters which are also in the Early Release SZ (ESZ) sample \citep{ESZ}. 
The ESZ mass estimates are systematically higher on average by about $12$ percent. This results in systematically lower pressure measurements (where there is actual signal), but
it is only at the 1$\sigma$ level (see Appendix \ref{massChange} for details). We will return to this issue in Section \ref{fGas}, because this effect is no longer small when determining gas mass fraction.

The results of our analysis as applied to the cluster subsamples introduced in Table \ref{tab:massBin} are included in Appendix \ref{app:ExactNums}. Because of the large statistical uncertainties, comparing the pressure profile of different mass bins is not terribly illuminating. We will, however, discuss the implications for gas mass fraction in the next Section.

%
%


\subsection{Gas Mass Fraction}
\label{fGas}
The density of gas $\rho^{(a)}_g(r)$ in the $a^{\text{th}}$ cluster with temperature profile $T^{(a)}(r)$ takes the form
\beq
\rho^{(a)}_g(r)=\frac{\mu_e m_pP^{(a)}_e(r)}{k_BT^{(a)}(r)},
\eeq
where $m_p$ is the proton mass and $\mu_e=\frac{2}{X+1}\simeq1.14$ is the mean molecular weight per free electron for a cosmic hydrogen abundance of $X\simeq0.76$.\footnote{
The free electron number density is $n_e=n_H+2n_{He}$, where $n_H$ and $n_{He}$ are the Hydrogen and Helium number density. The cosmic hydrogen abundance is
$X=n_H/(n_H+4n_{He})$. It then follows that $\rho_{b}=m_p(n_H+4n_{He})=\frac{2m_p}{X+1}n_e=\frac{2m_p}{X+1}\frac{P_e}{k_BT}$, where $m_p$ is the proton mass.
}
As it has been the case for the electron pressure profile (see Equation \eqref{univP}), we assume a universal temperature profile
\beq
T^{(a)}(r)=T^{(a)}_c\Tu(r/r^{(a)}_c),
\label{univTT}
\eeq
which in turn implies 
\begin{subequations}
	\begin{align}
		\rho^{(a)}_g(r) &=\rho^{(a)}_c\frac{\Pu(r/r_c^{(a)})}{\Tu(r/r_c^{(a)})}, \\
		\rho^{(a)}_c &\equiv\frac{\mu_e m_pP_c^{(a)}}{k_BT_c^{(a)}}.
	\end{align}
\end{subequations}
The volume-averaged gas density at radius $r$ takes the form
\begin{align}
	\bar{\rho}^{(a)}_g(<r) &= \frac{1}{\frac{4}{3}\pi r^3}\int_0^{r}\rho^{(a)}_g(r')4\pi r'^2dr' \nonumber\\
	&=3\rho^{(a)}_c(r/r_c^{(a)})^{-3}\int_0^{r/r_c^{(a)}}\frac{\Pu(x)}{\Tu(x)}x^2dx.
	\label{rhob}
\end{align}
Given that we have considered radial bins throughout which $\Pu$ is constant, we shall also bin the temperature profile:
\beq
\Tu_k=\frac{1}{\frac{4\pi}{3}(n_k^3-n_{k-1}^3)}\int_{n_{k-1}}^{n_k}\Tu(x)x^2dx,
\eeq
where $n_k$ is the value of the $k^{\text{th}}$ radial bin in units of $r_c^{(a)}$, with $n_0\equiv0$ (see Section \ref{UPP}). 
\footnote{
We will discretize all continuous profiles over our radial bins, because our pressure measurements are discrete by construction. In the case of temperature, it might seem more natural from 
Equation \eqref{rhob} to discretize $1/\Tu$ instead of $\Tu$. We have checked that the difference between these discretization schemes is insignificant. 
}
It then follows that
\beq
\bar{\rho}^{(a)}_g(<r)=\rho^{(a)}_c\sum_{k=1}^{N_b}V_k(r/r_c^{(a)})\Pu_k,
\eeq
where $N_b$ is the total number of radial bins and
\begin{equation}
 V_k(x)=
    \begin{cases}
      \frac{1}{\Tu_kx^3}\left[n_k^3-n_{k-1}^3\right] & \text{if } k\le k_*\\
      \frac{1}{\Tu_kx^3}\left[x^3-n_{k_*}^3\right] & \text{if } k=k_*+1\\ 
      0 & \text{if } k>k_*+1.
    \end{cases}
\end{equation}
Here $k_*$ is an integer defined via $n_{k_*}\le x<n_{k_*+1}$.
We assume the total matter density of the $a^{\text{th}}$ cluster to be of the NFW form:
\beq
\rho^{(a)}_{m}(r)=\frac{\rho_s^{(a)}}{r/r_s^{(a)}\left(1+r/r_s^{(a)}\right)^2}.
\eeq
Defining $c^{(a)}_{\Delta}\equiv R_{\Delta}^{(a)}/r_s^{(a)}$ and $\delta^{(a)}=\rho_s^{(a)}/\rho_{crit}(z_a)$, it may be checked that
the total mass enclosed within a radius $r$ is equal to
\footnote{
Here we have adopted the notation of \cite{bhattacharya2011dark}.
}
\begin{subequations}
  \begin{align}
    M^{(a)}_{m}(<r) &=\frac{m(c^{(a)}_\Delta r/R^{(a)}_\Delta)}{m(c^{(a)}_\Delta)}M^{(a)}_{\Delta}, \\
	m(x)&\equiv \ln(1+x)-\frac{x}{1+x},
  \end{align}
\end{subequations}
where $R^{(a)}_\Delta$ and $M^{(a)}_{\Delta}$ were defined in Equation \eqref{DeltaDef}. Also, it follows from Equation \eqref{DeltaDef} that
\beq
\frac{m(c^{(a)}_\Delta)}{[c^{(a)}_\Delta]^3}=\frac{\Delta}{3\delta^{(a)}}.
\label{m-delta}
\eeq
Therefore, knowing $c^{(a)}_\Delta$ (for any $\Delta$) determines $\delta^{(a)}$, or equivalently $\rho_s^{(a)}$.
We estimate the concentration parameter from the $c_{200}-M_{200}$ relation of \cite{bhattacharya2011dark}:
\begin{subequations}
  \begin{align}
	c_{200} &=5.9D(z)^{0.54}\nu(M_{200},z)^{-0.35}, \\
	\nu(M,z) &\simeq\frac{1}{D(z)}\left[1.12\left(\frac{M}{5\times10^{13}h^{-1}M_\odot}\right)^{0.3}+0.53\right],
  \end{align}
\end{subequations}
where $D(z)$ is the linear growth factor normalized to $1$ at $z=0$.
\footnote{
A good approximation to $D(z)$ is given by $D(z)=\frac{D_1(z)}{D_1(0)}$, where \citep[see][]{1992ARA&A..30..499C}
\begin{multline*}
D_1(z)\simeq
\frac{5\Omega_m(z)}{2(1+z)}\times\\
\left\{\Omega_m(z)^{4/7}-\Omega_{\Lambda}(z)+ 
\left[1+\Omega_m(z)/2\right]\left[1+\Omega_{\Lambda}(z)/70\right]\right\}^{-1}.
\end{multline*}
Here $\Omega_m(z)=\Omega_m(1+z)^3/E(z)^2$ and $\Omega_{\Lambda}(z)=\Omega_{\Lambda}/E(z)^2$.
}
As was the case with the temperature of baryons, we similarly bin $\rho^{(a)}_{m}(r)$ 
\beq
\rho^{(a)}_{m,k}=\frac{M^{(a)}_{m}(<r^{(a)}_k)-M^{(a)}_{m}(<r^{(a)}_{k-1})}{\frac{4\pi}{3}\left\{[r^{(a)}_k]^3-[r^{(a)}_{k-1}]^3\right\}},
\eeq
where as before $r^{(a)}_k=n_kr^{(a)}_c$.
Finally, the volume-averaged matter density $\bar{\rho}^{(a)}_{m}$ up to radius $r$ is
\beq
\bar{\rho}^{(a)}_{m}(<r)=\sum_{k=1}^{N_b}\tilde{V}_k(r/r_c^{(a)})\rho^{(a)}_{m,k},
\eeq
where ($k_*$ being defined as above)
\begin{equation}
 \tilde{V}_k(x)=
    \begin{cases}
      \frac{1}{x^3}\left[n_k^3-n_{k-1}^3\right] & \text{if } k\le k_*\\
      \frac{1}{x^3}\left[x^3-n_{k_*}^3\right] & \text{if } k=k_*+1\\ 
      0 & \text{if } k>k_*+1.
    \end{cases}
\end{equation}
The average gas mass-fraction up to radius $r$ in the $a^{\text{th}}$ cluster takes the form:
\begin{align}
	f^{(a)}_{gas}(<r)&\equiv\frac{\bar{\rho}^{(a)}_{g}(<r)}{\bar{\rho}^{(a)}_{m}(<r)} \nonumber \\
		&=\frac{\rho^{(a)}_c}{\bar{\rho}^{(a)}_{m}(<r)}\sum_{k=1}^{N_b}V_k(r/r_c^{(a)})\Pu_k.
\end{align}
In order to make a meaningful comparison with the universal gas mass-fraction, we average this quantity over all clusters 
\begin{subequations}
	\begin{align}
		f_{gas}(<x) &\equiv\frac{1}{N_c}\sum_{a=1}^{N_c}f^{(a)}_{gas}(<x r_c^{(a)}),\nonumber\\
		&=\mathcal{A}(x)\sum_{k=1}^{N_b}V_k(x)\Pu_k,\\
		\mathcal{A}(x) &\equiv\frac{1}{N_c}\sum_{a=1}^{N_c}\frac{\rho^{(a)}_c}{\bar{\rho}^{(a)}_{m}(<xr_c^{(a)})}.
	\end{align}
	\label{fgasAvg}
\end{subequations}
Since $f_{gas}(<x)$ is a linear combination of $\{\Pu_k\}$, it is a Gaussian random variable
with mean and variance
\begin{align}
	\left\langle f_{gas}(<x)\right\rangle &=\mathcal{A}(x)\sum_{k=1}^{N_b}V_k(x)\hat{\Pu}_k \label{fgasMean}\\
	\sigma^2_{f_{gas}(<x)} &\equiv \left\langle\left[f_{gas}(<x)-\left\langle f_{gas}(<x)\right\rangle\right]^2\right\rangle \notag\\
	                      &=\mathcal{A}(x)^2\sum_{k,k'=1}^{N_b}V_k(x)V_{k'}(x)[\matf{C}_{\Pu}]_{kk'}. \label{fgasErr}
\end{align}
The averaging scheme we have adopted in Equation \eqref{fgasAvg} may seem arbitrary. One could, for instance, assign different weights to different clusters.
If $P_c^{(a)}$ and $T_c^{(a)}$ scale similarly with mass, different averaging schemes would differ by a negligible amount.
This is because 
the only variation in $f_{gas}$ amongst different clusters would be
due to the scaling of $c_{500}$ with mass, which is fairly mild. As a result, given the temperature profile we have adopted (see Equation \eqref{tempVikh}),
we use the standard self-similar scaling of
$P_c^{(a)}$ with mass ($\delta=0$ in Equation \eqref{Pc} below) to compute $f_{gas}$.

We use the average temperature profile of \cite{vikhlinin2008chandra}:
\begin{subequations}
	\begin{align}
		\Tu(x) &=1.35\frac{(x/0.045)^{1.9}+0.45}{(x/0.045)^{1.9}+1}\frac{1}{[1+(x/0.6)^2]^{0.45}}, \\
		\frac{T^{(a)}_c}{5\text{ keV}} &=\left[\frac{M^{(a)}_{500}E(z)}{3.41\times10^{14}h^{-1}M_{\odot}}\right]^{1/1.51}.
	\end{align}
	\label{tempVikh}
\end{subequations}
This is an approximation to the averaged profile of about a dozen low-redshift X-ray clusters, with measurements obtained for $r<R_{500}$ \citep{vikhlinin2008chandra}.
The scatter about the mean profile is about $15$ percent. 
The assumption of universality (i.e. Equation \eqref{univTT}) may be easily relaxed if temperature measurements for individual clusters are available.
In the case of our present work, however, this option is not viable since a large cluster sample
is required to compensate for WMAP's insufficient sensitivity. 
To get an estimate for how this assumption affects our $f_{gas}$ measurements, consider an average $100\epsilon$ percent scatter about the universal profile.
For a relatively large sample of clusters, as is the case with our measurements, changing the individual temperature profiles by $\sim100\epsilon$ percent would introduce a systematic uncertainty of order $\epsilon^2$ in $f_{gas}$.
\footnote{
This is because $f_{gas}\propto\sum_{a}1/T^{(a)}$, where $T^{(a)}$ is the temperature of the $a^{\text{th}}$ cluster. Let $\{\epsilon_{a}\}$ be realizations of a gaussian random variable with zero mean and standard deviation $\epsilon$. Changing the temperature $T^{(a)}\to T^{(a)}(1+\epsilon_{a})$ is equivalent to $\sum_{a}1/T^{(a)}\to\sum_{a}1/T^{(a)}(1-\epsilon_{a}+\epsilon_{a}^2+\cdots)$. When $T^{(a)}$ doesn't change drastically from cluster to cluster, the contribution of the term linear in $\epsilon_{a}$ is on the order of $\sqrt{N}\epsilon$, while the second order term contributes about $N\epsilon^2$. When $N>1/\epsilon^2$, the $\epsilon^2$ term dominates.
}
Given the large statistical uncertainties in our pressure measurements, this effect is small.

\begin{figure}
	\begin{subfigure}[b]{\hsize}
        		\includegraphics[width=\hsize]{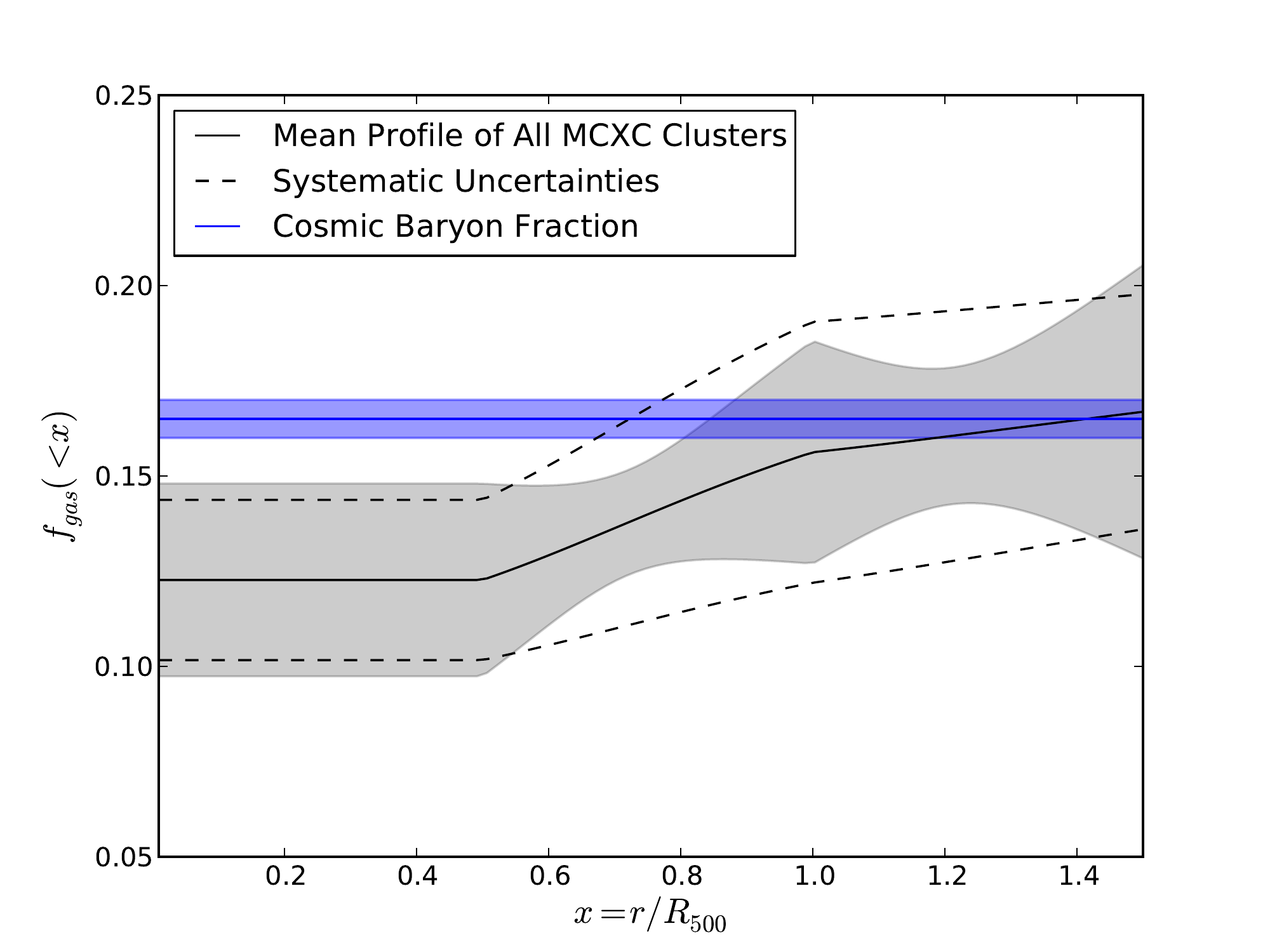}
		\caption{Gas mass fraction for all MCXC clusters. }		
    	\end{subfigure} %
	\begin{subfigure}[b]{\hsize}
        		\includegraphics[width=\hsize]{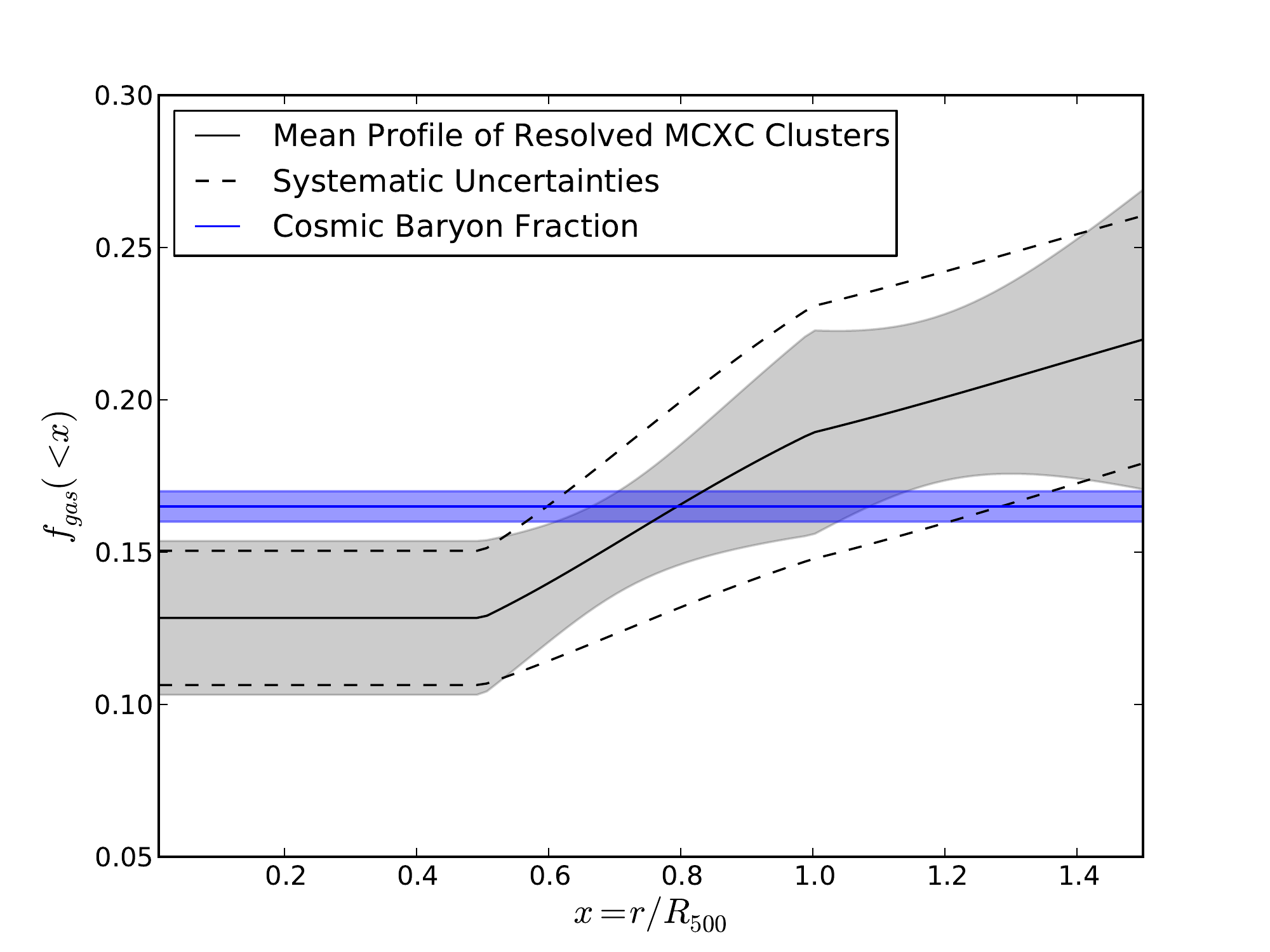}
		\caption{Gas mass fraction for resolved MCXC clusters.}
    	\end{subfigure} %
	\caption{
	Gas mass fraction of all MCXC, as well as the resolved subsample. A cluster is considered resolved if its first radial bin subtends a solid angle larger than the effective beam area of the W 	frequency channel (see Section 3.2). The standard self-similar model of pressure is assumed for both measurements ($\delta=0$ in Equation \eqref{Pc}).
	The black solid curves show the average gas mass fraction, computed using Equation \eqref{fgasMean}. The shaded areas represent the standard deviation in the measurement of $f_{gas}$ as given by Equation \eqref{fgasErr}. The dashed black lines show the expected systematic uncertainty about the mean gas mass profile, mostly due to cluster mass estimates.
	The cosmic gas mass fraction is obtained by fitting $\Lambda$CDM to WMAP9$+$SPT$+$ACT data and is equal to  $\Omega_b/\Omega_m=0.165\pm0.005$ \protect\citep{WMAP9_COSMO}.}
	\label{fgas}
\end{figure}

Systematic uncertainties associated with estimating masses of clusters have a more significant effect on gas mass fraction. 
As mentioned in the previous section, we have investigated this issue in Appendix \ref{massChange}
by considering 62 MCXC clusters which are also in the ESZ sample \citep{ESZ}. 
The ESZ mass estimates are systematically higher by about $12$ percent, which causes lower pressure measurements at the 1$\sigma$ level.
Because of the scaling of temperature with mass (Equation \eqref{tempVikh}), this decreases $f_{gas}$ by about $\bf 20$ percent
(see Fig. \ref{fig: massChange}). Repeating our analysis with \textit{Planck} CMB data is expected to reduce statistical errors significantly (see Section \ref{discussion} and \cite{collaboration2012planck}). In that case, $f_{gas}$ measurements would be solely dominated by systematic uncertainties associated with determining masses of clusters.

\begin{figure}
	\begin{subfigure}[b]{\hsize}
        		\includegraphics[width=\hsize]{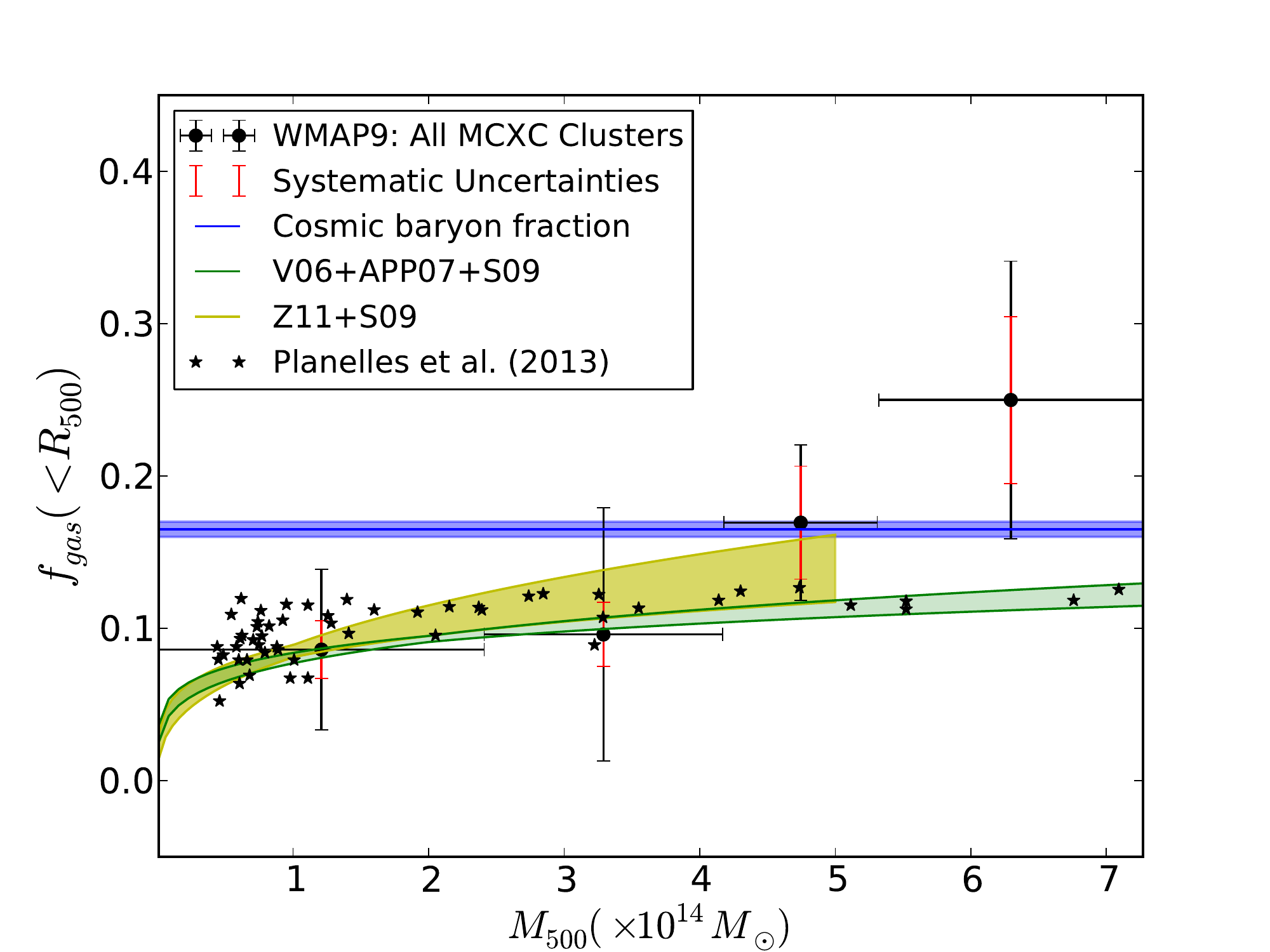}
		\caption{Binning the entire MCXC sample according to mass, as shown in Table \ref{tab:massBin_all}.}		
    	\end{subfigure} %
	\begin{subfigure}[b]{\hsize}
        		\includegraphics[width=\hsize]{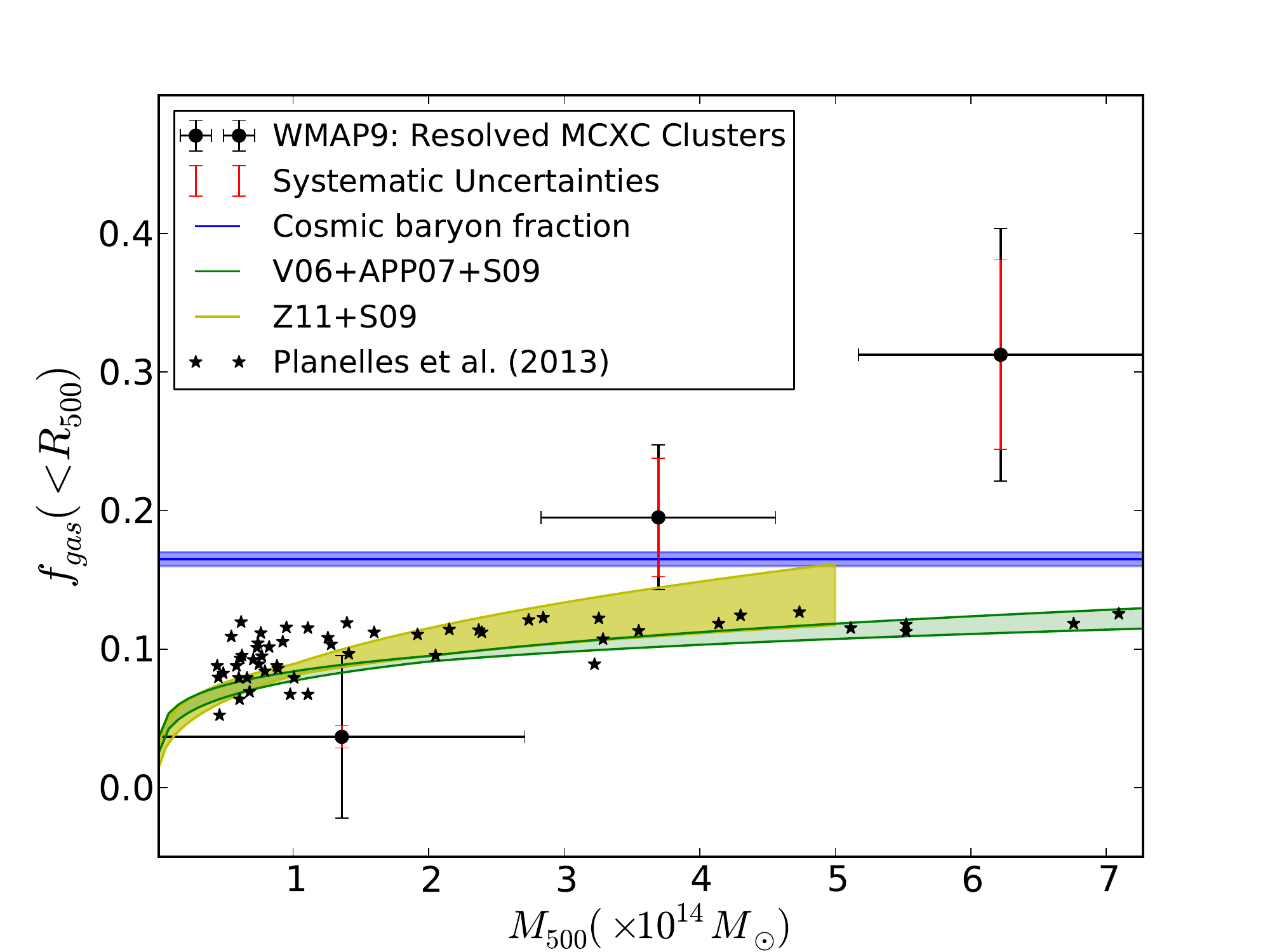}
		\caption{Binning the resolved MCXC sample according to mass, as shown in Table \ref{tab:massBin_162}.}
    	\end{subfigure} %
	\caption{Dependence of $f_{gas}$ on the mass of clusters. All mass bins are defined in Table \ref{tab:massBin}. The black dots show $f_{gas}$ measurements up to $R_{500}$, with black vertical bars denoting statistical errors. The red vertical bars show systematic uncertainties expected due to cluster mass estimates.
	Following \protect\cite{Planelles13}, we compare
	our measurements with two different observational samples: a combined sample of 41 clusters and groups from \protect\cite{vikhlinin2008chandra}, \protect\cite{Arnaud07} and \protect\cite{Sun09} (V06+APP07+S09), shown as the green region, and the sample obtained from the combination of the data by \protect\cite{Zhang11} and \protect\cite{Sun09} (Z11+S09), shown as the yellow area (see Table 1 of \protect\cite{Planelles13}). The black stars show $f_{gas}$ obtained from hydrodynamical simulations carried out by \protect\cite{Planelles13}, which include radiative cooling, star formation and feedback from supernovae and active galactic nuclei.  
	}
	\label{fgasVsMass}
\end{figure}
Fig. \ref{fgas} shows the result of our analysis, as applied to all MCXC clusters, as well as just the resolved ones. 
Given the large statistical and systematic uncertainties, we extrapolated the temperature profile of \cite{vikhlinin2008chandra} out to $R_{200}$. 
The black solid curves show the average gas mass fraction, computed using Equation \eqref{fgasMean}. The shaded areas represent the standard deviation in the measurement of $f_{gas}$ as given by Equation \eqref{fgasErr}. The dashed black lines show the expected systematic uncertainty about the mean gas mass profile ($\sim 20$ percent), mostly due to cluster mass estimates.
Considering both statistical and systematic errors, our results are fully consistent with the cosmic baryonic fraction up to $R_{200}$. 
Given the large error bars, accounting for all baryons in stars does not change this conclusion.

Fig. \ref{fgasVsMass} shows the dependence of gas mass fraction (up to
$R_{500}$) on the cluster subsamples which we have binned according to mass (see Table \ref{tab:massBin}). 
A general trend of increasing $f_{gas}$ with $M_{500}$ can be seen. Due to the large error bars, however, we cannot make any statistically significant statement about this dependence. 
Following \protect\cite{Planelles13}, we compare our measurements with two different observational samples: a combined sample of 41 clusters and groups from \protect\cite{vikhlinin2008chandra}, \protect\cite{Arnaud07} and \protect\cite{Sun09} (V06+APP07+S09), shown as the green region, and the sample obtained from the combination of the data by \protect\cite{Zhang11} and \protect\cite{Sun09} (Z11+S09), shown as the yellow area (see Table 1 of \protect\cite{Planelles13}). The black stars show $f_{gas}$ obtained from hydrodynamical simulations carried out by \protect\cite{Planelles13}, which include radiative cooling, star formation and feedback from supernovae and active galactic nuclei.  
Considering both statistical and systematic errors, our measurements are consistent with both the observational and numerical results.

\section{Discussion and Future Work}\label{discussion}
In a companion paper, we will repeat our entire analysis using the recently released \textit{Planck} CMB data.
To get an idea for how much our measurements will improve,
we \textit{estimate} here the pressure covariance matrix $\matf{C}_{\Pu}$ expected from \textit{Planck}. 
We consider the six \textit{Planck}-HFI channels, which have central frequencies $100, 143,  217, 353, 545$, and $857$ GHz, at HEALPix resolution $N_{side}=2048$. 
We assume a homogeneous detector noise constructed by averaging the noise variance of all pixels for a given frequency channel:
\beq
[\matf{C_N}]_{i\nu,i'\nu'}=n_{\nu}^2\delta_{\nu\nu'}\delta_{ii'},
\eeq
where $n_{100}=50.6$ $\mu$K, $n_{143}=20.1$ $\mu$K, $n_{217}=27.1$ $\mu$K, $n_{353}=0.1$ mK, $n_{545}=28.1$ mK, $n_{857}=27.9$ mK.
We assume Gaussian instrumental beams with Full-Width-Half-Maximum of $9.5, 7.1, 5.0, 5.0, 5.0, 5.0$ arcmins, for the six HFI channels in order of increasing frequency \citep{refId0}.
Finally, we assume no masking (i.e. $\matf{M}=\matf{1}$) but remove all clusters that are masked out from our templates.

Because we have assumed a homogeneous detector noise and no masking, the matrix $\matf{D}$ introduced in Equation \eqref{finalEq} can be inverted analytically, resulting in the pressure covariance matrix:
\begin{multline}
	[\matf{C}_{\Pu}^{-1}]_{kk'}=
	\sum_{\ell=0}^{\ell_{max}}\sum_{m=-\ell}^{\ell}
	\sum_{\nu}(\bar{t}_k^{(\nu)})_{lm}(t_{k'}^{(\nu)})_{lm}\xi^{(\nu)}_{\ell}\\
	-\frac{C_{\ell}}{1+C_{\ell}\xi_{\ell}}\left[\sum_{\nu}(\bar{t}_k^{(\nu)})_{lm}\xi^{(\nu)}_{\ell}\right]\left[\sum_{\nu}(t_{k'}^{(\nu)})_{lm}\xi^{(\nu)}_{\ell}\right],
	\label{estPlanck}
\end{multline}
where
\begin{subequations}
	\begin{align}
		\xi^{(\nu)}_{\ell}&=\frac{(B_{\ell\nu}W_{\ell})^2}{N_{\ell}^{(\nu)}}\\
		\xi_{\ell}&=\sum_{\nu}\xi^{(\nu)}_{\ell}\\
		N^{(\nu)}_{\ell}&=A_{pix}n_{\nu}^2.
	\end{align}
\end{subequations}
Here $A_{pix}$ denotes the pixel area and all other quantities are defined in Sections \ref{Statistical Methods} and \ref{Numerical Methods}.

\begin{figure}
	\begin{center}
		\includegraphics[width=\hsize]{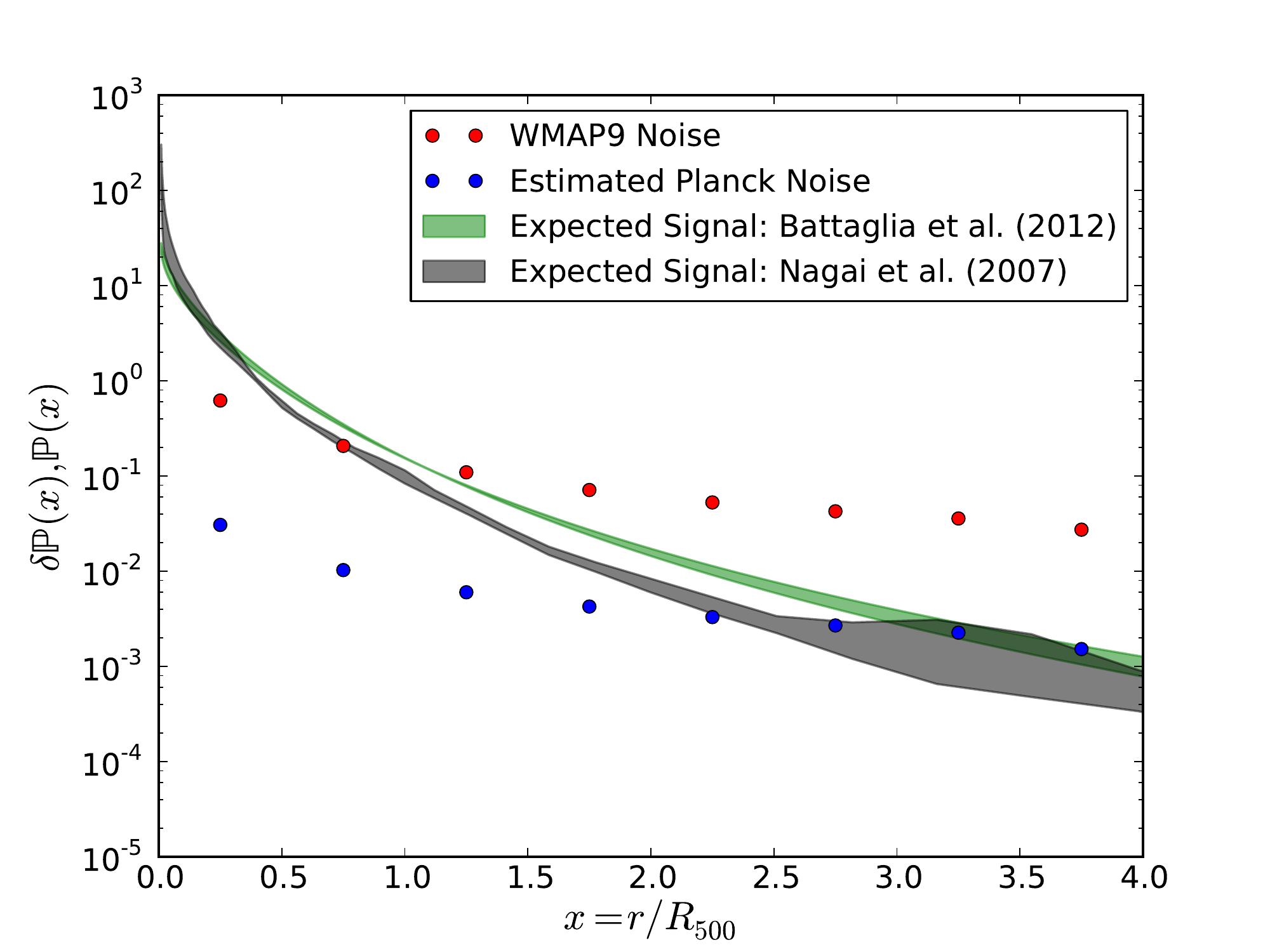}
	\end{center}
	\caption{Comparison of the statistical noise expected from analysis with \textit{Planck} CMB data vs. WMAP9. The blue data points are the estimated noise expected from repeating our analysis with \textit{Planck} CMB data (see Equation \eqref{estPlanck}). The red points correspond to the same quantity for WMAP9 measurements, applied to the resolved MCXC clusters with standard self-similar pressure-dependent scaling ($\delta=0$ in Equation \eqref{Pc}). The green and grey shaded areas show the expected pressure signal from simulated clusters of \protect\cite{Battaglia} and \protect\cite{Nagai}, respectively.}
	\label{planckEst}
\end{figure}

We generate our templates $\vecf{t_k^{(\nu)}}$ using the resolved MCXC clusters which are not masked (total of $122$ clusters), with a standard self-similar pressure-dependent scaling ($\delta=0$ in Equation \eqref{Pc}).
In order to make sure this estimate is reasonable, we computed the same quantity with WMAP9 data, using the $Q,V$, and $W$ channels. Assuming the best-fit pressure values $\hat{\Pu}$ remain the same, this leads to a null chi-squared of $\chi_0^2=124.551$, which is reasonably close to the actual value $\chi_0^2=115.626$. (We use the same 8 radial bins as for our pressure measurements, i.e. there are 8 degrees of freedom here.) Estimating the covariance matrix for \textit{Planck} using the same best-fit pressure values, we obtain $\chi_0^2=66154.8$. Therefore, assuming that the signal does not change, we expect the statistical uncertainties to reduce by a factor of $\sim \sqrt{66154.8/124.551}=\bf 23.9$. This is a \textit{significant} improvement, which will allow us to consider finer bins and possibly probe the ICM pressure to larger radii. Fig. \ref{planckEst} compares the expected error for different bins with those of WMAP9. 
Our analysis does not account for the uncertainty present in modelling of beams. In the case of \textit{Planck}, the beam uncertainty is modelled by 
\beq
B_{\ell\nu}=B^{\text{mean}}_{\ell\nu}\exp\left[\sum_{k=1}^{n_{modes}}g^kE^{k}_{\ell\nu}\right],
\eeq
where  $\{g^k\}$ are independent Gaussian random variables with unit variance, and $E^{k}_{\ell\nu}$ is the $k^\text{th}$ eigenmode of the beam covariance matrix  \citep{planckCMB13}.
\footnote{
The coefficients $E^{k}_{\ell\nu}$ are contained in the RIMO beam files of \textit{Planck}. Also, we use $n_{modes}=5$.
}
In order to see how this uncertainty affects our results, we have computed the best-fit pressure profile $\{\hat{\Pu}_k\}$ for 100 different realizations of the beams. To do so, we created mock CMB skies which contain the SZ signal, primary CMB and noise, and repeated the analysis outlined above for our \textit{Planck} forecast. 
\footnote{
The input pressure profile needed to create the SZ signal is taken to be the best-fit pressure values used in our \textit{Planck} forecast analysis.
}
We find $\delta\hat{\Pu}_k\sim0.01\sqrt{[\matf{C}_{\Pu}]_{kk}}$, where $\delta\hat{\Pu}_k$ denotes the standard deviation of the $100$ values of $\hat{\Pu}_k$ obtained through our simulations. Therefore, effects of beam modeling are quite small relative to the statistical uncertainty due to primary CMB and instrumental noise.

We have also ignored the impact of foreground residuals in our formalism. Our current model is sufficient for WMAP foreground cleaned maps but not for \textit{Planck}, because dust emission dominates at high frequencies and there are other emissions (e.g. CIB, radio and infrared point sources) which are not negligible and can not be modelled easily. To get an estimate for this effect, we consider foreground templates created by taking the difference of low and high frequency sky maps. More specifically, we created four templates by taking the difference between $030-044$, $044-070$, $353-545$, and $545-857$ \textit{Planck} sky maps. Because different frequency channels have different beam and noise properties, we smoothed these maps using a Gaussian window function with FWHM$=0.006$ radians $= 21'$. Considered as a template, each difference-map contributes a different coefficient to the total temperature anisotropy, depending on the frequency band. We then estimated the expected pressure covariance matrix as above, this time using only $100, 143, 217$ GHz frequency channels. Assuming the same best-fit pressure values, 
accounting for foreground residuals decreases the null chi-squared by about $0.5$ percent. 

We have not addressed the issue of point source contamination so far. In our framework, it is not feasible to fit locally for contribution of point sources, given the large number of clusters. 
We did try to account for them by assigning a constant absolute luminosity per frequency channels to all clusters. The results, however, change only by a negligible amount. 
We will consider a more detailed modelling of point source contamination in future work.
\section{Conclusions}
\label{conclusions}
We have introduced a statistically-optimal and model-independent framework for extracting the universal pressure profile of the hot gas in the intracluster medium.
The thermal Sunyaev-Zeldovich effect makes this possible because it is linearly proportional to the integral of the electron pressure along the line of sight.
We use the principle of maximum likelihood to find best-estimate values for the radially binned values of the pressure profile, as well as the full covariance matrix governing
their uncertainties. Once reformulated in the proper mathematical framework, the main technical challenge is solving a very large system of linear equations, which we do numerically by employing the conjugate gradient method.

We applied our methodology to WMAP9 data and various subsamples of the MCXC catalogue.
In the case of all MCXC clusters, we extract the pressure profile with a high accuracy at $\sim15\sigma$ confidence, with possible systematics uncertainties dragging our detection down to $\sim14\sigma$.
We also considered a subsample of the MCXC clusters completely resolved by the $W$ frequency channel of WMAP, resulting in a $\sim9\sigma$ detection.
In an upcoming companion to this paper, we apply the same methodology to the recently released {\it Planck} CMB maps.  
An estimation of the pressure covariance matrix expected from \textit{Planck} suggests that the current signal-to-noise will improve by a factor of $\sim24$. 

Assuming a temperature profile motivated by X-ray observations, we computed the average gas mass fraction as a function of radius. 
We argued that systematic uncertainties associated with estimating mass of clusters could have a drastic effect ($\sim20$ percent) on gas mass fraction. 
Considering both statistical and systematic errors, our results are fully consistent with the cosmic baryonic fraction and the expected gas mass fraction in halos, up to $R_{200}$.  

We also made a first attempt at studying the dependence of gas mass fraction on the mass of clusters. Due to the large error bars, we cannot make any statistically significant statements about this dependence. Nevertheless, our measurements suggest that gas mass fraction increases with the mass of clusters, which is consistent with findings from X-ray measurements \citep{vikhlinin2008chandra, Arnaud07, Sun09, Zhang11} and numerical simulations \citep{Planelles13} .

\section*{Acknowledgements}
We thank Daisuke Nagai for providing us with pressure profiles of simulated clusters, as well as helpful comments on the paper.
We would also like to thank Nick Battaglia, Colin Hill, Eiichiro Komatsu, Neelima Sehgal, and David Spergel for providing useful comments on our manuscript.
We acknowledge discussions on temperature profile of clusters with Helen Russell and Brian McNamara.
SA would like to thank Etienne Pointecouteau for helpful discussions on \textit{Planck} pressure measurements.
We thank Jean-Baptiste Melin for pointing out potential shortcomings of our formalism with regards to beam uncertainties and foreground residuals.
GL acknowledges support from CITA National Fellowship and financial support from the Government of Canada Post-Doctoral Research Fellowship.
Research at Perimeter Institute is supported by the Government of Canada through Industry Canada and by the Province of Ontario through the Ministry of Research and Innovation.
This work was made possible by the facilities of the Shared Hierarchical Academic Research Computing Network (SHARCNET:\url{http://www.sharcnet.ca}) and Compute/Calcul Canada.
We acknowledge the use of the Legacy Archive for Microwave Background Data Analysis (LAMBDA). Support for LAMBDA is provided by the NASA Office of Space Science.
Some of the results in this paper have been derived using the HEALPix\footnote{\url{http://HEALPix.jpl.nasa.gov}} \citep{HEALPIX} package.

\bibliography{SZ}
\FloatBarrier
\clearpage
\onecolumn
\appendix

\section{Technical Details}

\subsection{Likelihood Function}
\label{Appendix Statistical methods}
Given an underlying temperature field 
\footnote{
We follow the HEALPix conventions for spherical harmonic transforms \citep{healpixP}.
}
\beq
\delta T(\uvec{n})=\sum_{\ell=0}^{\ell_{max}}\sum_{m=-\ell}^{\ell}\delta T_{\ell m}Y_{\ell m}(\uvec{n}),
\eeq
the value of the discretized temperature map at pixel $i$ and frequency band $\nu$ is given by 
\beq
\delta T_{i\nu}=\sum_{\ell=0}^{\ell_{max}}\sum_{m=-\ell}^{\ell}\delta T_{\ell m}B_{\ell\nu}W_\ell Y_{\ell m}(\uvec{n}_i),
\eeq
where $B_{\ell\nu}$ is the isotropicized beam transfer function for the mode $\ell$ and frequency channel $\nu$, and $W_\ell$ is the isotropicized pixel transfer function.
In case of the primordial anisotropies, because $\dTpa_{i\nu}$ are linear functionals of $\dTpa(\uvec{n})$, they are correlated 
Gaussian random variables with zero mean and two-point function
\beq
[\matf{C_S}]_{i\nu,i'\nu'}\equiv\langle\dTpa_{i\nu}\dTpa_{i'\nu'}\rangle=\sum_{\ell=0}^{\ell_{max}}\left(\frac{2\ell+1}{4\pi}\right)C_\ell B_{\ell\nu}B_{\ell\nu'}W_\ell^2P_\ell(\uvec{n}_i\cdot\uvec{n}_{i'}),
\eeq
where $P_\ell$ is the $\ell$-th degree Legendre polynomial and we have used $\langle\dTpa_{lm}\dTpa_{l'm'}\rangle=C_l\delta_{ll'}\delta_{mm'}$.

Finally, the log-likelihood probability $-\chi^2/2$ of jointly measuring the CMB temperature values $\{\delta T_{i\nu}\}_{i\in L_{p}}^{\nu\in L_{\nu}}$ given the 
tSZ contribution $\{\dTsz_{i\nu}\}_{i\in L_{p}}^{\nu\in L_{\nu}}$
and the primary CMB fluctuations $\{\dTpa_{i\nu}\}_{i\in L_{p}}^{\nu\in L_{\nu}}$ is
\begin{multline}
 -\frac{1}{2}\chi^2(\{\delta T_{i\nu}\}_{i\in L_{p}}^{\nu\in L_{\nu}}|\{\dTsz_{i\nu}\}_{i\in L_{p}}^{\nu\in L_{\nu}},\{\dTpa_{i\nu}\}_{i\in L_{p}}^{\nu\in L_{\nu}})= \\
-\frac{1}{2}\vecf{\dTpa}^T {\matf{C}_S}^{-1} \vecf{\dTpa} 
-\frac{1}{2}(\vecf{\delta T}-\vecf{\delta T}^{PA}-\vecf{\dTsz})^T\matf{C}_N^{-1}(\vecf{\delta T}-\vecf{\delta T^{PA}}-\vecf{\dTsz}),
\end{multline}
where $\matf{C_N}$ is the noise covariance matrix and $L_p$ ($L_{\nu}$) is the set containing pixels (frequency channels) we wish to use in our analysis. 
\footnote{
The matrix notation used here is explicitly:
\beq
{\dTpa}^{T}\matf{C_S}^{-1}\dTpa=\sum_{\nu,\nu'\in L_{\nu}}\sum_{i,i'\in L_p}[\matf{C_S}^{-1}]_{i\nu,i'\nu'}\dTpa_{i\nu}\dTpa_{i'\nu'},
\eeq
where we are considering $\matf{C_S}$ as a matrix whose rows and columns are labeled by $i\nu$.
}
After integrating over all possible primary fluctuations $\vecf{\dTpa}$, which can be done analytically, the log-likelihood
takes the form
\begin{equation}
    -\frac{1}{2}\chi^2(\{\delta T_{i\nu}\}_{i\in L_{p}}^{\nu\in L_{\nu}}|\{\dTsz_{i\nu}\}_{i\in L_{p}}^{\nu\in L_{\nu}})=\\
      -\frac{1}{2}(\vecf{\delta T}-\vecf{\dTsz})^T\matf{C}^{-1}(\vecf{\delta T}-\vecf{\dTsz}), 
\end{equation}
with
\begin{equation}
  \matf{C} =\matf{C}_S+\matf{C}_N.
\end{equation}
This proves Equation~\eqref{eq:log_likelihood}.

\subsection{Masking}
\label{masking}
All quantities used (and not defined) here are introduced in Sections \ref{Statistical Methods} and \ref{Numerical Methods}.

Let $L_{\bar{p}}$ ($L_{p}$ respectively) denotes the set of all masked (unmasked respectively) pixels, so that $L=L_{\bar{p}}\cup L_{p}$ contains all pixels on the sky. 
In what will follow, $\matf{C_{S_f}}$ ($\matf{C_{S}}$) will denote the signal covariance matrix defined on $L$ ($L_p$). The same notation will be used for the noise covariance matrix.
Consider now the full covariance matrix $\matf{C_f}=\matf{C_{S_f}}+\matf{C_{N_f}}$.
It is related to $\matf{C}=\matf{C_{S}}+\matf{C_{N}}$
via $\matf{C}=\matf{C_P}^T \matf{C_f} \matf{C_P}$, 
where $[\matf{C_P}]_{i\nu,j\nu'}=P_{ij}\delta_{\nu\nu'}$ is a projection matrix with $i\in L$ and $j\in L_p$, such that all components of $\matf{P}$ are zero except
$P_{ii}=1$ for $i\in L_p$.
Construct from $\matf{C_f}$ another matrix $\matf{\tilde{C}_f}$ with the same entries, except that $[\matf{\tilde{C}_f}]_{i\nu,i\nu}=x$ for all $i\in L_{\bar{p}}$. Then, it can be shown from the definition of the inverse of a matrix that
\footnote{
As a simple example, consider the case where there are only two pixels, one frequency channel, and one of the pixels is masked out: $L_p=\{1\}$, $L_{\bar{p}}=\{2\}$. In this case 
$
\matf{C_P}=\matf{P}=
 \begin{bmatrix}
  1  \\
  0  
 \end{bmatrix},
$
$\matf{C}=[\matf{C_f}]_{11}$, 
$
\matf{\tilde{C}_f}=
 \begin{bmatrix}
  [\matf{C_f}]_{11}&  [\matf{C_f}]_{12} \\
  [\matf{C_f}]_{21}&  x 
 \end{bmatrix}
$
and 
$
\matf{\tilde{C}_f}^{-1}=\frac{1}{x [\matf{C_f}]_{11}- [\matf{C_f}]_{12}^2}
 \begin{bmatrix}
  x &  - [\matf{C_f}]_{12} \\
  - [\matf{C_f}]_{12}&   [\matf{C_f}]_{11} 
 \end{bmatrix}.
$
Equality \eqref{maskTheorem} can now be easily verified:
\beq
\lim_{x\to\infty}\matf{C_P}^T\matf{\tilde{C}_f}^{-1}\matf{C_P}=\lim_{x\to\infty}\frac{x}{x[\matf{C_f}]_{11}-[\matf{C_f}]_{12}^2}=\frac{1}{[\matf{C_f}]_{11}}=\matf{C}^{-1}.
\eeq
}
\beq
\matf{C}^{-1}=\lim_{x\to\infty}\matf{C_P}^T \matf{\tilde{C}_f}^{-1} \matf{C_P}.
\label{maskTheorem}
\eeq
Because $[\matf{C_{N_f}}]_{i\nu,i'\nu'}=[\matf{\tilde{N}}_{\nu}]_{ii'}\delta_{\nu\nu'}$, we may construct $\matf{\tilde{C}_f}$ as follows:
$[\matf{\tilde{C}_f}]_{i\nu,i'\nu'}=[\matf{\tilde{N}}_{\nu}]_{ii'}\delta_{\nu\nu'}+\matf{C_{S_f}}$, where $\matf{\tilde{N}}_{\nu}$ has the same elements as $\matf{N}_{\nu}$ except $[\matf{\tilde{N}}_{\nu}]_{ii}=x$ for all $i\in L_{\bar{p}}$. 
Since $[\matf{N}_{\nu}]_{ii'}=n_{i\nu}^2\delta_{ii'}$ is diagonal, it follows that $[\matf{\tilde{N}}_{\nu}^{-1}]_{ii'}=1/[\matf{\tilde{N}}_{\nu}]_{ii'}$, which then implies
\beq
\lim_{x\to\infty}\matf{\tilde{N}}_{\nu}^{-1}=\matf{M}\matf{N}_{\nu}^{-1}\matf{M},
\eeq
where $\matf{M}=\matf{P}\matf{P}^{T}$ is the masking matrix defined in Section \ref{Numerical Methods}. 
Moreover, we show in Appendix \ref{proof} that
\footnote{Note that the proof presented in Appendix \ref{proof} proceeds in the exact same fashion when $\matf{N}_{\nu}$ is replaced by $\matf{\tilde{N}}_{\nu}$.}
\beq
[\matf{\tilde{C}_f}^{-1}]_{i\nu,i'\nu'}=\left[\matf{\tilde{N}}_{\nu}^{-1}\delta_{\nu,\nu'}-\matf{\tilde{N}}_{\nu}^{-1}\matf{B}_{\nu}\matf{S}^{1/2}\left\{1+\matf{S}^{1/2}\left(\sum_{\mu\in L_{\nu}}\matf{B}_{\mu}\matf{\tilde{N}}_{\mu}^{-1} \matf{B}_{\mu}\right)\matf{S}^{1/2}\right\}^{-1}\matf{S}^{1/2}\matf{B}_{\nu'}\matf{\tilde{N}}_{\nu'}^{-1}\right]_{ii'}.
\label{conv}
\eeq
Using the above two equations we find
\beq
\lim_{x\to\infty}[\matf{\tilde{C}_f}^{-1}]_{i\nu,i'\nu'}=[\matf{G}_{\nu,\nu'}]_{ii'},
\label{conv2}
\eeq
where $\matf{G}_{\nu,\nu'}$ is given by \eqref{Gdef}.

Finally, let $[\vecf{V_f}]_{i\nu}$ and $[\vecf{W_f}]_{i\nu}$ be two vectors  defined on every pixel on the sky (i.e. $i\in L$). Also, let $V$ and $W$ be the corresponding vectors 
defined only on the unmasked pixels: $\vecf{V}=\matf{C_P}^T\vecf{V_f}$, $\vecf{W}=\matf{C_P}^T\vecf{W_f}$. Then
\begin{align}
\vecf{V}^T\matf{C}^{-1}\vecf{W}&=\lim_{x\to\infty}\vecf{V_f}^T\matf{C_P}\matf{C_P}^T\matf{\tilde{C}_f^{-1}}\matf{C_P}\matf{C_P}^T\vecf{W_f}\\
&=\sum_{i,i'\in L}\sum_{\nu,\nu'\in L_{\nu}}[\matf{M}\matf{G}_{\nu,\nu'}\matf{M}]_{ii'}[\vecf{V_f}]_{i\nu}[\vecf{W_f}]_{i'\nu'}\\
&=\sum_{i,i'\in L}\sum_{\nu,\nu'\in L_{\nu}}[\matf{G}_{\nu,\nu'}]_{ii'}[\vecf{V_f}]_{i\nu}[\vecf{W_f}]_{i'\nu'},
\end{align}
where we have used the fact that $[\matf{C_P}\matf{C_P}^T]_{i\nu,i'\nu'}=M_{ii'}\delta_{\nu\nu'}$, as well as $\matf{M}^2=\matf{M}$, which combined with \eqref{conv2} implies $\matf{M}\matf{G}_{\nu,\nu'}\matf{M}=\matf{G}_{\nu,\nu'}$.
This justifies the equality between the corresponding equations in \eqref{bestFit} and \eqref{bestFit2}.

\subsection{Covariance Matrix Re-loaded}
\label{proof}
All quantities used (and not defined) here are introduced in Sections \ref{Statistical Methods} and \ref{Numerical Methods}. Also, 
all matrices are defined on the entire sky.

Using the definition of matrices $\matf{B}_{\nu}$ and $\matf{S}$ as given in Section \ref{Numerical Methods}, it may be checked that
\beq
[\matf{C_{S}}]_{i\nu,i'\nu'}\simeq\left[\matf{B}_{\nu}\matf{S}\matf{B}_{\nu'}\right]_{ij}, 
\eeq
where we have used 
\beq
A_{pix}\sum_{i\in L}Y_{lm}(\uvec{n}_i)\bar{Y}_{l'm'}(\uvec{n}_i)\simeq\int Y_{lm}(\uvec{n})\bar{Y}_{l'm'}(\uvec{n})d^2n=\delta_{ll'}\delta_{mm'}.
\label{integApp}
\eeq
It then follows that
\beq
[\matf{C}]_{i\nu,i'\nu'}=\left[\matf{B}_{\nu}\matf{S}\matf{B}_{\nu'}\right]_{ii'}+\left[\matf{N}_{\nu}\right]_{ii'}\delta_{\nu\nu'}.
\eeq
Let us now prove the following
\beq
[\matf{C}^{-1}]_{i\nu,i'\nu}=[\matf{G}_{\nu,\nu'}]_{ii'},
\label{reload1}
\eeq
where
\beq
\matf{G}_{\nu,\nu'}=\matf{N}_{\nu}^{-1}\delta_{\nu,\nu'}-\matf{C_P}\matf{N}_{\nu}^{-1}\matf{B}_{\nu}\matf{S}^{1/2}\matf{D}^{-1}\matf{S}^{1/2}\matf{B}_{\nu'}\matf{N}_{\nu'}^{-1},\qquad
\matf{D}=1+\matf{S}^{1/2}\left(\sum_{\nu\in L_{\nu}}\matf{B}_{\nu}\matf{N}_{\nu}^{-1}\matf{B}_{\nu}\right)\matf{S}^{1/2}.
\label{reload2}
\eeq
That this is true may be easily checked:
\bea
\sum_{j\in L}\sum_{\mu\in L_{\nu}}[\matf{C}]_{i\nu,j\mu}[\matf{G}_{\mu,\rho}]_{jk}
&=&\sum_{j\in L}\sum_{\mu\in L_{\nu}}[\matf{B}_{\nu}\matf{S}\matf{B}_{\nu}]_{ij}[\matf{N}_{\mu}^{-1}]_{jk}\delta_{\mu,\rho}+[\matf{N}_{\nu}]_{ij}[\matf{N}_{\mu}^{-1}]_{jk}\delta_{\mu,\rho}\delta_{\nu\mu}\notag\\
&-&\sum_{j\in L}\sum_{\mu\in L_{\nu}}[\matf{N}_{\nu}]_{ij}[\matf{N}_{\mu}^{-1}\matf{B}_{\nu}\matf{S}^{1/2}\matf{D}^{-1}\matf{S}^{1/2}\matf{B}_{\nu}\matf{N}_{\rho}^{-1}]_{jk}\delta_{\nu\mu}\notag\\
&-&\sum_{j\in L}\sum_{\mu\in L_{\nu}}[\matf{B}_{\nu}\matf{S}\matf{B}_{\nu}]_{ij}[\matf{N}_{\mu}^{-1}\matf{B}_{\nu}\matf{S}^{1/2}\matf{D}^{-1}\matf{S}^{1/2}\matf{B}_{\nu}\matf{N}_{\rho}^{-1}]_{jk}\notag\\
&=&[\matf{B}_{\nu}\matf{S}\matf{B}_{\nu}\matf{N}_{\rho}^{-1}]_{ik}+[\matf{N}_{\nu}\matf{N}_{\rho}^{-1}]_{ik}\delta_{\nu\rho}-[\matf{B}_{\nu}\matf{S}^{1/2}\matf{D}^{-1}\matf{S}^{1/2}\matf{C_f}\matf{N}_{\rho}^{-1}]_{ik}\notag\\
&-&\left[\matf{B}_{\nu}\matf{S}^{1/2}\left\{1+\matf{S}^{1/2}\left(\sum_{\mu\in L_{\nu}}\matf{B}_{\nu}\matf{N}_{\rho}^{-1}\matf{B}_{\nu}\right)\matf{S}^{1/2}\right\}\matf{D}^{-1}\matf{S}^{1/2}\matf{B}_{\nu}\matf{N}_{\rho}^{-1}\right]_{ik}\notag\\
&+&[\matf{B}_{\nu}\matf{S}^{1/2}\matf{D}^{-1}\matf{S}^{1/2}\matf{B}_{\nu}\matf{N}_{\rho}^{-1}]_{ik}\notag\\
&=&[\matf{B}_{\nu}\matf{S}\matf{B}_{\nu}\matf{N}_{\rho}^{-1}]_{ik}+\delta_{ik}\delta_{\nu\rho}-[\matf{B}_{\nu}\matf{S}^{1/2}\matf{D}\matf{D}^{-1}\matf{S}^{1/2}\matf{B}_{\nu}\matf{N}_{\rho}^{-1}]_{ik}\notag\\
&=&\delta_{ik}\delta_{\nu\rho},
\eea
where we have used $\matf{\matf{S}^{1/2}}\matf{\matf{S}^{1/2}}=\matf{S}$, which may be checked using \eqref{integApp}.

\subsection{Fitting the Monopole and the Dipole}
\label{monDipFit}
Let us briefly discuss how any possible residual monopole and dipole CMB components can be accounted for in our framework. 
We denote monopole and dipole contributions by $\delta T^{\ell=0}$ and $\delta T^{\ell=1}$, respectively. They take the form
\begin{subequations}
  \begin{align}
\delta T^{\ell=0}(\hat{n}) &=a_{00}Y_{00}(\hat{n}), \\
\delta T^{\ell=1}(\hat{n}) &=a_{10}Y_{10}(\hat{n})+2\text{Re}(a_{11})\text{Re}(Y_{11}(\hat{n}))-2\text{Im}(a_{11})\text{Im}(Y_{11}(\hat{n})).
   \end{align}
\end{subequations}
These components should be added to the tSZ signal: $\dTsz(\vecf{\hat{n}},\nu)\to\dTsz(\vecf{\hat{n}},\nu)+\delta T^{\ell=0}(\hat{n})+\delta T^{\ell=1}(\hat{n})$. This may be conveniently done by making the following definitions:
\begin{subequations}
\beq
\Pu_{N_b+1}=a_{00}, \qquad
\Pu_{N_b+2}=a_{10}, \qquad
\Pu_{N_b+3}=\text{Re}(a_{11}), \qquad
\Pu_{N_b+4}=\text{Im}(a_{11}),
\eeq
\beq
t^{(\nu)}_{N_b+1}(\hat{n})=Y_{00}(\hat{n}), \qquad
t^{(\nu)}_{N_b+2}(\hat{n})=Y_{10}(\hat{n}), \qquad
t^{(\nu)}_{N_b+3}(\hat{n})=2\text{Re}(Y_{11}(\hat{n})), \qquad
t^{(\nu)}_{N_b+4}(\hat{n})=-2\text{Im}(Y_{11}(\hat{n})),
\eeq
\end{subequations}
where $N_b$ is the total number of radial bins.
The statistical machinery developed in Section \ref{Statistical Methods} now goes through exactly the same way, except that $N_b\to N_b+4$. 
Once the matrix $\matf{\alpha}$ is found (see \eqref{bestFit}), which is now $(N_b+4)\times(N_b+4)$ dimensional, the pressure covariance matrix 
becomes the restriction of its inverse to the bins of physical interest: $[\matf{C_{\Pu}}]_{kk'}=[\matf{\alpha}^{-1}]_{kk'}$ where $k,k'\in\{1,\dots,N_b\}$.

\section{Robustness Tests}
\label{convergence}
As it was shown in Section \ref{Numerical Methods}, 
the most important part of our analysis is solving a linear equation of the form $\matf{A}\vecf{x}=\vecf{b}$, where $\matf{A}$ is a very large ($\sim10^6\times 10^6$) matrix. In order to be certain that our numerical methods are correct, we perform two tests. 
\subsection{Simulations}
\begin{figure*}
	\begin{center}
		\includegraphics[width=0.45\hsize]{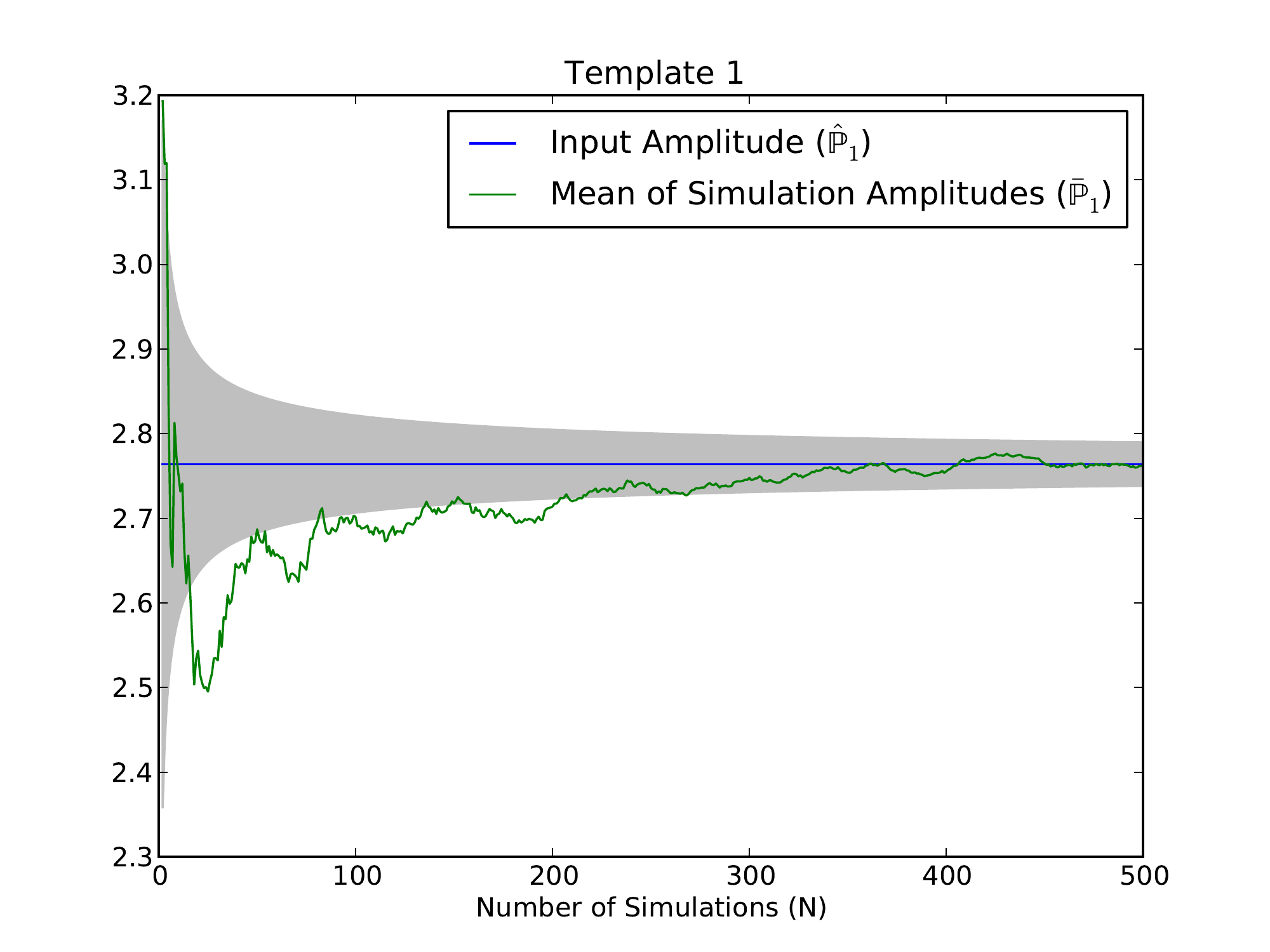}
		\includegraphics[width=0.45\hsize]{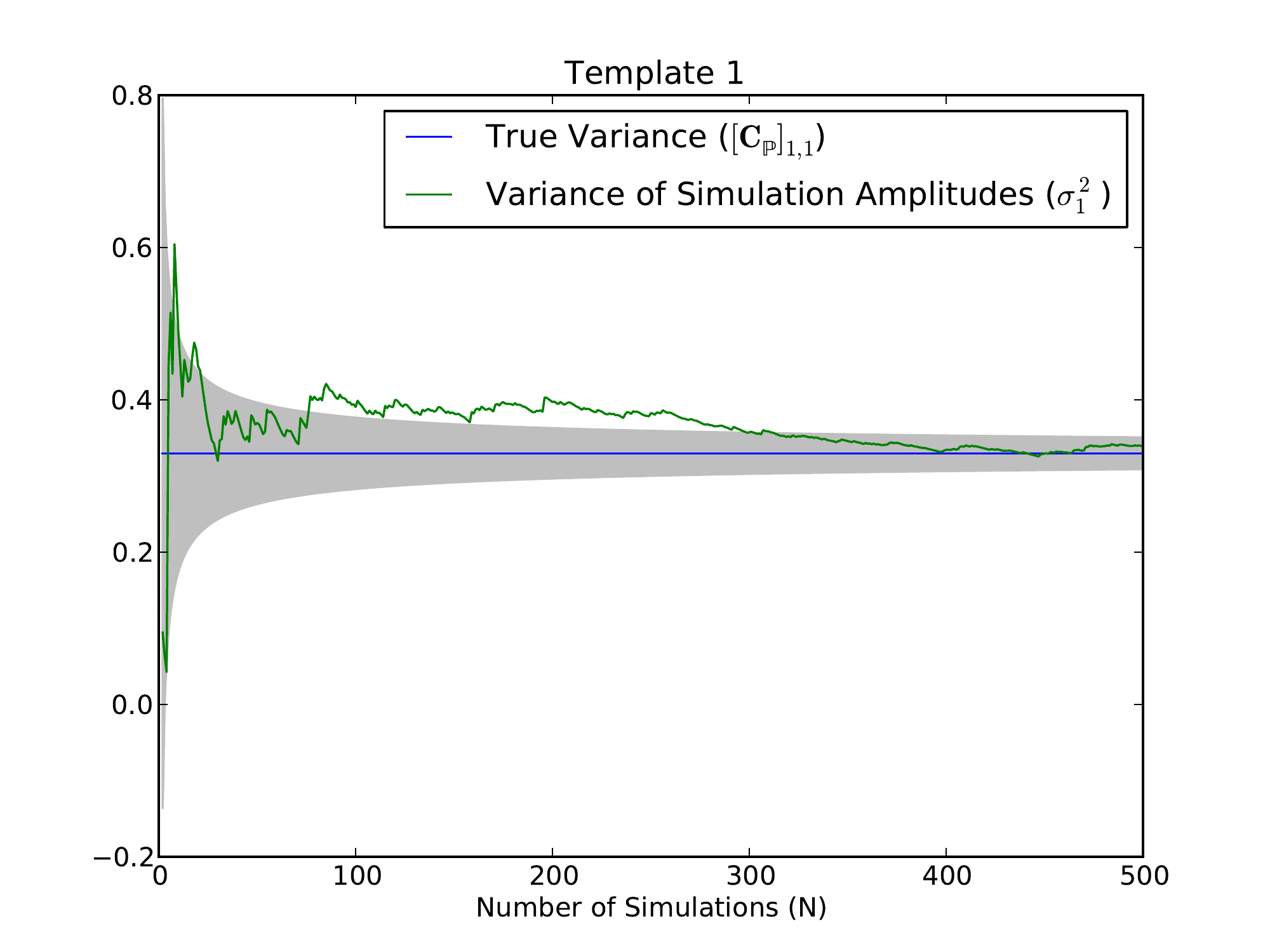}

		\includegraphics[width=0.45\hsize]{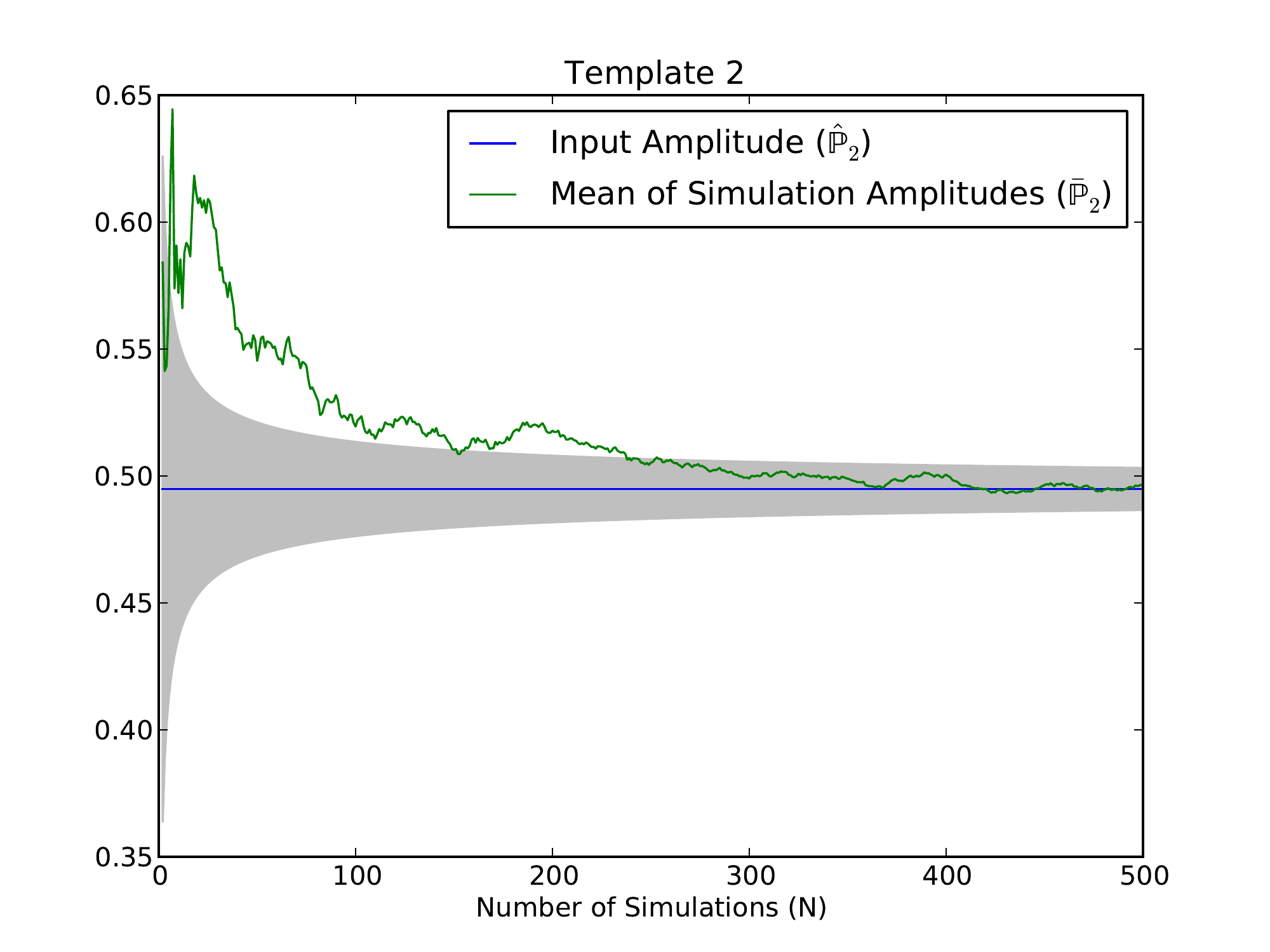}
		\includegraphics[width=0.45\hsize]{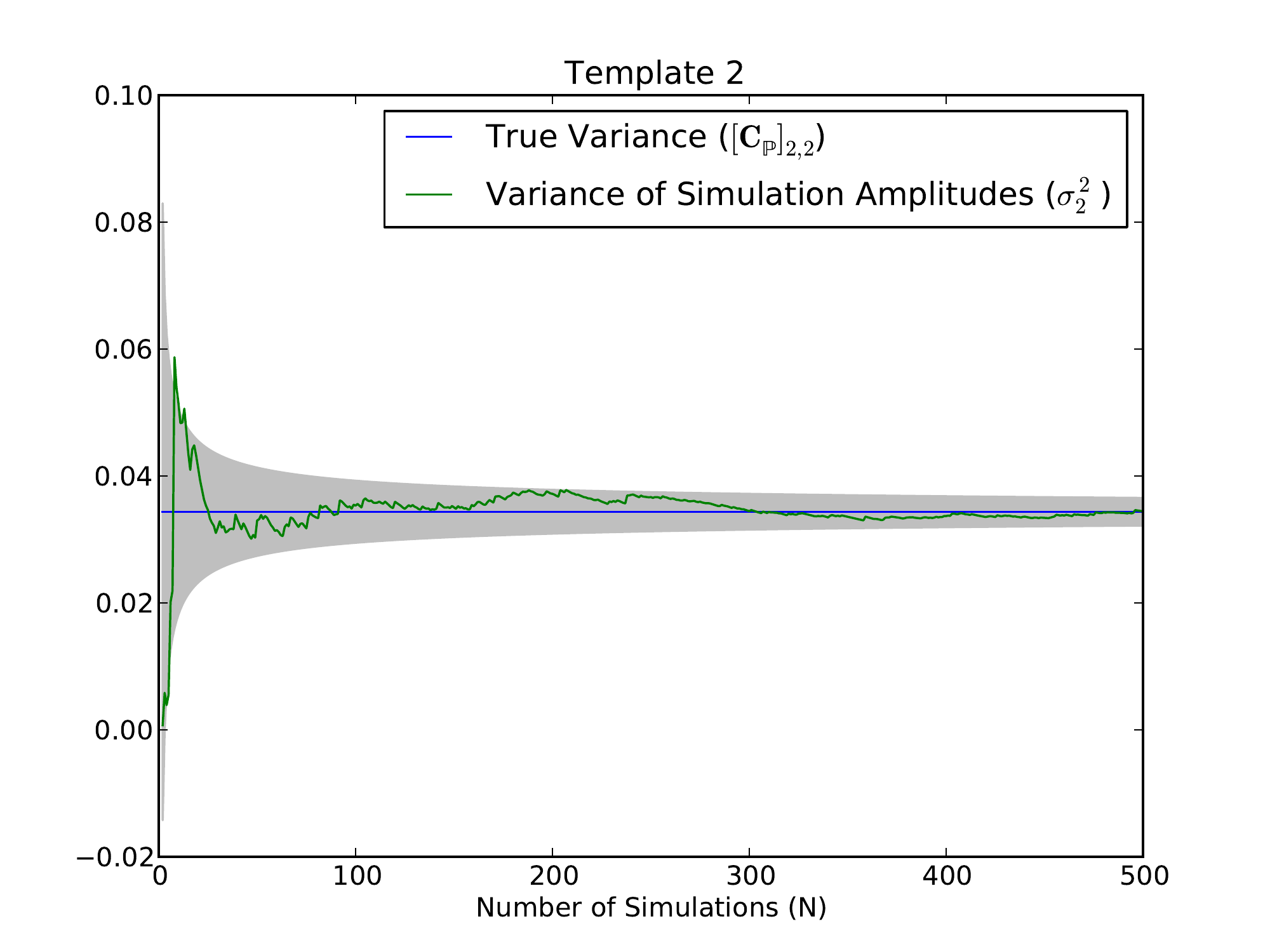}

		\includegraphics[width=0.45\hsize]{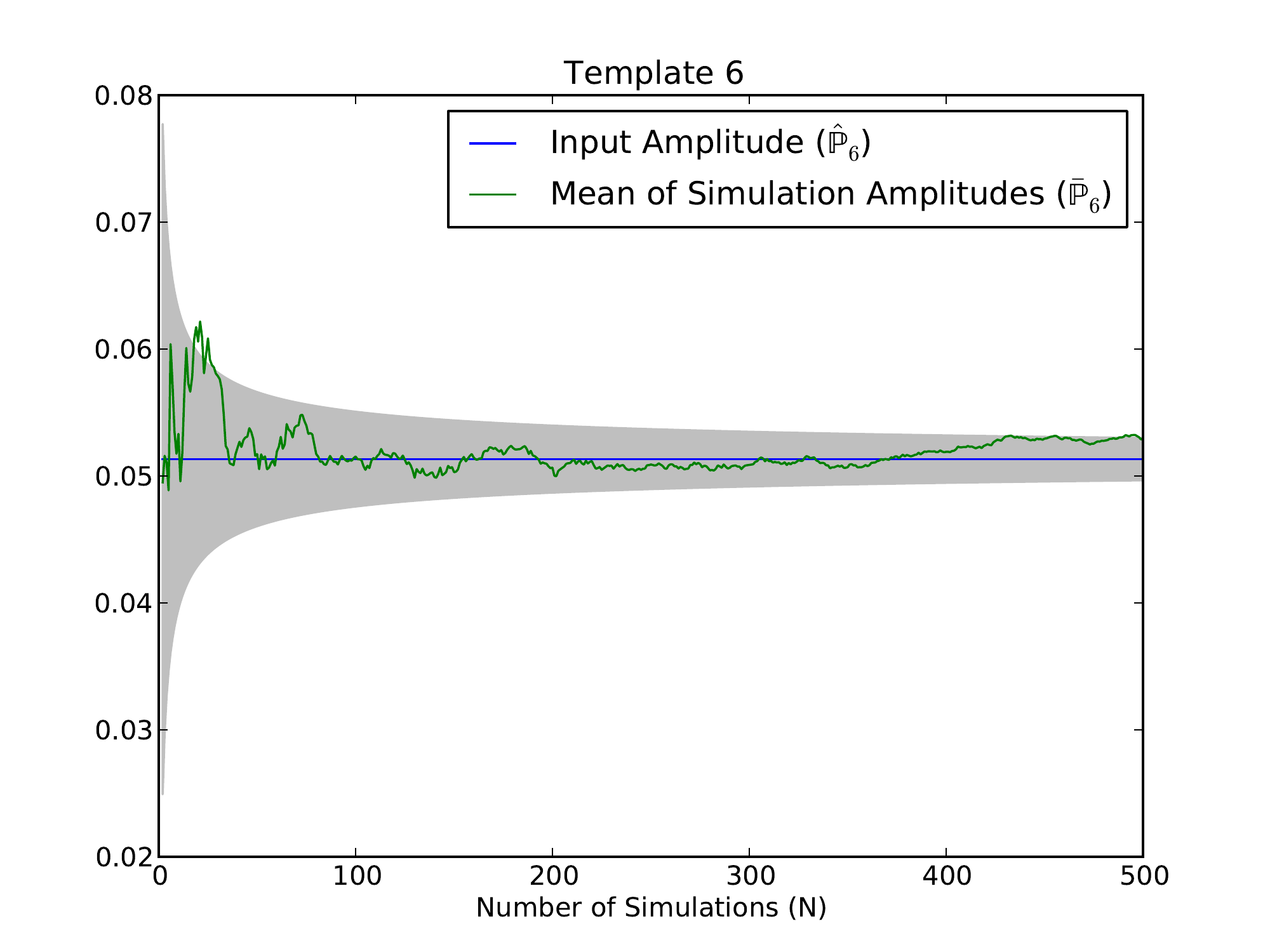}
		\includegraphics[width=0.45\hsize]{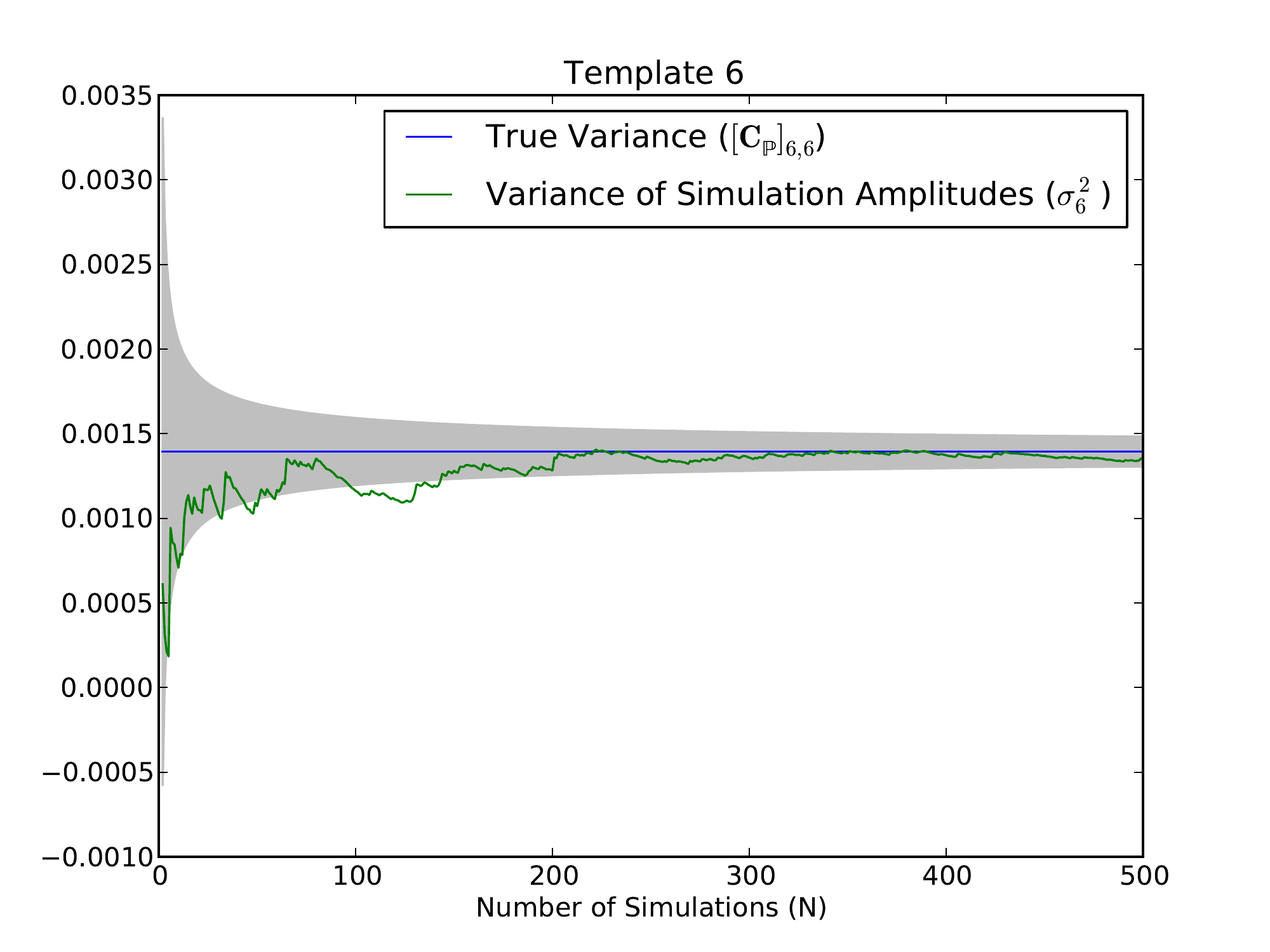}

	\end{center}
	\caption{Testing the pipeline with random primary CMB+noise simulations. The shaded area represents the expected error in the quantity of interest (see text for more details.) \label{simulations} }
\end{figure*}

We create sky maps with known tSZ amplitudes (i.e. the quantities of interest $\Pu_k$ and other parameters such as the monopole and dipole anisotropies) and see if the outcome of the pipeline matches with what is inputted. More specifically, we generate $N$ random realizations of the CMB primary anisotropies, add the tSZ signal with known amplitudes $\hat{\Pu}_k$, and finally add random detector noise. The outcome of every fitting procedure is the set of values $\Pu_k^{(i)}$, with $i\in\{1,\dots,N\}$ denoting the $i$-th simulation, and the covariance matrix $[\matf{C}_{\Pu}]_{k,k'}$ (see Section \ref{Statistical Methods}). The covariance matrix $[\matf{C}_{\Pu}]_{k,k'}$ does not change from one simulation to the other since it only depends on the tSZ templates, and the detector noise and primary CMB covariance matrices. If $\Pu_k^{(i)}$ really are realizations of a gaussian random variable with mean $\hat{\Pu}_k$ and variance $[\matf{C}_{\Pu}]_{k,k}$, then their mean $\overline{\Pu}_k=\frac{1}{N}\sum_{i=1}^{N}\Pu_k^{(i)}$ should converge to $\hat{\Pu}_k$ as $N$ becomes large. More specifically, the expected error in determining the true value of the mean is $\sqrt{<(\overline{\Pu}_k-\hat{\Pu}_k)^2>}=\sqrt{[\matf{C}_{\Pu}]_{k,k}/N}$. Similarly the best estimator of the variance $\sigma_k^2=\frac{1}{N-1}\sum_{i=1}^N(\Pu_k^{(i)}-\overline{\Pu}_k)^2$ should converge to $[\matf{C}_{\Pu}]_{k,k'}$, with an expected error of $\sqrt{<(\sigma_k^2-[\matf{C}_{\Pu}]_{k,k})^2>}=\sqrt{\frac{2}{N-1}}[\matf{C}_{\Pu}]_{k,k}$. Fig. \ref{simulations} shows the results of our simulations for a few templates (i.e. values of $k$). As it can be seen, all estimators converge to the values computed by our pipeline.

\subsection{Different Resolutions}
\label{DiffRes}
\begin{figure}
	\begin{center}
		\includegraphics[width=.6\hsize]{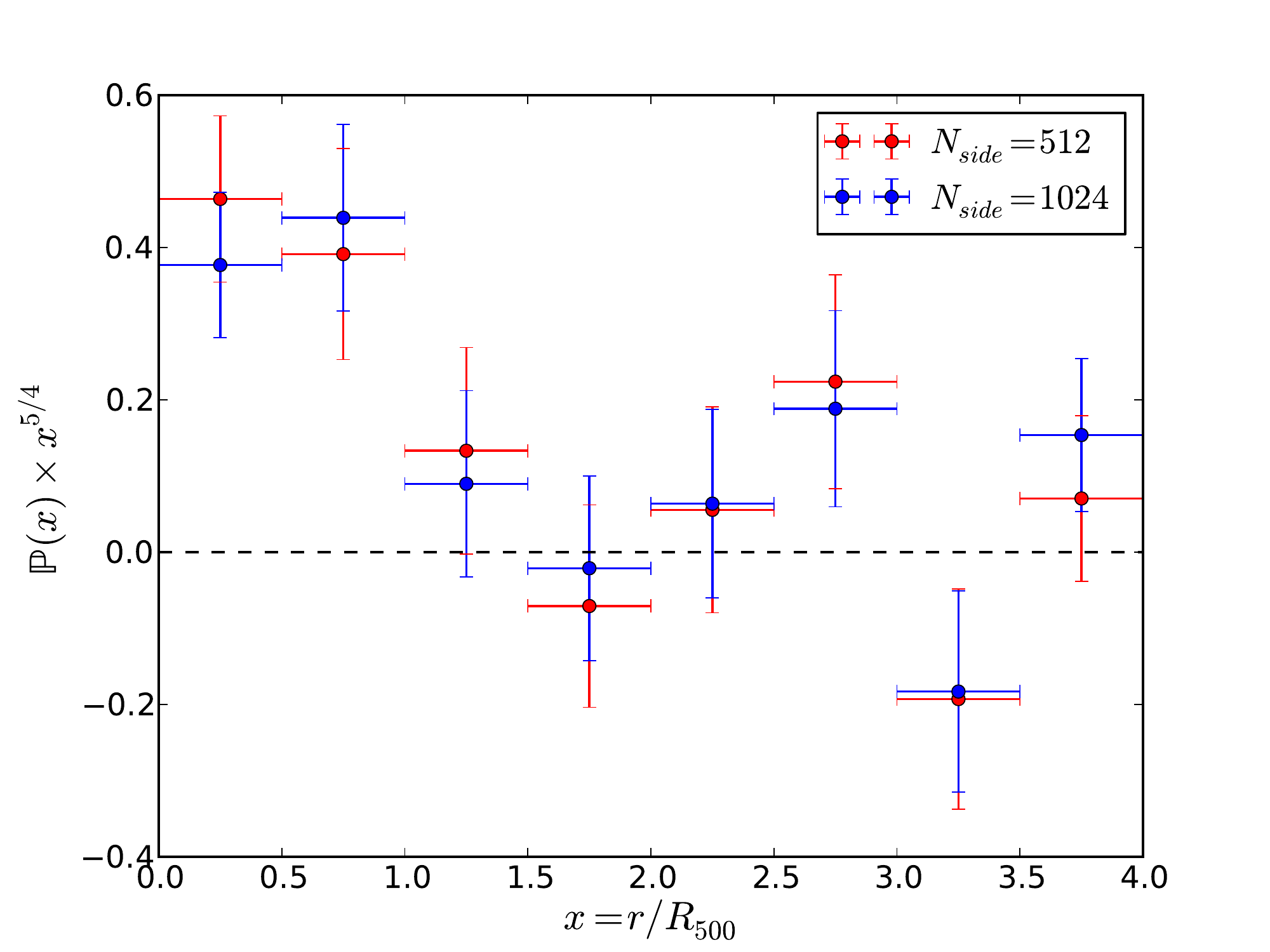}
	\end{center}
	\caption{Results of our analysis of WMAP7 sky maps at two different resolutions. \label{resComp}}
\end{figure}

We performed our analysis on all MCXC clusters using WMAP7 sky maps at two different HEALPix resolutions of $9$ ($N_{side}=512$) and $10$ ($N_{side}=1024$).
The modified self-similar scaling of pressure with mass is used for these measurements (i.e. $\delta=0.12$ in Equation \eqref{Pc}). 
Fig.~\ref{resComp} shows the results. The null chi-squared for the $N_{side}=512$ ($N_{side}=1024$) measurement is $\chi_0^2=246.8$ ($\chi_0^2=272.1$), corresponding to a 
$\bf 14.68\sigma$ ($\bf 15.50\sigma$) detection.

\section{Effect of Uncertainty in Mass of Clusters}
\label{massChange}
\begin{figure}
	\begin{subfigure}[t]{.48\hsize}
        		\includegraphics[width=\hsize]{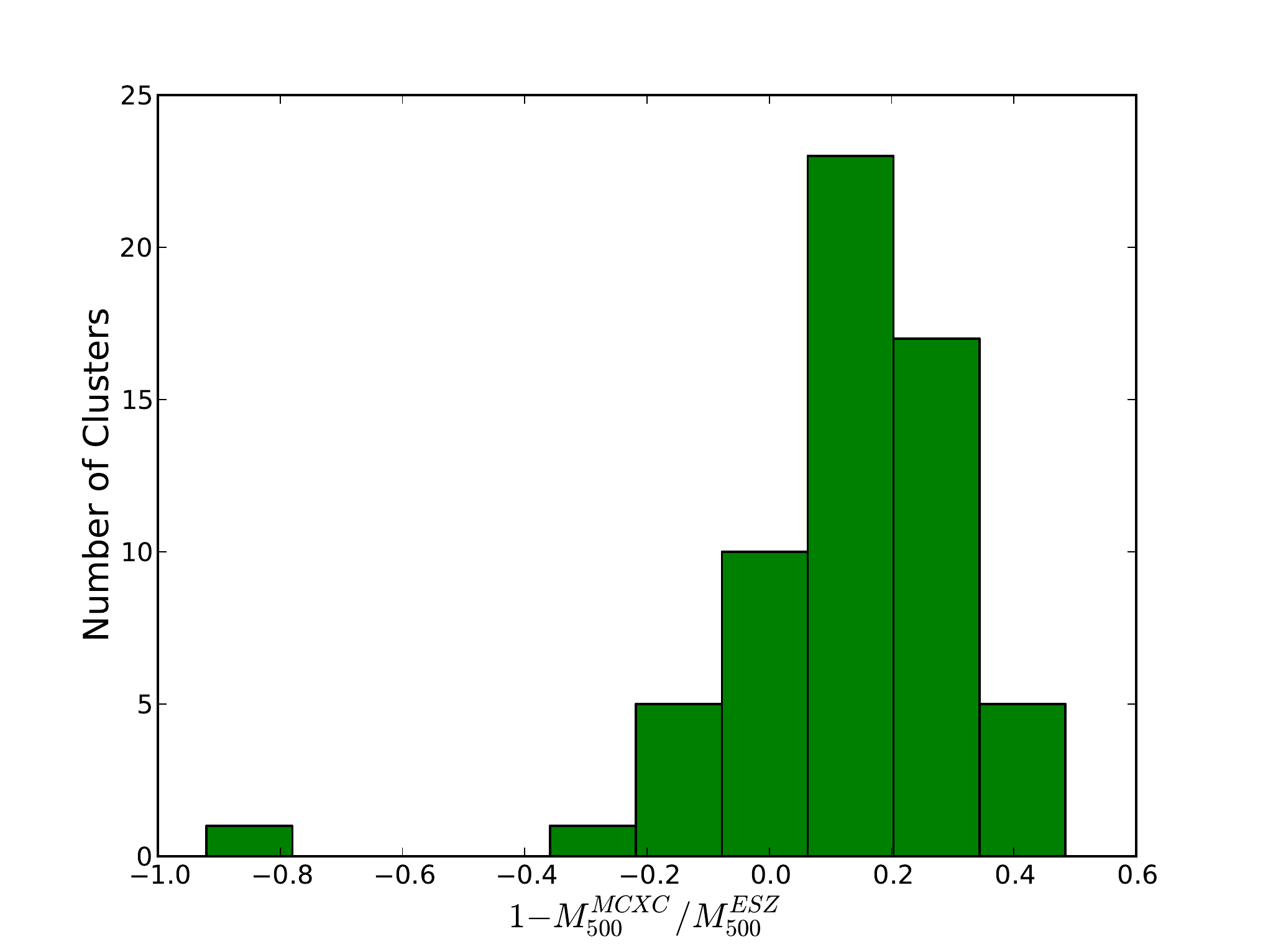}
		\caption{ESZ vs. MCXC mass estimates of 62 clusters common to both catalogues \citep{PlanckClusters}.}	
		\label{ESZvsMCXC}	
    	\end{subfigure} 
    \hfill
	\begin{subfigure}[t]{.48\hsize}
        		\includegraphics[width=\hsize]{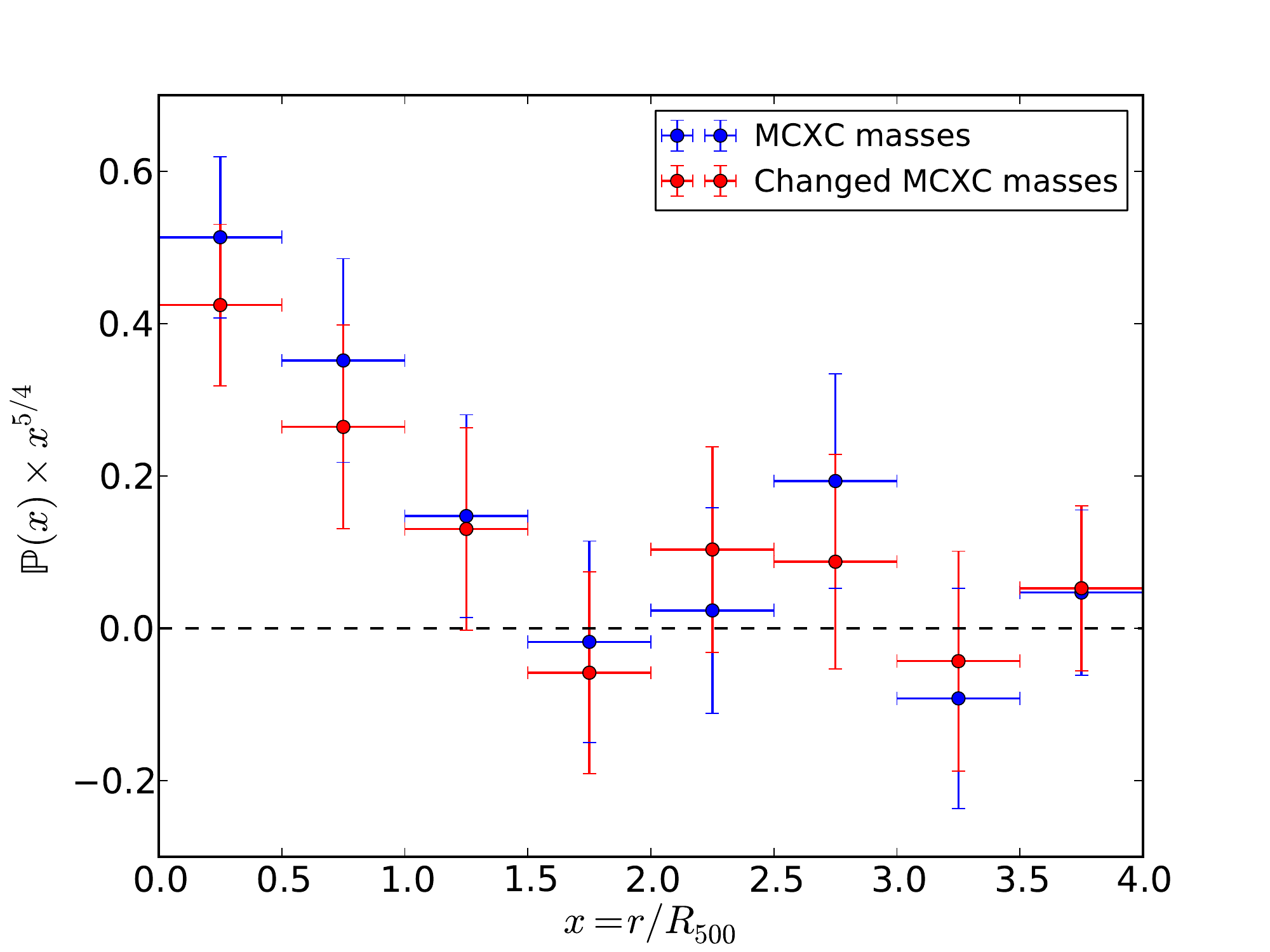}
		\caption{Effect of cluster mass uncertainties on the universal pressure profile. }
		\label{PmassChange}	
    	\end{subfigure} %
	\begin{center}
	\begin{subfigure}[t]{.48\hsize}
        		\includegraphics[width=\hsize]{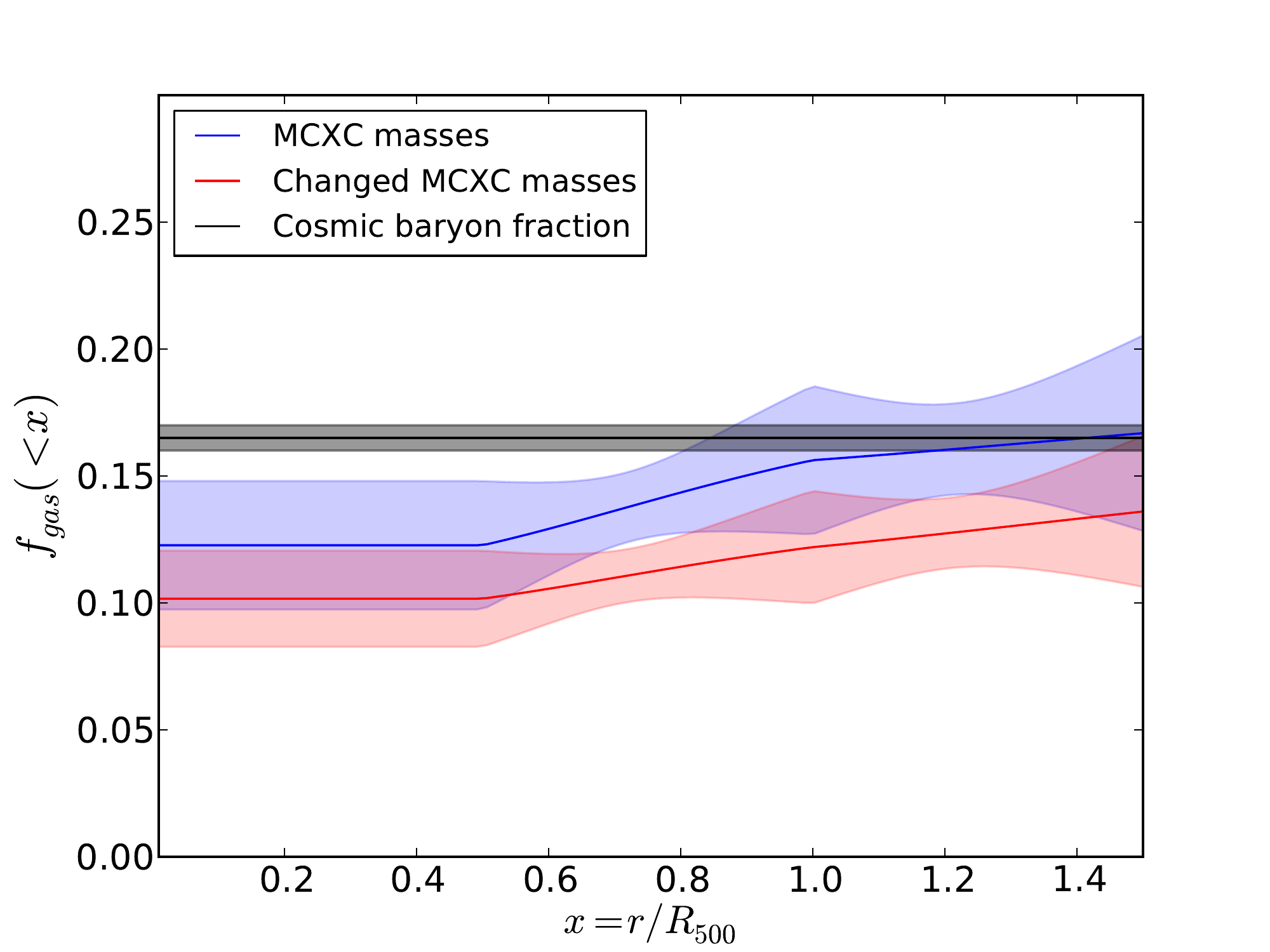}
		\caption{Effect of cluster mass uncertainties on gas mass profile. }
		\label{fGmassChange}	
    	\end{subfigure} %
	\end{center}
	\caption{
	Effect of systematic uncertainties associated with the mass of clusters. (a) Difference between ESZ and MCXC mass estimates of 62 clusters common to both catalogues 	\citep{PlanckClusters}.  (b) Effect of changing MCXC masses on the universal pressure profile. The entire MCXC sample is used and the standard self-similar model is assumed ($\delta=0$ in 	Equation \eqref{Pc}). The blue data points show the result of our measurements using MCXC mass estimates. The red points show measurements for which the MCXC masses are randomly changed according to the distribution shown in (a). (c) Gas mass fraction, computed using Equation \eqref{fgasMean}, corresponding to pressure measurements in (b). The shaded areas represent the standard deviation in the measurement of $f_{gas}$ as given by Equation \eqref{fgasErr}.
	}
	\label{fig: massChange}
\end{figure}

Our entire analysis depends crucially on the self-similarity length/mass scales of clusters. Therefore, it is important to investigate how
our measurements are affected by the uncertainty present in determining masses of clusters. 
\cite{collaboration2012planck} use 62 clusters
from the Early Release SZ (ESZ) sample \citep{ESZ} which also belong to the MCXC catalogue \citep{PlanckClusters}.
To get an idea for the degree of uncertainty present in mass estimates, we compare ESZ and MCXC masses of these clusters. This is shown in Fig. \ref{ESZvsMCXC}.
The ESZ mass estimates are systematically higher on average by about $12$ percent. To investigate how such systematics affect our pressure measurements, 
we randomly changed masses of all MCXC clusters according to the distribution in Fig. \ref{ESZvsMCXC}. 
(We used the standard self-similar scaling to create our templates, i.e. we set $\delta=0$ in Equation \eqref{Pc}.)
The resulting pressure profile is shown in Fig. \eqref{PmassChange}. In the first three bins, where there is signal, the pressure values decrease systematically. 
This difference, however, is at most at the 1$\sigma$ level. As shown in Fig. \ref{fGmassChange}, this is no longer the case for gas mass fraction, which decreases
by about $\bf 20$ percent on average. 

\section{All vs. Resolved MCXC clusters}
\label{EigenBusiness}
\begin{figure}
	\begin{subfigure}[b]{.48\hsize}
        		\includegraphics[width=\hsize]{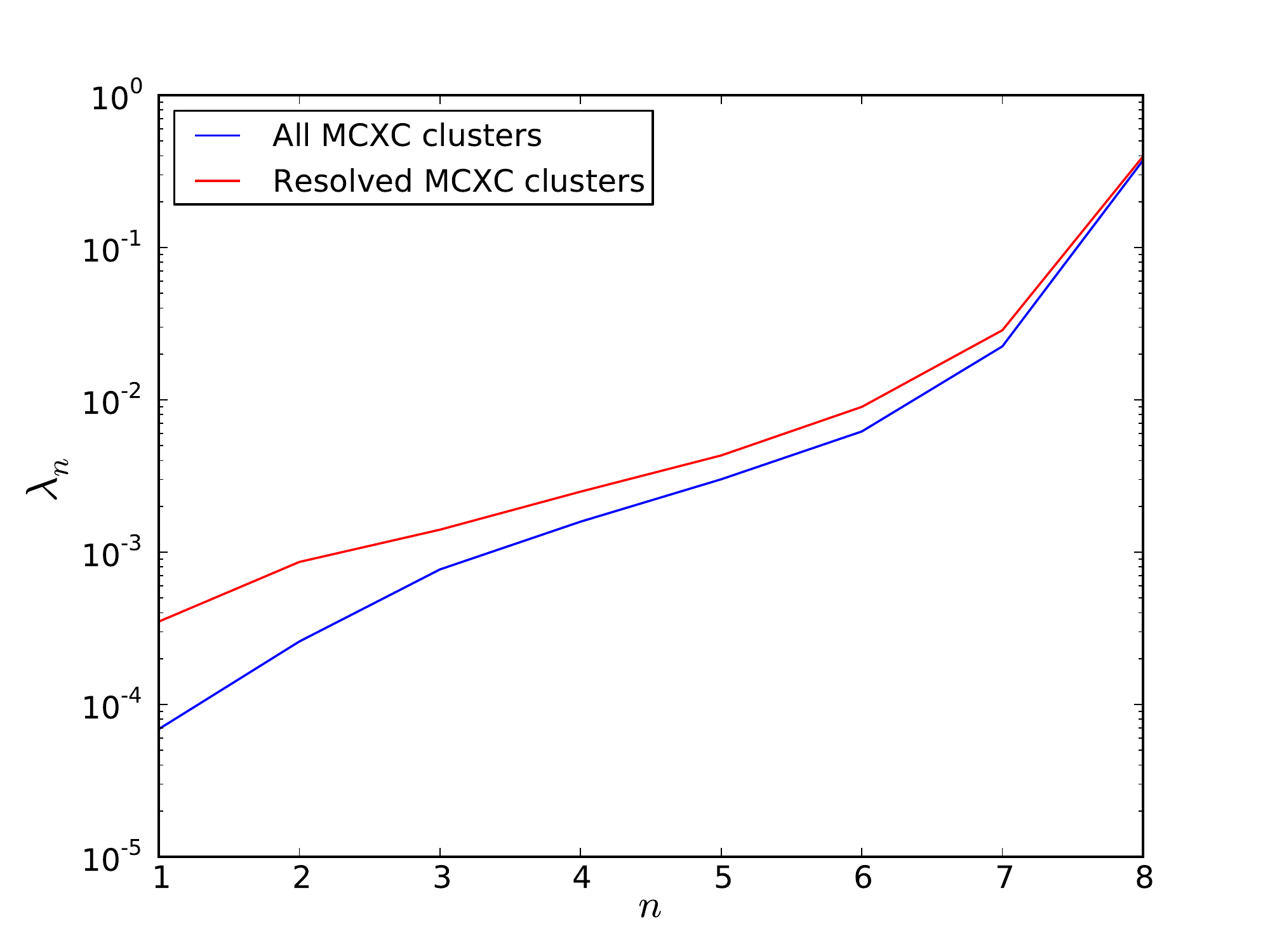}
		\caption{Eigenvalues of the covariance matrix $\matf{C}_{\Pu}$.}		
    	\end{subfigure} 
    \hfill
	\begin{subfigure}[b]{.48\hsize}
        		\includegraphics[width=\hsize]{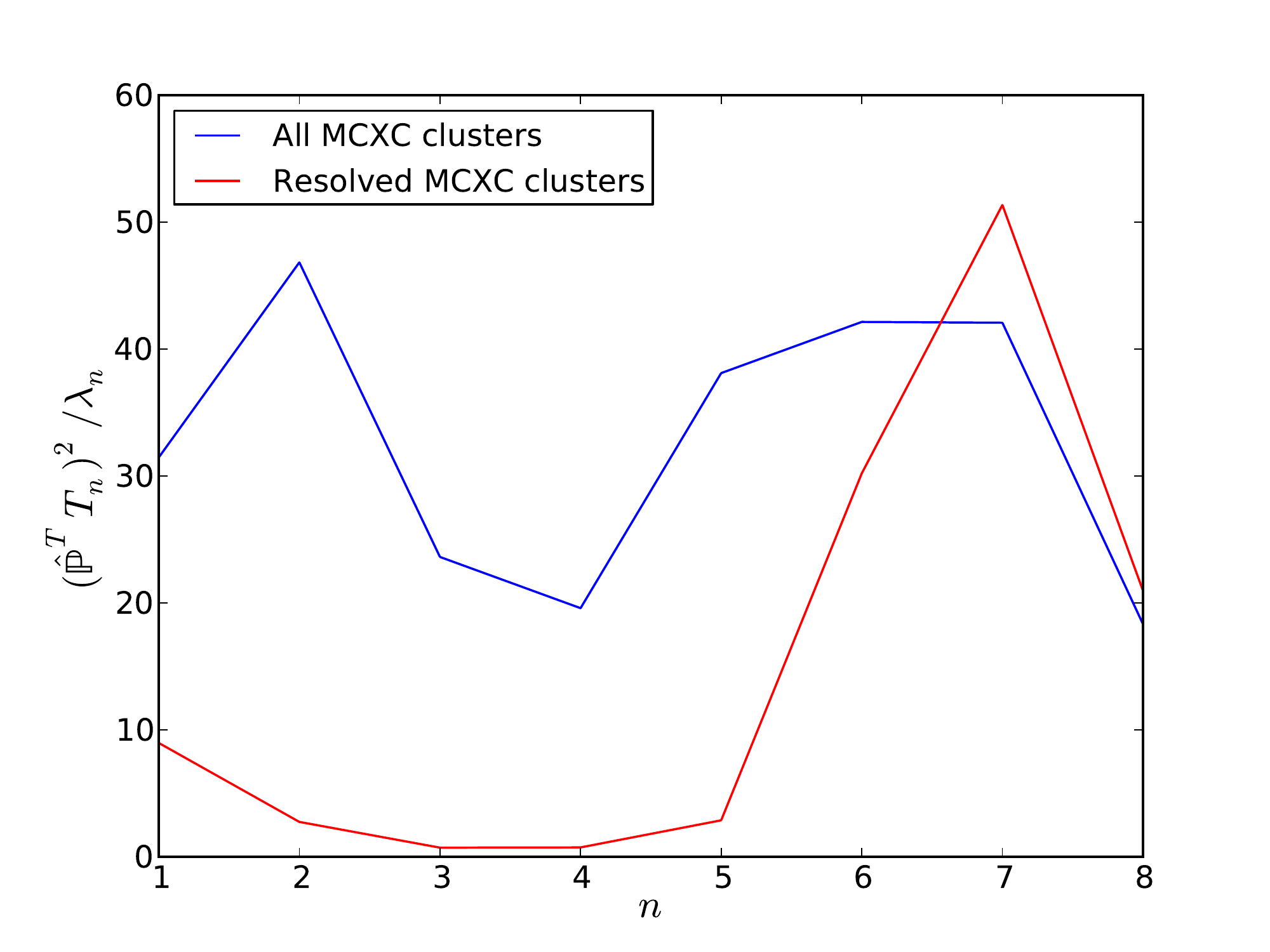}
		\caption{Contribution of eigenmodes of $\matf{C}_{\Pu}$ to $\chi_0^2$.}
    	\end{subfigure} %
	\caption{Spectrum of the covariance matrix and the contribution of  different eigenmodes to $\chi_0^2$.}
	\label{fig: eigValChiSq}
\end{figure}

\begin{figure}
	\begin{subfigure}[b]{.48\hsize}
       		\includegraphics[width=\hsize]{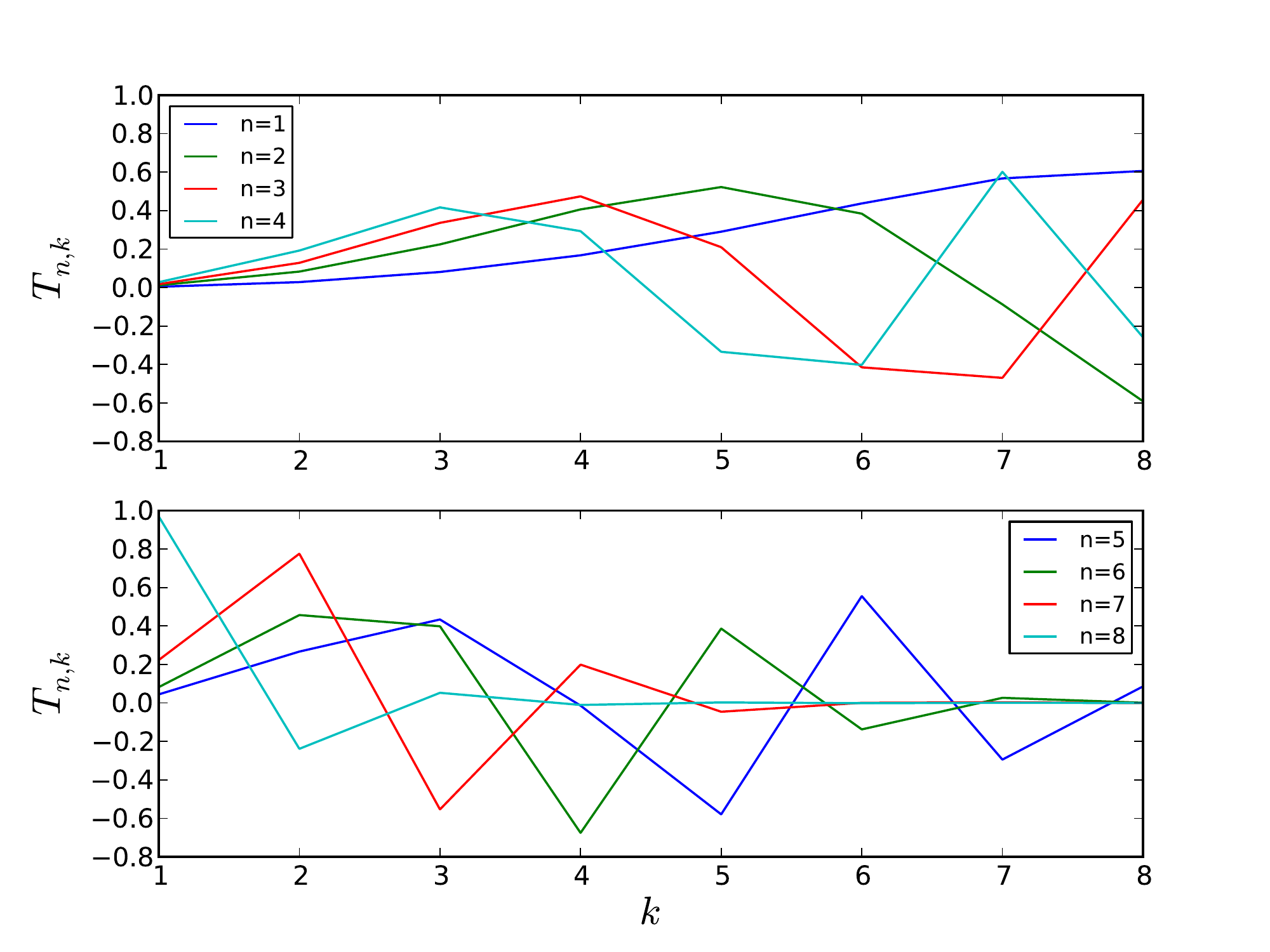}
		\caption{All MCXC clusters.}		
    	\end{subfigure} 
    \hfill
	\begin{subfigure}[b]{.48\hsize}
        		\includegraphics[width=\hsize]{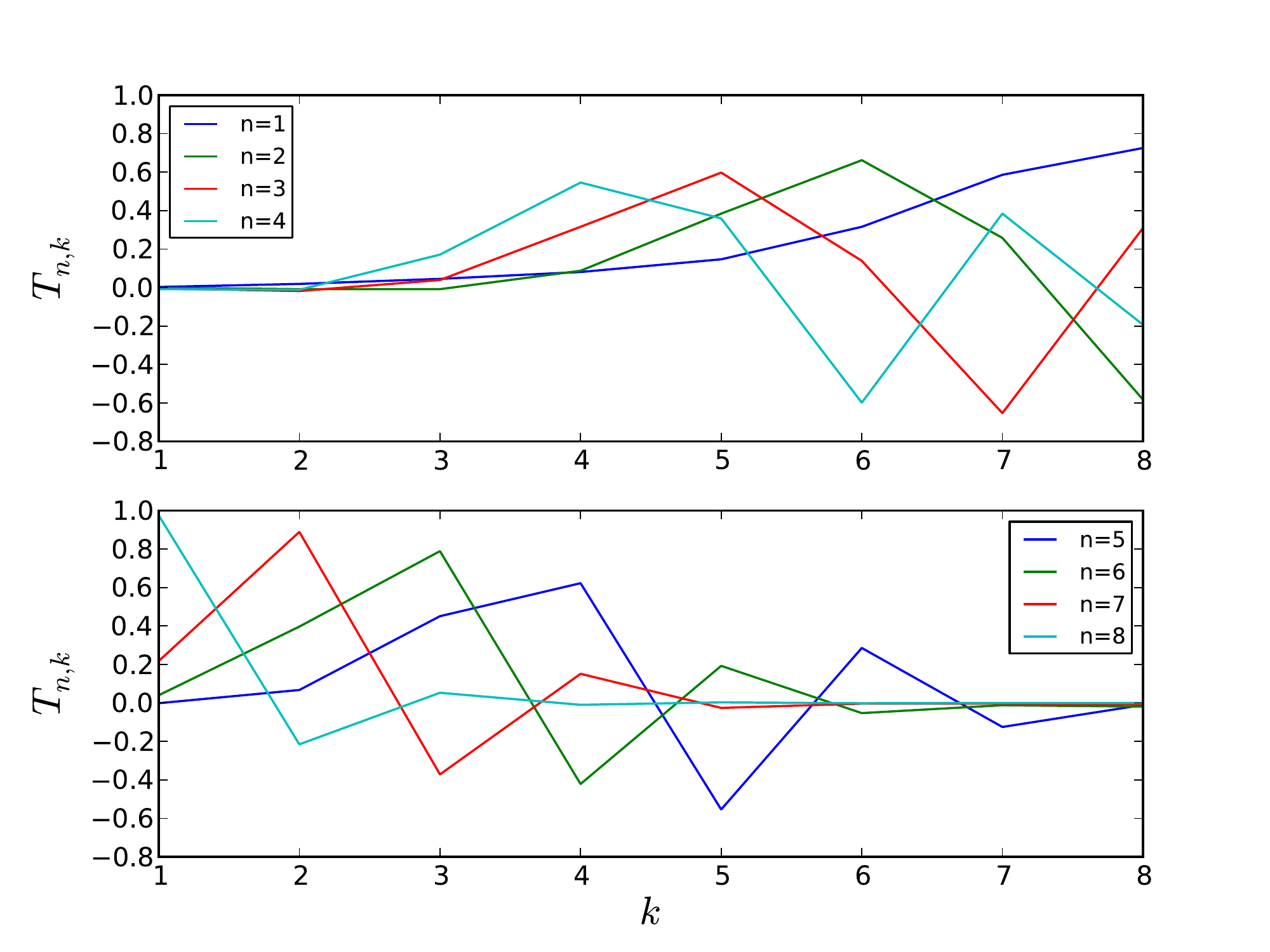}
		\caption{Resolved MCXC clusters.}
    	\end{subfigure} %
	\caption{Components $T_{n,k}$  of the eigenvectors $\vecf{T_n}$ of the covariance matrix. Here $n$ labels different eigenvectors, chosen so that the eigenvalues satisfy $\lambda_1<\lambda_2<\cdots<\lambda_{N_b}$. The label $k$ runs over all radial bins.}
	\label{fig: eigVecComp}
\end{figure}

Let us provide a quantitative explanation for how statistical information is lost when unresolved clusters are not accounted for. 
The discussion that will follow is based on measurements presented in Fig. \ref{delta12}, i.e. the modified scaling of pressure with mass ($\delta=0.12$ in Equation \eqref{Pc}). 
The same analysis for the standard self-similar scaling (i.e. measurements presented in Fig. \ref{delta12}) gives the same results.
Let $\lambda_n$ and $\vecf{T_n}$ denote the eigenvalues and eigenvectors of $\matf{C}_{\Pu}$, respectively. We choose the labels $n$ such that $\lambda_1<\lambda_2<\cdots<\lambda_{N_b}$, where $N_b=8$ is the total number of radial bins. Since $\matf{C}_{\Pu}$ is a positive symmetric matrix, its eigenvalues are positive and its eigenvectors are real. The null chi-squared can be re-expressed as
\beq
\chi_0^2=\hat{\vecf{\Pu}}^T\matf{C}_{\Pu}^{-1}\hat{\vecf{\Pu}}=\sum_{n=1}^{N_b}(\hat{\Pu}^T\vecf{T_n})^2/\lambda_n.
\eeq
Fig. \ref{fig: eigValChiSq} shows the eigenvalues $\lambda_n$ and the contribution $(\hat{\Pu}^T\vecf{T_n})^2/\lambda_n$ of the different eigenmodes to $\chi_0^2$.
In the case of the resolved clusters, the modes with the three largest eigenvalues are responsible for most of the contribution to $\chi_0^2$.
For the whole MCXC sample, however, all eigenmodes contribute more or less equally. Eigenvectors corresponding to larger eigenvalues carry most of their weight from the inner bins. To see this, we have plotted the components of all eigenvectors in Fig. \ref{fig: eigVecComp}. We denote the $k^{\text{th}}$ component of the eigenvector $\vecf{T_n}$ by $T_{n,k}$. As in the text,  $k=1\dots N_b$ labels the radial bins around clusters with $k=1$ and $k=N_b$ corresponding to the inner and outer-most bins, respectively. It is clear from Fig. \ref{fig: eigVecComp} that for eigenvectors with larger eigenvalues, the components corresponding to the inner bins dominate, and vice versa. Therefore, in the case of resolved MCXC clusters, the fact that most of the contribution to $\chi_0^2$ comes from modes with the three largest eigenvalues indicates that mostly the inner bins are contributing to the signal. Whereas for the whole MCXC sample, there is also contribution from outer bins. This analysis reassures our intuition that even unresolved clusters contribute to the tSZ signal in the outskirts of the ICM.

\section{Pressure Measurements: the Exact Numbers}
\label{app:ExactNums}
Here we report exact numbers corresponding to the measurement of $\Pu$, i.e. the best fit universal pressure value $\hat{\Pu}$ and its associated covariance matrix $\matf{C}_\Pu$.
As before, $\delta$ quantifies deviation from the standard self similar model (see Equation \eqref{Pc}).
We give in Table \ref{allMCXC_no12} (\ref{resolvedMCXC_no12}) $\hat{\Pu}$ and $\matf{C}_\Pu$ in the case of all clusters (resolved clusters), with $\delta=0$. Similarly, Table \ref{allMCXC_12} (\ref{resolvedMCXC_12}) shows our measurements in the case of all clusters (resolved clusters), with $\delta=0.12$. 
In Tables \ref{allMCXC_M1}$-$\ref{allMCXC_M4}, we show the result of our analysis on mass bins $1-4$
introduced in Table \ref{tab:massBin_all}, respectively (with $\delta=0$). Similarly, Tables \ref{resolvedMCXC_M1}$-$\ref{resolvedMCXC_M3} provide the best fit universal pressure values and the corresponding covariance matrix for mass bins $1-3$
introduced in Table \ref{tab:massBin_162}, respectively (with $\delta=0$). 

\begin{table}
	\caption{Pressure measurement of all MCXC clusters with the standard self-similar model ($\delta=0$ in Equation \eqref{Pc}).}
	\centering
	\begin{subtable}{\linewidth}
		\centering
		\begin{tabular}{c|c|c|c|c|c|c|c|c}
			\hline
			 &  Bin 1 & Bin 2 & Bin 3 & Bin 4 & Bin 5 & Bin 6 & Bin 7 & Bin 8 \\
\hline
Pressure &  2904.845 & 503.878 & 111.528 & -8.831 & 8.515 & 54.610 & -21.088 & 8.990 \\

			\hline
		\end{tabular}
		\caption{Best fit pressure values $\hat{\Pu}$ ($\times 10^{-3}$)}
	\end{subtable}%
	\\
	\begin{subtable}{\linewidth}
		\centering
		\begin{tabular}{c|c|c|c|c|c|c|c|c}
			\hline
			 &  Bin 1 & Bin 2 & Bin 3 & Bin 4 & Bin 5 & Bin 6 & Bin 7 & Bin 8 \\
\hline
 Bin 1 &  359.047 & -82.359 & 16.095 & -2.989 & 0.872 & -0.540 & 0.220 & -0.016 \\
 Bin 2 &  -82.359 & 36.703 & -12.792 & 2.632 & -0.470 & 0.052 & -0.023 & 0.018 \\
 Bin 3 &  16.095 & -12.792 & 10.142 & -4.167 & 0.720 & -0.010 & -0.067 & 0.031 \\
 Bin 4 &  -2.989 & 2.632 & -4.167 & 4.320 & -1.928 & 0.298 & 0.010 & -0.017 \\
 Bin 5 &  0.872 & -0.470 & 0.720 & -1.928 & 2.405 & -1.162 & 0.200 & -0.008 \\
 Bin 6 &  -0.540 & 0.052 & -0.010 & 0.298 & -1.162 & 1.577 & -0.786 & 0.132 \\
 Bin 7 &  0.220 & -0.023 & -0.067 & 0.010 & 0.200 & -0.786 & 1.094 & -0.476 \\
 Bin 8 &  -0.016 & 0.018 & 0.031 & -0.017 & -0.008 & 0.132 & -0.476 & 0.432 \\

			\hline
		\end{tabular}
		\caption{Covariance matrix $\matf{C}_\Pu$ ($\times 10^{-3}$) }
	\end{subtable}%
	\label{allMCXC_no12}
\end{table}

\begin{table}
	\caption{Pressure measurement of resolved MCXC clusters with the standard self-similar model ($\delta=0$ in Equation \eqref{Pc}).}
	\centering
	\begin{subtable}{\linewidth}
		\centering
		\begin{tabular}{c|c|c|c|c|c|c|c|c}
			\hline
			 &  Bin 1 & Bin 2 & Bin 3 & Bin 4 & Bin 5 & Bin 6 & Bin 7 & Bin 8 \\
\hline
Pressure &  3156.051 & 652.465 & 163.366 & 2.700 & 48.387 & 78.075 & -14.598 & 14.782 \\

			\hline
		\end{tabular}
		\caption{Best fit pressure values $\hat{\Pu}$ ($\times 10^{-3}$)}
	\end{subtable}%
	\\
	\begin{subtable}{\linewidth}
		\centering
		\begin{tabular}{c|c|c|c|c|c|c|c|c}
			\hline
			 &  Bin 1 & Bin 2 & Bin 3 & Bin 4 & Bin 5 & Bin 6 & Bin 7 & Bin 8 \\
\hline
 Bin 1 &  385.004 & -76.888 & 18.073 & -2.839 & 1.073 & -0.664 & -0.018 & -0.199 \\
 Bin 2 &  -76.888 & 42.837 & -11.045 & 3.328 & -0.439 & -0.052 & -0.177 & -0.240 \\
 Bin 3 &  18.073 & -11.045 & 11.960 & -3.454 & 0.846 & -0.087 & -0.154 & -0.136 \\
 Bin 4 &  -2.839 & 3.328 & -3.454 & 5.082 & -1.641 & 0.291 & -0.054 & -0.141 \\
 Bin 5 &  1.073 & -0.439 & 0.846 & -1.641 & 2.785 & -1.022 & 0.214 & -0.084 \\
 Bin 6 &  -0.664 & -0.052 & -0.087 & 0.291 & -1.022 & 1.812 & -0.677 & 0.102 \\
 Bin 7 &  -0.018 & -0.177 & -0.154 & -0.054 & 0.214 & -0.677 & 1.281 & -0.466 \\
 Bin 8 &  -0.199 & -0.240 & -0.136 & -0.141 & -0.084 & 0.102 & -0.466 & 0.745 \\

			\hline
		\end{tabular}
		\caption{Covariance matrix $\matf{C}_\Pu$ ($\times 10^{-3}$) }
	\end{subtable}%
	\label{resolvedMCXC_no12}
\end{table}
\begin{table}
	\caption{Pressure measurement of all MCXC clusters with the modified self-similar model ($\delta=0.12$ in Equation \eqref{Pc}).}
	\centering
	\begin{subtable}{\linewidth}
		\centering
		\begin{tabular}{c|c|c|c|c|c|c|c|c}
			\hline
			 &  Bin 1 & Bin 2 & Bin 3 & Bin 4 & Bin 5 & Bin 6 & Bin 7 & Bin 8 \\
\hline
Pressure &  2819.487 & 515.355 & 97.656 & -15.799 & 4.248 & 50.901 & -22.110 & 8.053 \\

			\hline
		\end{tabular}
		\caption{Best fit pressure values $\hat{\Pu}$ ($\times 10^{-3}$)}
	\end{subtable}%
	\\
	\begin{subtable}{\linewidth}
		\centering
		\begin{tabular}{c|c|c|c|c|c|c|c|c}
			\hline
			 &  Bin 1 & Bin 2 & Bin 3 & Bin 4 & Bin 5 & Bin 6 & Bin 7 & Bin 8 \\
\hline
 Bin 1 &  352.226 & -82.197 & 16.578 & -3.117 & 0.820 & -0.496 & 0.211 & -0.018 \\
 Bin 2 &  -82.197 & 36.333 & -12.687 & 2.618 & -0.467 & 0.035 & -0.018 & 0.023 \\
 Bin 3 &  16.578 & -12.687 & 9.842 & -4.026 & 0.683 & 0.001 & -0.075 & 0.036 \\
 Bin 4 &  -3.117 & 2.618 & -4.026 & 4.114 & -1.830 & 0.269 & 0.019 & -0.019 \\
 Bin 5 &  0.821 & -0.467 & 0.683 & -1.830 & 2.273 & -1.093 & 0.179 & -0.004 \\
 Bin 6 &  -0.496 & 0.035 & 0.001 & 0.268 & -1.093 & 1.487 & -0.740 & 0.121 \\
 Bin 7 &  0.211 & -0.018 & -0.075 & 0.019 & 0.179 & -0.740 & 1.034 & -0.451 \\
 Bin 8 &  -0.018 & 0.023 & 0.036 & -0.019 & -0.004 & 0.121 & -0.451 & 0.405 \\

			\hline
		\end{tabular}
		\caption{Covariance matrix $\matf{C}_\Pu$ ($\times 10^{-3}$) }
	\end{subtable}%
	\label{allMCXC_12}
\end{table}

\begin{table}
	\caption{Pressure measurement of resolved MCXC clusters with the modified self-similar model ($\delta=0.12$ in Equation \eqref{Pc}).}
	\centering
	\begin{subtable}{\linewidth}
		\centering
		\begin{tabular}{c|c|c|c|c|c|c|c|c}
			\hline
			 &  Bin 1 & Bin 2 & Bin 3 & Bin 4 & Bin 5 & Bin 6 & Bin 7 & Bin 8 \\
\hline
Pressure &  3095.361 & 672.793 & 159.082 & 0.880 & 46.378 & 75.526 & -16.262 & 8.843 \\

			\hline
		\end{tabular}
		\caption{Best fit pressure values $\hat{\Pu}$ ($\times 10^{-3}$)}
	\end{subtable}%
	\\
	\begin{subtable}{\linewidth}
		\centering
		\begin{tabular}{c|c|c|c|c|c|c|c|c}
			\hline
			 &  Bin 1 & Bin 2 & Bin 3 & Bin 4 & Bin 5 & Bin 6 & Bin 7 & Bin 8 \\
\hline
 Bin 1 &  378.015 & -77.464 & 18.369 & -2.951 & 1.042 & -0.619 & -0.015 & -0.185 \\
 Bin 2 &  -77.463 & 42.440 & -11.029 & 3.331 & -0.412 & -0.058 & -0.166 & -0.237 \\
 Bin 3 &  18.369 & -11.029 & 11.612 & -3.340 & 0.811 & -0.060 & -0.153 & -0.126 \\
 Bin 4 &  -2.951 & 3.331 & -3.340 & 4.854 & -1.554 & 0.263 & -0.036 & -0.140 \\
 Bin 5 &  1.042 & -0.412 & 0.811 & -1.554 & 2.639 & -0.959 & 0.193 & -0.072 \\
 Bin 6 &  -0.619 & -0.058 & -0.060 & 0.263 & -0.959 & 1.713 & -0.638 & 0.094 \\
 Bin 7 &  -0.015 & -0.166 & -0.153 & -0.036 & 0.193 & -0.638 & 1.214 & -0.442 \\
 Bin 8 &  -0.185 & -0.237 & -0.126 & -0.140 & -0.072 & 0.094 & -0.442 & 0.711 \\

			\hline
		\end{tabular}
		\caption{Covariance matrix $\matf{C}_\Pu$ ($\times 10^{-3}$) }
	\end{subtable}%
	\label{resolvedMCXC_12}
\end{table}

\begin{table}
	\caption{Pressure measurement of Mass-bin $1$ of all MCXC clusters, as defined in Table \ref{tab:massBin_all} ($\delta=0$ in Equation \eqref{Pc}).}
	\centering
	\begin{subtable}{\linewidth}
		\centering
		\begin{tabular}{c|c|c|c|c|c|c|c|c}
			\hline
			 &  Bin 1 & Bin 2 & Bin 3 & Bin 4 & Bin 5 & Bin 6 & Bin 7 & Bin 8 \\
\hline
Pressure &  2864.897 & 143.372 & 165.443 & 82.563 & 57.910 & -22.757 & 53.503 & 37.140 \\

			\hline
		\end{tabular}
		\caption{Best fit pressure values $\hat{\Pu}$ ($\times 10^{-3}$)}
	\end{subtable}%
	\\
	\begin{subtable}{\linewidth}
		\centering
		\begin{tabular}{c|c|c|c|c|c|c|c|c}
			\hline
			 &  Bin 1 & Bin 2 & Bin 3 & Bin 4 & Bin 5 & Bin 6 & Bin 7 & Bin 8 \\
\hline
 Bin 1 &  1351.387 & -266.076 & 30.865 & 0.085 & 1.519 & -0.727 & -0.089 & 0.199 \\
 Bin 2 &  -266.075 & 117.062 & -37.249 & 5.581 & 0.115 & 0.093 & -0.088 & -0.025 \\
 Bin 3 &  30.866 & -37.249 & 34.247 & -13.720 & 2.080 & 0.136 & 0.038 & -0.086 \\
 Bin 4 &  0.084 & 5.582 & -13.720 & 16.293 & -7.307 & 1.239 & 0.070 & -0.055 \\
 Bin 5 &  1.520 & 0.115 & 2.080 & -7.307 & 9.706 & -4.636 & 0.856 & -0.025 \\
 Bin 6 &  -0.728 & 0.093 & 0.136 & 1.239 & -4.636 & 6.452 & -3.190 & 0.559 \\
 Bin 7 &  -0.089 & -0.088 & 0.038 & 0.070 & 0.856 & -3.190 & 4.424 & -1.869 \\
 Bin 8 &  0.199 & -0.025 & -0.086 & -0.055 & -0.025 & 0.559 & -1.869 & 1.832 \\

			\hline
		\end{tabular}
		\caption{Covariance matrix $\matf{C}_\Pu$ ($\times 10^{-3}$) }
	\end{subtable}%
	\label{allMCXC_M1}	
\end{table}

\begin{table}
	\caption{Pressure measurement of Mass-bin $2$ of all MCXC clusters, as defined in Table \ref{tab:massBin_all} ($\delta=0$ in Equation \eqref{Pc}).}
	\centering
	\begin{subtable}{\linewidth}
		\centering
		\begin{tabular}{c|c|c|c|c|c|c|c|c}
			\hline
			 &  Bin 1 & Bin 2 & Bin 3 & Bin 4 & Bin 5 & Bin 6 & Bin 7 & Bin 8 \\
\hline
Pressure &  2546.299 & 226.786 & 253.102 & 18.308 & -79.667 & 204.195 & -123.737 & -7.187 \\

			\hline
		\end{tabular}
		\caption{Best fit pressure values $\hat{\Pu}$ ($\times 10^{-3}$)}
	\end{subtable}%
	\\
	\begin{subtable}{\linewidth}
		\centering
		\begin{tabular}{c|c|c|c|c|c|c|c|c}
			\hline
			 &  Bin 1 & Bin 2 & Bin 3 & Bin 4 & Bin 5 & Bin 6 & Bin 7 & Bin 8 \\
\hline
 Bin 1 &  2504.452 & -719.122 & 230.700 & -84.783 & 37.468 & -19.648 & 10.328 & -3.445 \\
 Bin 2 &  -719.123 & 311.542 & -137.925 & 54.001 & -22.478 & 10.146 & -4.269 & 1.271 \\
 Bin 3 &  230.702 & -137.925 & 99.347 & -53.731 & 22.694 & -9.263 & 3.633 & -0.946 \\
 Bin 4 &  -84.785 & 54.002 & -53.731 & 46.988 & -27.135 & 11.806 & -4.740 & 1.355 \\
 Bin 5 &  37.469 & -22.479 & 22.695 & -27.135 & 26.089 & -15.997 & 6.810 & -1.871 \\
 Bin 6 &  -19.648 & 10.146 & -9.264 & 11.806 & -15.997 & 16.275 & -9.554 & 2.819 \\
 Bin 7 &  10.328 & -4.269 & 3.633 & -4.740 & 6.809 & -9.554 & 9.325 & -3.905 \\
 Bin 8 &  -3.445 & 1.271 & -0.946 & 1.355 & -1.871 & 2.819 & -3.905 & 2.673 \\

			\hline
		\end{tabular}
		\caption{Covariance matrix $\matf{C}_\Pu$ ($\times 10^{-3}$) }
	\end{subtable}%
	\label{allMCXC_M2}	
\end{table}

\begin{table}
	\caption{Pressure measurement of Mass-bin $3$ of all MCXC clusters, as defined in Table \ref{tab:massBin_all} ($\delta=0$ in Equation \eqref{Pc}).}
	\centering
	\begin{subtable}{\linewidth}
		\centering
		\begin{tabular}{c|c|c|c|c|c|c|c|c}
			\hline
			 &  Bin 1 & Bin 2 & Bin 3 & Bin 4 & Bin 5 & Bin 6 & Bin 7 & Bin 8 \\
\hline
Pressure &  2614.687 & 599.174 & -119.300 & -23.585 & 9.475 & 161.234 & -148.958 & 139.492 \\

			\hline
		\end{tabular}
		\caption{Best fit pressure values $\hat{\Pu}$ ($\times 10^{-3}$)}
	\end{subtable}%
	\\
	\begin{subtable}{\linewidth}
		\centering
		\begin{tabular}{c|c|c|c|c|c|c|c|c}
			\hline
			 &  Bin 1 & Bin 2 & Bin 3 & Bin 4 & Bin 5 & Bin 6 & Bin 7 & Bin 8 \\
\hline
 Bin 1 &  816.335 & -142.117 & 21.841 & -11.060 & 7.061 & -4.547 & 1.150 & 0.328 \\
 Bin 2 &  -142.117 & 90.898 & -31.204 & 6.198 & -3.173 & 1.336 & -0.726 & 0.394 \\
 Bin 3 &  21.841 & -31.204 & 33.339 & -13.358 & 2.133 & -0.887 & 0.478 & -0.212 \\
 Bin 4 &  -11.060 & 6.198 & -13.358 & 16.314 & -6.910 & 1.096 & -0.559 & 0.262 \\
 Bin 5 &  7.061 & -3.173 & 2.133 & -6.911 & 9.575 & -4.265 & 0.700 & -0.205 \\
 Bin 6 &  -4.547 & 1.336 & -0.887 & 1.096 & -4.265 & 6.279 & -2.804 & 0.305 \\
 Bin 7 &  1.150 & -0.726 & 0.478 & -0.559 & 0.700 & -2.804 & 4.368 & -1.827 \\
 Bin 8 &  0.328 & 0.394 & -0.212 & 0.262 & -0.205 & 0.305 & -1.827 & 1.997 \\

			\hline
		\end{tabular}
		\caption{Covariance matrix $\matf{C}_\Pu$ ($\times 10^{-3}$) }
	\end{subtable}%
	\label{allMCXC_M3}	
\end{table}

\begin{table}
	\caption{Pressure measurement of Mass-bin $4$ of all MCXC clusters, as defined in Table \ref{tab:massBin_all} ($\delta=0$ in Equation \eqref{Pc}).}
	\centering
	\begin{subtable}{\linewidth}
		\centering
		\begin{tabular}{c|c|c|c|c|c|c|c|c}
			\hline
			 &  Bin 1 & Bin 2 & Bin 3 & Bin 4 & Bin 5 & Bin 6 & Bin 7 & Bin 8 \\
\hline
Pressure &  4522.423 & 811.544 & 140.014 & -63.051 & -16.867 & 10.731 & 60.649 & -94.528 \\

			\hline
		\end{tabular}
		\caption{Best fit pressure values $\hat{\Pu}$ ($\times 10^{-3}$)}
	\end{subtable}%
	\\
	\begin{subtable}{\linewidth}
		\centering
		\begin{tabular}{c|c|c|c|c|c|c|c|c}
			\hline
			 &  Bin 1 & Bin 2 & Bin 3 & Bin 4 & Bin 5 & Bin 6 & Bin 7 & Bin 8 \\
\hline
 Bin 1 &  4322.774 & -1224.992 & 307.886 & -40.992 & -10.662 & 8.444 & -1.387 & -0.802 \\
 Bin 2 &  -1224.992 & 434.752 & -126.640 & 19.097 & 2.756 & -3.787 & 0.775 & 0.135 \\
 Bin 3 &  307.888 & -126.641 & 59.055 & -15.870 & 0.103 & 1.643 & -0.966 & 0.207 \\
 Bin 4 &  -40.993 & 19.097 & -15.870 & 13.835 & -4.773 & -0.305 & 0.650 & -0.317 \\
 Bin 5 &  -10.663 & 2.756 & 0.103 & -4.773 & 6.955 & -2.669 & -0.122 & 0.219 \\
 Bin 6 &  8.445 & -3.788 & 1.643 & -0.305 & -2.669 & 4.461 & -1.854 & 0.077 \\
 Bin 7 &  -1.387 & 0.775 & -0.966 & 0.650 & -0.122 & -1.854 & 3.298 & -1.417 \\
 Bin 8 &  -0.803 & 0.135 & 0.207 & -0.316 & 0.219 & 0.077 & -1.417 & 1.642 \\

			\hline
		\end{tabular}
		\caption{Covariance matrix $\matf{C}_\Pu$ ($\times 10^{-3}$) }
	\end{subtable}%
	\label{allMCXC_M4}
\end{table}

\begin{table}
	\caption{Pressure measurement of Mass-bin $1$ of resolved MCXC clusters, as defined in Table \ref{tab:massBin_162} ($\delta=0$ in Equation \eqref{Pc}).}
	\centering
	\begin{subtable}{\linewidth}
		\centering
		\begin{tabular}{c|c|c|c|c|c|c|c|c}
			\hline
			 &  Bin 1 & Bin 2 & Bin 3 & Bin 4 & Bin 5 & Bin 6 & Bin 7 & Bin 8 \\
\hline
Pressure &  2970.143 & -124.936 & 58.889 & 102.331 & 97.592 & 54.978 & 75.993 & 94.229 \\

			\hline
		\end{tabular}
		\caption{Best fit pressure values $\hat{\Pu}$ ($\times 10^{-3}$)}
	\end{subtable}%
	\\
	\begin{subtable}{\linewidth}
		\centering
		\begin{tabular}{c|c|c|c|c|c|c|c|c}
			\hline
			 &  Bin 1 & Bin 2 & Bin 3 & Bin 4 & Bin 5 & Bin 6 & Bin 7 & Bin 8 \\
\hline
 Bin 1 &  1375.649 & -236.216 & 47.033 & -2.420 & 2.119 & -1.271 & -1.240 & -0.716 \\
 Bin 2 &  -236.214 & 129.290 & -31.665 & 8.192 & -0.684 & -0.393 & -0.512 & -0.764 \\
 Bin 3 &  47.036 & -31.667 & 38.119 & -11.362 & 2.634 & -0.653 & -0.387 & -0.557 \\
 Bin 4 &  -2.420 & 8.192 & -11.361 & 17.521 & -6.171 & 1.401 & -0.481 & -0.385 \\
 Bin 5 &  2.120 & -0.685 & 2.634 & -6.171 & 10.149 & -4.066 & 1.023 & -0.461 \\
 Bin 6 &  -1.271 & -0.392 & -0.653 & 1.401 & -4.066 & 6.674 & -2.656 & 0.468 \\
 Bin 7 &  -1.240 & -0.512 & -0.387 & -0.481 & 1.023 & -2.656 & 4.630 & -1.700 \\
 Bin 8 &  -0.716 & -0.764 & -0.557 & -0.385 & -0.461 & 0.468 & -1.700 & 2.578 \\

			\hline
		\end{tabular}
		\caption{Covariance matrix $\matf{C}_\Pu$ ($\times 10^{-3}$) }
	\end{subtable}%
	\label{resolvedMCXC_M1}
\end{table}

\begin{table}
	\caption{Pressure measurement of Mass-bin $2$ of resolved MCXC clusters, as defined in Table \ref{tab:massBin_162} ($\delta=0$ in Equation \eqref{Pc}).}
	\centering
	\begin{subtable}{\linewidth}
		\centering
		\begin{tabular}{c|c|c|c|c|c|c|c|c}
			\hline
			 &  Bin 1 & Bin 2 & Bin 3 & Bin 4 & Bin 5 & Bin 6 & Bin 7 & Bin 8 \\
\hline
Pressure &  2482.111 & 747.796 & -29.912 & -32.605 & 2.269 & 180.980 & -115.173 & 74.383 \\

			\hline
		\end{tabular}
		\caption{Best fit pressure values $\hat{\Pu}$ ($\times 10^{-3}$)}
	\end{subtable}%
	\\
	\begin{subtable}{\linewidth}
		\centering
		\begin{tabular}{c|c|c|c|c|c|c|c|c}
			\hline
			 &  Bin 1 & Bin 2 & Bin 3 & Bin 4 & Bin 5 & Bin 6 & Bin 7 & Bin 8 \\
\hline
 Bin 1 &  682.399 & -103.800 & 28.794 & -8.645 & 4.760 & -4.283 & 0.648 & -0.521 \\
 Bin 2 &  -103.799 & 88.006 & -19.527 & 7.560 & -3.040 & 0.400 & -1.156 & -0.276 \\
 Bin 3 &  28.794 & -19.527 & 31.489 & -8.677 & 2.714 & -1.348 & -0.072 & -0.722 \\
 Bin 4 &  -8.645 & 7.560 & -8.676 & 15.190 & -4.710 & 1.212 & -0.803 & -0.205 \\
 Bin 5 &  4.760 & -3.040 & 2.714 & -4.710 & 8.745 & -3.022 & 0.850 & -0.533 \\
 Bin 6 &  -4.283 & 0.400 & -1.348 & 1.212 & -3.022 & 5.652 & -1.930 & 0.398 \\
 Bin 7 &  0.648 & -1.156 & -0.072 & -0.803 & 0.850 & -1.930 & 3.906 & -1.247 \\
 Bin 8 &  -0.521 & -0.276 & -0.722 & -0.205 & -0.533 & 0.398 & -1.247 & 2.349 \\

			\hline
		\end{tabular}
		\caption{Covariance matrix $\matf{C}_\Pu$ ($\times 10^{-3}$) }
	\end{subtable}%
	\label{resolvedMCXC_M2}
\end{table}

\begin{table}
	\caption{Pressure measurement of Mass-bin $3$ of resolved MCXC clusters, as defined in Table \ref{tab:massBin_162} ($\delta=0$ in Equation \eqref{Pc}).}
	\begin{subtable}{\linewidth}
		\centering
		\begin{tabular}{c|c|c|c|c|c|c|c|c}
			\hline
			 &  Bin 1 & Bin 2 & Bin 3 & Bin 4 & Bin 5 & Bin 6 & Bin 7 & Bin 8 \\
\hline
Pressure &  3737.228 & 1220.556 & 334.228 & 7.300 & 34.974 & 23.814 & 7.113 & -101.074 \\

			\hline
		\end{tabular}
		\caption{Best fit pressure values $\hat{\Pu}$ ($\times 10^{-3}$)}
	\end{subtable}%
	\\
	\begin{subtable}{\linewidth}
		\centering
		\begin{tabular}{c|c|c|c|c|c|c|c|c}
			\hline
			 &  Bin 1 & Bin 2 & Bin 3 & Bin 4 & Bin 5 & Bin 6 & Bin 7 & Bin 8 \\
\hline
 Bin 1 &  3842.297 & -1117.928 & 295.139 & -49.881 & -6.830 & 7.859 & -1.586 & -0.158 \\
 Bin 2 &  -1117.928 & 417.687 & -116.036 & 26.214 & 2.461 & -3.534 & 0.417 & -1.168 \\
 Bin 3 &  295.139 & -116.036 & 58.291 & -14.939 & 1.537 & 1.550 & -1.007 & -0.033 \\
 Bin 4 &  -49.882 & 26.214 & -14.939 & 14.707 & -4.168 & 0.113 & 0.563 & -0.664 \\
 Bin 5 &  -6.829 & 2.461 & 1.537 & -4.168 & 6.989 & -2.360 & 0.175 & 0.093 \\
 Bin 6 &  7.858 & -3.534 & 1.550 & 0.113 & -2.360 & 4.449 & -1.665 & 0.203 \\
 Bin 7 &  -1.586 & 0.417 & -1.007 & 0.563 & 0.175 & -1.665 & 3.262 & -1.232 \\
 Bin 8 &  -0.158 & -1.168 & -0.033 & -0.664 & 0.093 & 0.203 & -1.232 & 2.008 \\

			\hline
		\end{tabular}
		\caption{Covariance matrix $\matf{C}_\Pu$ ($\times 10^{-3}$) }
	\end{subtable}%
	\label{resolvedMCXC_M3}	
\end{table}

\end{document}